\documentclass[twocolumn ]{aastex62}

\newcommand{\CIV}{C~{\sc iv}}

\newcommand{\OIII}{[O~{\sc iii}]}

\newcommand{\simgt}{\lower 2pt \hbox{$\, \buildrel {\scriptstyle >}\over {\scriptstyle\sim}\,$}}
\newcommand{\simlt}{\lower 2pt \hbox{$\, \buildrel {\scriptstyle <}\over {\scriptstyle\sim}\,$}}

\newcommand{\xmm}{\emph{XMM-Newton}}
\newcommand{\hs}{HS~0810+2554}

\newcommand{\apm}{APM~08279+5255}

\newcommand{\aox}{\hbox{$\alpha_{\rm ox}$}}
\newcommand{\Am}{\hbox{\tiny \AA}}

\received{January 23, 2021}
\accepted{June 25, 2021}

\submitjournal{ApJ}

\shorttitle{Quasar Winds}
\shortauthors{Chartas et al.}
\begin{document}

%\runningpagewiselinenumbers
%\linenumbers

\def\sarc{$^{\prime\prime}\!\!.$}

%\title{Relativistic Quasar Accretion Disk Winds}
\title{Multiphase Powerful Outflows Detected in High-z Quasars}

\correspondingauthor{G. Chartas}
\email{chartasg@cofc.edu}

\author[0000-0003-1697-6596]{G. Chartas}
\affil{Department of Physics and Astronomy, College of Charleston, Charleston, SC, 29424, USA}

\author{M. Cappi}
\affiliation{INAF, Osservatorio di Astrofisica e Scienza dello Spazio di Bologna, via P. Gobetti 93/3, 40129 Bologna, Italy}

\author{C. Vignali}
\affiliation{Dipartimento di Fisica e Astronomia dell'Universit\`{a} degli Studi di Bologna, via P. Gobetti 93/2, 40129 Bologna, Italy}
\affiliation{INAF, Osservatorio di Astrofisica e Scienza dello Spazio di Bologna, via P. Gobetti 93/3, 40129 Bologna, Italy}

\author{M. Dadina}
\affiliation{INAF, Osservatorio di Astrofisica e Scienza dello Spazio di Bologna, via P. Gobetti 93/3, 40129 Bologna, Italy}

\author{V. James}
\affiliation{Department of Physics and Astronomy, College of Charleston, Charleston, SC, 29424, USA}

\author{G. Lanzuisi}
\affiliation{INAF, Osservatorio di Astrofisica e Scienza dello Spazio di Bologna, via P. Gobetti 93/3, 40129 Bologna, Italy}

%\affiliation{Dipartimento di Fisica e Astronomia dell'Universit\`{a} degli Studi di Bologna, via P. Gobetti 93/2, 40129 Bologna, Italy}

\author{M. Giustini}
\affiliation{Centro de Astrobiolog\'ia (CSIC-INTA), Camino Bajo del Castillo s/n, Villanueva de la Ca\~nada, E-28692 Madrid, Spain}

\author{M. Gaspari}
\affiliation{INAF, Osservatorio di Astrofisica e Scienza dello Spazio di Bologna, via P. Gobetti 93/3, 40129 Bologna, Italy}
\affiliation{Dept. of Astrophysical Sciences, Princeton University, 4 Ivy Lane, Princeton, NJ 08544, USA }
%Lyman Spitzer Jr. Fellow

\author{S. Strickland}
\affiliation{Department of Physics and Astronomy, College of Charleston, Charleston, SC, 29424, USA}

\author{E. Bertola}
\affiliation{INAF, Osservatorio di Astrofisica e Scienza dello Spazio di Bologna, via P. Gobetti 93/3, 40129 Bologna, Italy}
\affiliation{Dipartimento di Fisica e Astronomia dell'Universit\`{a} degli Studi di Bologna, via P. Gobetti 93/2, 40129 Bologna, Italy}
%\affiliation{}

%\author{S. Strickland}
%\affiliation{}

%\author{C.~Saez}
%\affiliation{}

%% Note that the \and command from previous versions of AASTeX is now
%% depreciated in this version as it is no longer necessary. AASTeX 
%% automatically takes care of all commas and "and"s between authors names.

%% AASTeX 6.2 has the new \collaboration and \nocollaboration commands to
%% provide the collaboration status of a group of authors. These commands 
%% can be used either before or after the list of corresponding authors. The
%% argument for \collaboration is the collaboration identifier. Authors are
%% encouraged to surround collaboration identifiers with ()s. The 
%% \nocollaboration command takes no argument and exists to indicate that
%% the nearby authors are not part of surrounding collaborations.

%% Mark off the abstract in the ``abstract'' environment. 
\begin{abstract}

We present results from a comprehensive study of ultrafast outflows (UFOs) detected in a sample of fourteen quasars, twelve of which are gravitationally lensed,  in a redshift range of 1.41~--~3.91, near the peak of the AGN and star formation activity. 
New {\sl XMM-Newton} observations are presented for six of them which were selected 
to be lensed and contain a narrow absorption line (NAL) in their UV spectra. Another lensed quasar was added to the sample, albeit already studied, because it was not searched for UFOs.
The remaining seven quasars of our sample are known to contain UFOs.  The main goals of our study are to infer the outflow properties of high-$z$ quasars, constrain their outflow induced feedback, study the relationship between the outflow properties  and the properties of the ionizing source, and compare these results to those of nearby AGN. Our study adds six new detections ($>$~99\% confidence) of UFOs at $z~>~1.4$, almost doubling the current number of cases.
Based on our survey of six quasars selected to contain a NAL and observed with {\sl XMM-Newton}, the coexistence of intrinsic UV NALs and UFOs is found to be significant in $>$~83\% of these quasars suggesting a link between multiphase AGN feedback properties of the meso- and micro-scale.
The kinematic luminosities of the UFOs of our high-$z$ sample are large compared to their bolometric luminosities (median of $L_{\rm K}/L_{\rm Bol}$~$\simgt$~50\%).
This suggests they provide efficient feedback to influence the evolution of their host galaxies and that magnetic driving may be a significant contributor to their acceleration. 
\end{abstract}


%% Keywords should appear after the \end{abstract} command. 
%% See the online documentation for the full list of available subject
%% keywords and the rules for their use.

%% From the front matter, we move on to the body of the paper.
%% Sections are demarcated by \section and \subsection, respectively.
%% Observe the use of the LaTeX \label
%% command after the \subsection to give a symbolic KEY to the
%% subsection for cross-referencing in a \ref command.
%% You can use LaTeX's \ref and \label commands to keep track of
%% cross-references to sections, equations, tables, and figures.
%% That way, if you change the order of any elements, LaTeX will
%% automatically renumber them.
%%
%% We recommend that authors also use the natbib \citep
%% and \citet commands to identify citations.  The citations are
%% tied to the reference list via symbolic KEYs. The KEY corresponds
%% to the KEY in the \bibitem in the reference list below. 

\section{Introduction} \label{sec:intro}
Systematic studies of the X-ray spectra of a sample of z $\simlt$ 0.1 Seyfert galaxies showed that about 40\% of these Active Galactic Nuclei (AGN) have highly-ionized ultrafast outflows (UFOs) %with velocities exceeding 10,000 km~s$^{-1}$ and 
with average velocities ranging between 0.1$c$ and 0.3$c$ \citep[e.g.,][]{2006AN....327.1012C,2010A&A...521A..57T,2013MNRAS.430...60G}.
The two main proposed mechanisms responsible for the acceleration of the X-ray absorbing material to near-relativistic velocities are radiation and magnetic driving \citep[e.g.,][]{1995ApJ...451..498M,2000ApJ...543..686P,2004ApJ...616..688P,1994ApJ...434..446K,2007Ap&SS.311..269E,2010ApJ...715..636F,2014ApJ...780..120F,2010MNRAS.404.1369S,2012MNRAS.426.2859S}.
Theoretical models have been proposed to link the kinematics and energetics of these small scale ultrafast outflows originating at $\sim$10$-$100 $r_{\rm g}$ to the larger kpc-scale cold molecular outflows \citep[e.g.,][]{1998A&A...331L...1S,2010MNRAS.402.1516K,2012MNRAS.425..605F,2012ApJ...745L..34Z,2013ApJ...763L..18W,2020NatAs...4...10G}.

These models consider two distinct phases of the interaction of the wind-angle outflows with the interstellar medium (ISM).
According to these models, in the early phase of the AGN's evolution, the wind is considered to be momentum-conserving and relatively narrow shocks are formed within the ISM where substantial loss of energy occurs through inverse Compton cooling. In a later phase, when the mass of the supermassive black hole reaches a critical value, the wind becomes energy-conserving and adiabatically expands into the ISM and reaches a terminal velocity of a few 1,000~km~s$^{-1}$ \citep[e.g.,][]{2017ApJ...837..149G}.

The influence of AGN outflows on the star formation rate of their host galaxies is not clear from current observations.
By comparing the \OIII$\lambda$5007 line kinematics of outflows with the specific star formation rates of the host galaxies in a sample of 110,000 galaxies at $z < 0.3$ \cite{2017ApJ...839..120W} % Woo et. al. 2017 
find no evidence of negative feedback from AGN with current strong outflows.
As one plausible explanation, they propose a delay between the onset of outflows and the impact they have on the star formation over the entire host galaxy.

\cite{2015ApJ...799...82C} %Cresci, G. et al, 2015 
analyzed the velocity maps of the \OIII$\lambda$5007 and H${\alpha}$ lines of a radio-quiet z = 1.59 quasar
and found that the ionized outflow occupies a space of reduced star formation (negative feedback) and the edges of the outflow
show enhanced star formation possibly triggered by increased pressure at the edges of the outflow. 

\cite{2016A&A...591A..28C} %Carniani et al.(2016)
analyzed the velocity maps of the \OIII$\lambda$5007 and H${\alpha}$ lines of two quasars at $z$ $\sim$ 2.4 and find a spatial anti-correlation between the ionized outflow (traced by the blueshifted component of the OIII line) and star formation  (traced by the narrow component of H${\alpha}$). However, they find that in regions outside the ionized outflow, the star formation rates are high suggesting that 
negative feedback is only significant along the outflow or that it takes several outflow episodes directed along different paths before star formation is quenched in the entire galaxy.  

\cite{2018A&A...617A..81V} studied a sample of WISE/SDSS selected hyper-luminous (WISSH) quasars at $z$ $\approx$ 2$-$4 and obtained constraints on the
properties of AGN winds as traced by blueshifted or skewed OIII and CIV emission lines. The study found one population that exhibits powerful OIII outflows and modest CIV outflows ($v^{\rm peak}_{\rm CIV}~\simlt$~2000~km~s$^{-1}$), and a second population that has weak or absent OIII emission and an extremely large blueshifted CIV emission ($v^{\rm peak}_{\rm CIV} \simgt$ 8000~km s$^{-1}$).
The observed dependence $v^{\rm peak}_{\rm CIV}$  $\propto$ $L^{0.28 \pm 0.04}$ is consistent with a radiatively-driven-winds scenario for the outflows in the WISSH quasars. 

In addition to star formation quenching, AGN feedback outflows are found to be crucial in quenching the cooling flows emerging out of the extended hot halos \citep[e.g.,][]{2017ApJ...837..149G}.

Observations of molecular outflows as traced by CO lines emitting in the mm waveband, in galaxies that host obscured AGN, indicate lower gas fractions in these galaxies compared to star forming galaxies \citep[e.g.,][]{2015A&A...578A..11B,2017MNRAS.468.4205K,2019A&A...630A..59B}.
 
It is important to study the properties of outflows in galaxies at redshifts near the peaks of the AGN and star formation activity where most of the feedback is thought to have taken place \citep[e.g.,][]{2014ARA&A..52..415M}. Detections of relativistic outflows of X-ray absorbing material in distant quasars are rare primarily due to their X-ray weakness. The few cases where relativistic X-ray absorbing outflows have been detected mostly correspond to observations of gravitationally lensed quasars with relatively large (\simgt 5) magnification factors. 

Optical and UV absorption lines in quasars are commonly classified by their widths into broad (BALs; FWHM $ > $ 2000 km s$^{-1}$), narrow (NALs; FWHM $\simlt$ 500 km s$^{1}$), and mini-BALs with absorption line widths ranging between those of BALs and NALs. These class definitions are considered somewhat arbitrary. 
The definition of NALs for example was chosen such that the C~IV doublet can be resolved \citep[e.g.,][]{2004ASPC..311..203H}.

We have initiated a program of increasing the current number of X-ray detected ultrafast outflows in quasars  by targeting gravitationally lensed narrow absorption line (NAL) quasars.  NAL quasars are targeted because they contain an outflow of UV absorbing material and are likely not heavily absorbed in X-rays \citep[e.g.,][]{2009NewAR..53..128C,2011MNRAS.410.1957H}.
This indicates that on average NAL quasars are less X-ray absorbed/weak than BAL quasars and observations with current X-ray missions are more likely to provide medium-to-high signal-to-noise (S/N) spectra of gravitationally lensed NAL quasars.

The main goals of our study are to:
(a) infer the outflow properties of a sample of quasars near the peak of AGN activity, 
(b) determine the significance of such outflows in regulating black hole growth and in influencing structure formation, and
(c) study the relation between the outflow properties of these winds with bolometric luminosity and the spectral energy distribution of the ionizing radiation, and compare these results to those of nearby AGN.

In $\S$2 we present the sample of studied quasars, and in $\S$3 we describe the X-ray observations, the data analysis, and provide estimates of the energetics of the outflows.
Finally, in $\S$4 we present a discussion of our results and in $\S$5 we summarize our conclusions. 
Throughout this paper we adopt a flat $\Lambda$ cosmology with $H_{0}$ = 68~km~s$^{-1}$~Mpc$^{-1}$ $\Omega_{\rm \Lambda}$ = 0.69, and  $\Omega_{\rm M}$ = 0.31 \citep{2016A&A...594A..13P}. 

\section{The Sample}
Gravitational lensing does not produce a bias against the type of background object lensed, however, the fraction of detected lensed QSOs of a certain type (e.g., BAL, mini-BAL, NAL QSOs)  will depend on the attenuation of the objects and the magnification \citep{1997ApJ...474..606G,2000ApJ...531...81C}. One advantage of selecting lensed quasars is that it allows us to investigate the properties of quasars with luminosities that are substantially lower than those of unlensed ones.

Recent SDSS surveys have uncovered a significant number of new gravitational lenses \citep{2012AJ....143..119I,2014AJ....147..153I,2016MNRAS.456.1595M}. From these surveys we identified lensed quasars with blueshifted \CIV\ troughs having widths in the range of   $\simlt$  500 km~s$^{-1}$. 
The X-ray brightest of these lensed NAL quasars were recently observed with {\it XMM-Newton} and constitute six out of fourteen objects of our sample.  We included in our sample seven $z>1$ quasars with reported ultrafast outflows 
\citep{2002ApJ...579..169C,2003ApJ...595...85C,2007AJ....133.1849C,2009ApJ...706..644C,2014ApJ...783...57C,2016ApJ...824...53C,2020MNRAS.496..598C,2012A&A...544A...2L,2015A&A...583A.141V,2018A&A...610L..13D,2020A&A...638A.136B}.

We also included the $z =  2.197$ lensed quasar SDSS~J1029$+$2623 in our sample. \cite{2012ApJ...758...26O} presented results from the analysis of a {\it Chandra} observation of SDSS~J1029$+$2623, however, their study did not include an investigation of a possible outflow in this quasar. 

Our sample of fourteen $z > 1$ quasars is relatively small, however, we note that an identification of an ultrafast wind in a quasar has only been reported in 
seven objects with $z > 1$. One of the goals of our study is to determine the properties of UFOs in $z > 1$ quasars and compare them to the properties of 
UFOs detected in $z \simlt 0.1$ AGN.  Six quasars of our sample, further referred to as our subsample, were selected to contain a UV NAL without prior knowledge of the existence of a UFO. This subsample is therefore unbiased towards UFO detection and is used to infer the fraction of $z > 1$ NAL quasars that contain UFOs.

In Table \ref{tab:prop} we list the properties of the quasar sample including classifications, redshifts, and black hole masses.
Black hole masses were obtained from values published in the literature (see Table \ref{tab:prop} for references) when available. For the other sources we used the C IV line in the SDSS spectra
following the \cite{2017MNRAS.465.2120C} prescription that decreases the uncertainty of the black hole mass estimates from $\sim$ 0.4 to $\sim$0.2 dex  \citep[see Figure 11 of][]{2017MNRAS.465.2120C}.
To summarize, our sample contains three QSOs, ten intrinsic narrow absorption line QSOs (NALQSOs), and one BALQSO.
All the sources in our sample are radio-quiet quasars with the exception of MG~J0414+0534  which has a radio loudness parameter	
$R = f_{\rm 5GHz}/f_{\rm 4400\AA}$  $\approx$ 780 \citep{2018A&A...610L..13D}.
Twelve of the 14 quasars of our sample are gravitationally lensed. Detecting ultrafast outflows in distant quasars is challenging partly because the X-ray spectra of most  $z > 1$ quasars obtained with current X-ray telescopes are low S/N.  The magnification of gravitationally lensed quasars provides a boost in the detected X-ray flux by up to $\sim$ 100 and makes it possible to study ultra-fast outflows in these distant objects.  

Even if unresolved, we can provide a hint of the host circum-AGN/ISM properties: by leveraging the BH mass versus X-ray scaling relations \citep{2019ApJ...884..169G}, we estimate that our SMBHs ($M_{\rm bh} \sim 10^9 - 10^{10}\ M_\odot$) reside in galactic hot halos with temperature $T_{\rm x}~\sim~1-2$~keV

\begin{figure}[ht!]
\epsscale{1.}
\plotone{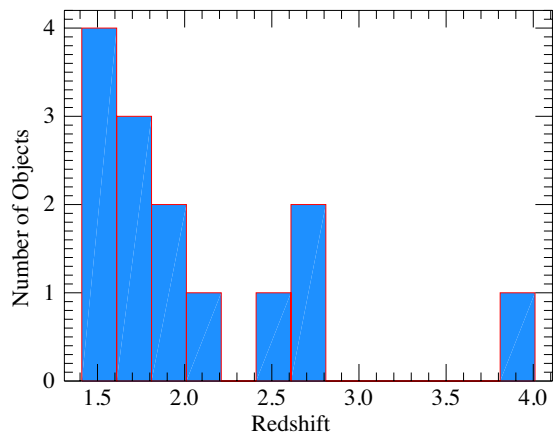}
\caption{Redshift distribution of the quasars in our sample.  \label{fig:redshift}}
\end{figure}

In Figure \ref{fig:redshift} we show the redshift distribution of the quasars of our sample. Previous statistical studies of ultrafast outflows have focused on AGN with redshifts of up to 0.25 (e.g., see  Figure 1 of \cite{2013MNRAS.430...60G}). The AGN in our sample lie near the peak of quasar activity in the history of our Universe.

\section{X-ray Observations and Data Analysis} \label{sec:obs}
In Table \ref{tab:log} we list the observation dates, exposure times and number of background-subtracted source counts of the quasars in our sample.
For several objects we included published results. Specifically, for \apm\ we used the results published in 
\cite{2002ApJ...579..169C,2009ApJ...706..644C}, for HS~1700$+$6416 we incorporated the results published in \cite{2012A&A...544A...2L}, 
for MG J0414+0534 we included the results from \cite{2018A&A...610L..13D},  for PG~1115+080
we included the results from \cite{2003ApJ...595...85C,2007AJ....133.1849C} 
for PID352 we included the results from \cite{2015A&A...583A.141V},
and for the 16 Dec 2013 observation of  \hs\  we used the results published in \cite{2016ApJ...824...53C}. 
For Q2237+030, \cite{2020A&A...638A.136B} find significant spectral variability and indications of UFOs in several observations of this object,
with outflow velocities of up to 0.5$c$. However, the main goal of the \cite{2020A&A...638A.136B} study was to assess the recurrence of UFOs
in this source. Since our study focuses on the physical properties of UFOs detected at high significance, we only include our recent results in the analysis. 
For the objects SDSS~J1029$+$2623, SDSS~J1529+1038, SDSS~J0904$+$1512, SDSS~J1353+1138, SDSS~J1128+2402,  SDSS~J0921$+$2854, and Q2237+030 we reduced the X-ray observations using the following procedures. 

For the reduction of the \xmm\ observations we used the Science Analysis System software version 18.
We filtered the pn \citep{2001A&A...365L..18S} and MOS \citep{2001A&A...365L..27T} data by selecting events corresponding to instrument \verb+PATTERNS+ in the 0--4 (single and double pixel events) and 0--12 (up to quadruple pixel events) ranges, respectively. Moderate-amplitude background flares were present during several of the \xmm\ observations. The pn and MOS data were filtered to exclude times when these flares occurred resulting in the effective exposure times listed in Table \ref{tab:log}. 

The contribution of the background spectra to identified absorption and/or emission features is determined by over-plotting both source and background spectra and determining the significance of the inferred spectral features while adjusting the source and background extraction regions.  We note that for the {\sl Chandra} spectra of the quasars in our sample the background contribution is negligible, however, the background can become comparable  to the source spectra for the {\it XMM-Newton} pn spectra, especially at high energies \citep[e.g.,][]{2018ApJ...867..103C}.

To test for sensitivity to background non-uniformity we also tried multiple background extraction regions. We did not find any differences in the spectral shapes and features using  more conservative threshold cuts or selecting different background extraction regions.  We selected extraction regions to optimize the S/N and
ensure that the background spectra were significantly below the source spectra, especially near the energies of detected absorption lines.

The energy ranges used for fitting the pn and MOS spectra were 0.3--11~keV and 0.4--10~keV, respectively. 
We performed spectral fits to the pn spectra alone, and to the pn and MOS data simultaneously.
Both approaches resulted in values for the fitted parameters that were consistent within the errors, however, in most cases the fits to the higher quality pn data alone yielded lower reduced $\chi^{2}$ values compared to the combined fits. We therefore consider the results from the fits to the pn data alone better suited for characterizing the properties of the X-ray absorption features. 
There are two exceptions where our analysis relied mostly on the MOS cameras. Specifically, for the May 2018 XMM-Newton observation of SDSS~J0921 only MOS1 and MOS2 data are available. For the Oct 2018 observation of SDSS~J0921 significant background flaring is present throughout the observation and the flares are much more intense in the pn than in the MOS cameras  \citep{2020XMM-Newton..Users..Handbook}. 

For the reduction of the {\sl Chandra} observations we used the CIAO 4.12 software with CALDB version 4.9.1 provided by the {\sl Chandra X-ray Center} (CXC).
We used standard CXC threads to screen the data for status, grade, and time intervals of acceptable aspect solution and background levels. 
The energy ranges used for fitting the ACIS-S spectra were 0.5--8~keV.  

For the gravitationally lensed sources, we extracted spectra for the combined images unless mentioned otherwise.

The extracted spectra were grouped to obtain a minimum of 20 counts in each energy bin (with the exception of Q~2237$+$0305 that was grouped to obtain a minimum of 15 counts in each energy bin), allowing use of $ \chi^{2}$ statistics. This grouping was chosen for $\chi^2$ to be statistically valid (e.g., Cash 1979; Bevington \& Robinson 2003),
and to allow the maximum spectral resolution for the spectra. To further test the validity of the use of $\chi^2$ statistics in our analysis for our selected grouping of the data we
also used the $C$-statistic (Cash 1979) on the same datasets and binned the data to have at least one count per bin.
Background spectra were extracted from source-free regions.

\subsection{Spectral Analysis Results}
Spectra were fitted with a variety of models employing \verb+XSPEC+ version 12 \citep{1996ASPC..101...17A}. 
X-ray, UV, bolometric luminosities and Eddington ratios of the quasar sample are listed in Table \ref{tab:sample}. The UV luminosity densities at 1450{\AA} were obtained from analyzing available SDSS spectra of the sampled quasars. The bolometric luminosities were calculated using two different methods: first from the X-ray bolometric correction factors and second from the luminosity density at 1450{\AA}. Specifically, for the first method we apply a bolometric correction to the 2--10~keV luminosities based on the empirical relations presented in \cite{2020A&A...636A..73D}.
%\cite{2012MNRAS.425..623L}. 
The second independent estimate of $L_{\rm Bol}$ is provided from the monochromatic luminosities at 1450$\rm \AA$  based on the empirical equations of \cite{2012MNRAS.422..478R}. Our independent estimates of the bolometric luminosities are very similar and show no systematic offsets. We use the difference between the two estimates to calculate the uncertainty of $L_{\rm Bol}$. The Eddington ratios use the bolometric luminosities derived from the observed optical/UV flux densities. The Eddington ratios of our sample lie in the range of 0.03 -- 0.8 with a mean of 0.21 and a median of 0.15. We used the gravitational lens fitting code \verb+glafic+ version 1.1.6 \citep{2010ascl.soft10012O} to model the gravitational lens systems and obtain the magnification factors.  For all spectral models we included Galactic absorption due to neutral gas \citep{2016A&A...594A.116H}.% \citep{1990ARA&A..28..215D}.

We proceed in fitting the following models to the data (see Tables \ref{tab:fit1}, \ref{tab:fit2}, and \ref{tab:fit3} ) guided by the shape and location of  identified absorption and/or emission residuals:  

\noindent
1) power-law modified by neutral intrinsic absorption at the source. 

\noindent
2) power-law modified by neutral intrinsic absorption and a number of absorption and/or emission lines.

\noindent
3) power-law modified by neutral intrinsic absorption, outflowing intrinsic ionized absorption 
and an emission line if required. 

For the outflowing ionized absorber we used the \verb+XSTAR+ photoionization model \verb+warmabs+ \citep{2001ApJS..133..221K,1996ApJ...465..994K}. 
For improved accuracy and flexibility we use the analytic \verb+XSTAR+ versions of the \verb+warmabs+ 
model instead of the \verb+XSTAR+ table models. 
Our \verb+XSTAR+   \verb+warmabs+ model assumes a spherical, constant density photo-ionized outflowing optically thin absorber with a source at its center. 

The default atomic population file \verb+pops.fits+  provided in NASA's \verb+warmabs+ distribution uses a fixed value of the photon index of $\Gamma$~=~2. However, our spectral analysis indicates that the photon index $\Gamma$ differed from this default value for most observations of our quasar sample. We therefore used \verb+XSTAR+ to create new population files appropriate for photon indices of each observation. 

We model the velocity broadening of the absorption lines by introducing in the \verb+XSTAR+  models large turbulent velocities.
Several mechanisms may lead to velocity broadening, including velocity gradients along the radial direction of motion \citep[e.g.,][]{2007MNRAS.381.1413S,2011ApJ...737...91S,2018ApJ...864L..27F},
and velocity gradients along a transverse direction of motion of plasma around the black hole corona  \citep[e.g.,][]{2019ApJ...885L..38F}.
We propose additional explanations, such as relativistic effects, that may be important for outflows launched near the ISCO, and variability of the velocity of the outflow over timescales shorter than the total exposure time.
The spectra presented here, however,  do not have adequate S/N and/or spectral resolution to distinguish between these possible mechanisms.
We performed several fits where we allowed the turbulent velocity to vary and found the best-fit values.
Because of the low to moderate S/N of the {\sl Chandra} and {\it XMM-Newton} spectra, the
turbulent velocities, are not well constrained. For the error analysis of the remaining variables in spectral fits that used the \verb+XSTAR+  model
we froze the turbulent velocities at the best-fit values and list these values in Tables \ref{tab:fit1}, \ref{tab:fit2}, and \ref{tab:fit3}.

\begin{figure*}[t]
\plotone{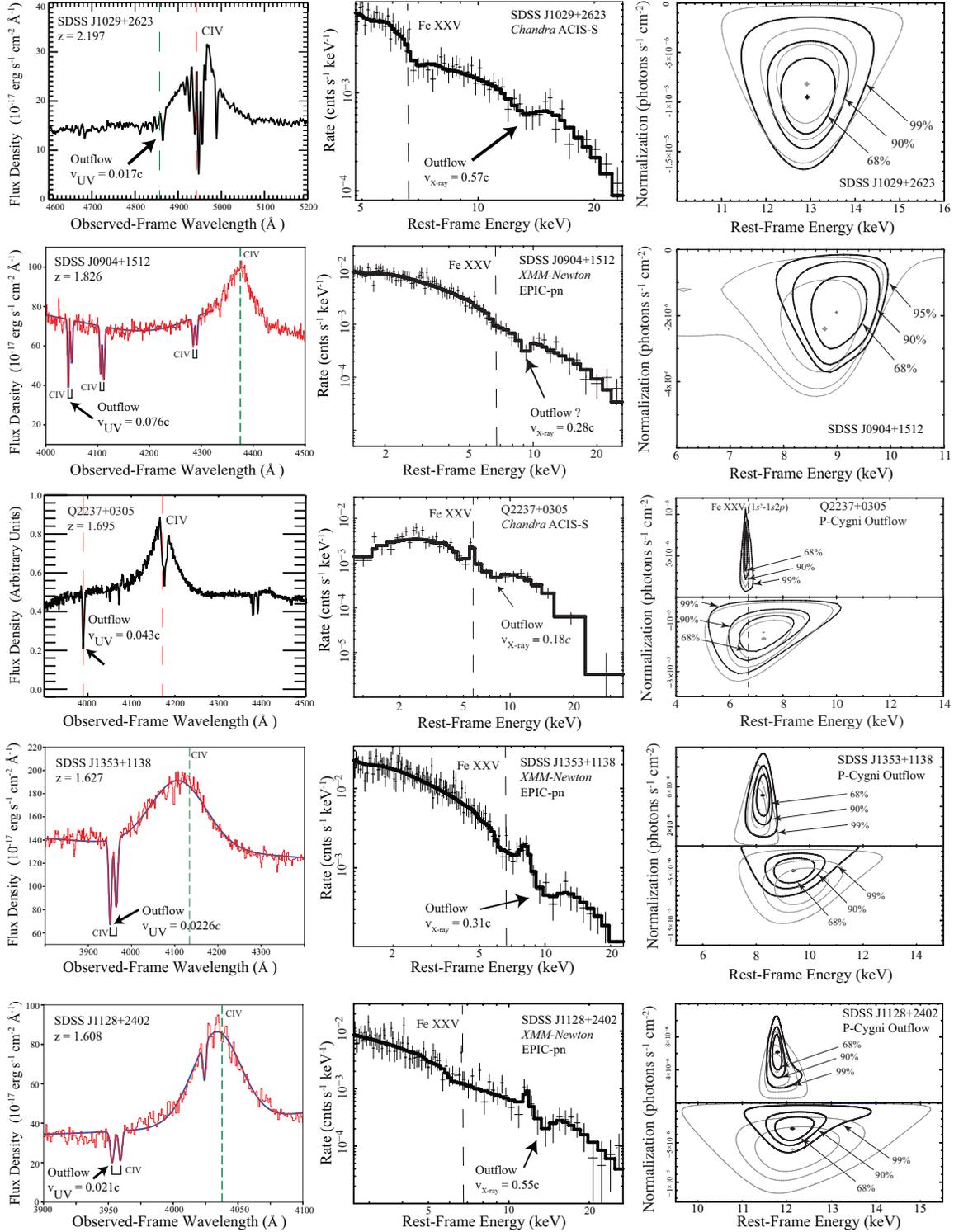}
\caption{The UV and X-ray spectra of the sources for which the X-ray spectra have not been previously published with claimed ultrafast outflows. (left) Rest-frame UV spectra showing NALs.
The spectrum of Q2237 is adapted from \cite{2011MNRAS.415.1985O}. (middle) X-ray spectra showing blueshifted absorption lines.
(right)  confidence contours of the absorption and/or emission lines of the ultrafast outflows. Black and thick contours are based on spectral fits that use ${\chi}^2$ statistics and
grey and thin contours are based on spectral fits that use $Cash$ statistics.
\label{fig:spectra}}
\end{figure*}

In Figure \ref{fig:spectra} we show the UV and X-ray spectra of the sources where significant ultrafast outflows are shown, and have not been previously published.
The UV spectra (left column) show the CIV broad emission lines and blueshifted absorption lines indicative of outflows with velocities of up to 0.075$c$.
The X-ray spectra (middle column) show highly blueshifted absorption lines which are occasionally accompanied by emission lines (P-Cygni profiles), indicative of ultrafast outflows with non-negligible (or large) covering factors. 
The emission line in a P-Cygni feature is thought to be produced by fluorescence from the entire outflow, whereas the expanding outflow along our line-of-sight produces the blueshifted absorption line.
We find that the X-ray spectra of 5(3) of the 14(6) quasars in our sample(subsample) contain P-Cygni profiles.
The energies and equivalent widths of the emission lines in these P-Cygni profiles indicate that they do not originate from reflection from the accretion disk or distant cold matter such as a molecular torus. Specifically, the energies of the detected emission lines in the P-Cygni profiles detected in our sample lie in the range of $6.6 - 11.7$~keV and their equivalent widths lie in the range of $0.75 - 2.3$~keV. Conversely, studies of the X-ray spectra of quasars \citep[e.g.,][]{2007ApJ...662..860I,2010A&A...524A..50D} indicate that a large fraction of
them contain emission lines due to reflection with energies of  $\sim$ 6.4~keV and with equivalent widths that lie in the range of $130 - 280$ eV.

The confidence levels (right column) of the ultrafast outflow detections, calculated using $\chi^2$ and $\sl Cash$ statistics are found to be  $>$ 99\%, with the exception of SDSSJ0904 that is detected with a significance of  $>$ 90\% confidence.
We emphasize that these results are independent of the statistic used in the analysis of the spectra. In particular, the absorption and emission line parameters are always consistent (within the 68\% confidence level) using the $\sl Cash$ or the $\chi^2$ statistic (see Figure 2).

We followed a more robust approach of estimating the significance of the blueshifted absorption and emission lines in the X-ray spectra based on
Monte Carlo simulations to determine the distribution of the F-statistic between models \citep{2002ApJ...571..545P}. 
We considered a null model that included a simple absorbed power law and an alternative model that in addition included
one or two Gaussian absorption and/or emission lines. 
For each observed spectrum we simulated 1000 data sets using the XSPEC \verb+fakeit+ command. We fit the null and alternative models to the 1000 simulated data sets and computed the $F$-statistic for each fit. We computed the probability, $P_{F\rm }$, for the F value to exceed the value determined from the fits of the null and alternative models to the observed spectra. The probabilities $P_{\rm F}$ for all objects are listed in Tables \ref{tab:fit1}, \ref{tab:fit2}, and \ref{tab:fit3} and are found to be $< $0.01 with the exception of SDSSJ0904 for which $P_{\rm F}$ = 0.035.  

\renewcommand{\thefigure}{\arabic{figure} (Cont.)}
\addtocounter{figure}{-1}
\begin{figure*}[t]
\plotone{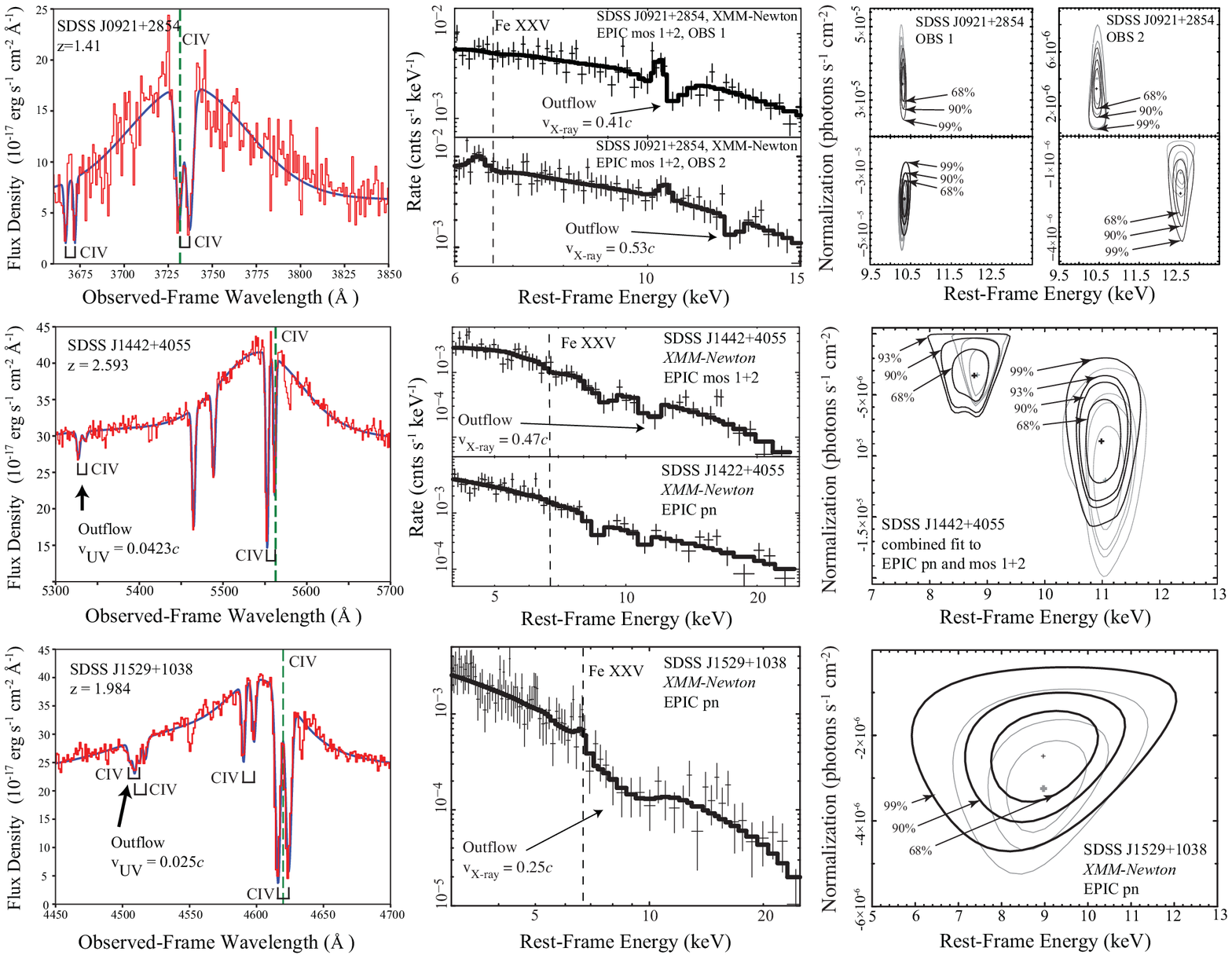}
\caption{The UV and X-ray spectra of the sources for which the X-ray spectra have not been previously published with claimed ultrafast outflows. (left) Rest-frame UV spectra showing NALs. (middle) X-ray spectra showing blueshifted absorption lines. (right) confidence contours of the absorption and/or emission lines of the ultrafast outflows. Black and thick contours are based on spectral fits that use ${\chi}^2$ statistics and grey and thin contours are based on spectral fits that use $Cash$ statistics.}
\end{figure*}
\renewcommand{\thefigure}{\arabic{figure}}

\begin{figure*}
\epsscale{1.1}
\plotone{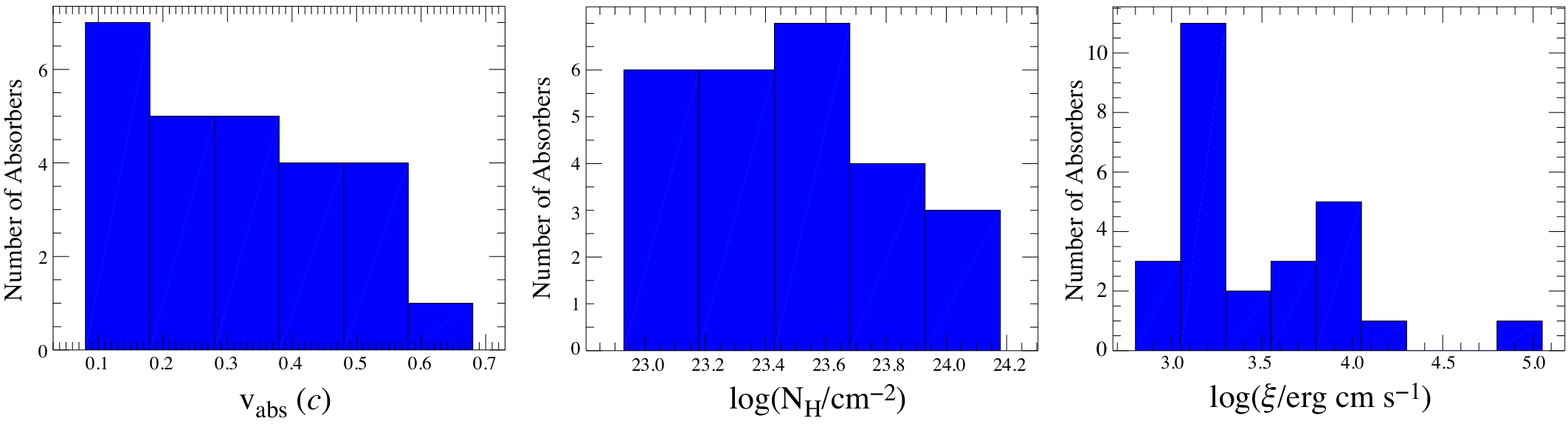}
\caption{Distributions of the velocities, absorber column densities, and ionization parameters of our sample of high-$z$ quasars.  \label{fig:distributions}}
\end{figure*}

Due to an operational problem, only MOS 1,2 data were obtained during the {\sl XMM-Newton} observation of SDSSJ0921 in May 2018.
The observation of SDSSJ0921 was rescheduled in Oct 2018 during which both EPIC pn and MOS 1,2 data were obtained. In Figure \ref{fig:spectra} 
we show the variability of the outflow in SDSSJ0921 between the two observations. The projected velocity of the outflow along our line of sight has increased from 0.41$c$ to 0.53$c$ between observations. The outflow is detected at the same velocity in both the MOS12 and pn spectra of the Oct 2018 observation.
The emission line detected in SDSSJ0921, which possibly originates from iron fluorescence from the entire outflow, shows no significant change in energy between the two observations.
The {\sl XMM-Newton} observation of SDSSJ1442 was also significantly affected by flares and the background was especially elevated in the pn detector.  Two blueshifted absorptions lines are detected in both MOS 1+2 and pn spectra of  SDSSJ1442 indicating an outflow with two velocity components.

We also detect an emission line with a rest-frame energy of  $\sim$16.44~keV in the spectrum of SDSSJ1442. The 16.44~keV line is detected at the $>$ 99\% confidence level and our Monte Carlo simulations confirm it to be significant ($>$ 99.9\% confidence level) and not a random fluctuation.
We interpret the large energy shift of the line with respect to the energy of the expected Fe K$\alpha$ fluorescence line as possibly being the result of microlensing in one of the lensed images of SDSS J1442. Similar blueshifted lines have been detected in several lensed quasars \citep[e.g.,][]{2017ApJ...837...26C}. Future observations with the	{\sl Chandra} X-ray Observatory would	be required to resolve the spectra of the images to confirm the microlensing interpretation.

In Figure \ref{fig:distributions} we show the distributions of the velocities, absorber column densities, and  ionization parameters\footnote{Throughout this paper we adopt the definition of the ionization parameter of \cite{1969ApJ...156..953T} given by $\xi=\frac{L_{\rm ion}}{n_H r^2}=\frac{4 \pi}{n_H} \int_{1Rdy}^{1000Rdy}F_{\nu}d\nu$, where $n_H$ is the hydrogen number density, and $r$ is the source-cloud separation.} of the outflowing X-ray absorbers of our sample. These outflow properties are taken from spectral fits with models that incorporate the outflowing photoionized absorber listed as model 3 in Tables \ref{tab:fit1}, \ref{tab:fit2}, and \ref{tab:fit3}.
Several of the objects in our sample were observed over multiple epochs (see Table \ref{tab:log}) and several of the objects contained multiple outflowing absorbers. The projected values of the outflow velocities of the absorber lie in the range of $\sim$ 0.1$-$0.6$c$, the absorber column densities lie in the range of $\sim$ 9 $\times$ 10$^{22}$ $-$  1 $\times$ 10$^{24}$~cm$^{-2}$, and the ionization parameters lie in the range of $\sim$ 10$^{2.8}$ $-$ 10$^{5}$~erg~cm~s$^{-1}$.

\subsubsection{Energetics of Quasar Outflows}
For estimating the energetics of the wind we assume a spherically symmetric outflow with a covering factor of $f_{\rm c}$ \citep[e.g.,][]{1999isw..book.....L}. We approximate the hydrogen column density $N_{\rm H} \sim n(r){\Delta}r$, where $n(r)$ is the number density (particles~cm$^{-3}$) of the gas at radius $r$. We use the following expressions to estimate the mass-outflow rate (equation \ref{eq:1}), the kinetic power (equation \ref{eq:2}), and the rate of change of momentum of the outflow (equation \ref{eq:3}):

\begin{equation}\label{massoutflow}
 \dot{M} = 4{\pi}r(r/{\Delta}{r})N_{\rm H}m_{\rm p}v_{\rm wind}f_{\rm c} \label{eq:1}
\end{equation}

\begin{equation}\label{kinetic}
\dot{E}_{\rm K} = {{1}\over{2}}{\dot{M}{v^{2}_{\rm wind}}} \label{eq:2}
\end{equation}

\begin{equation}\label{power}
\dot { { p } }  = \dot {{M}} v_{\rm wind} \label{eq:3}
\end{equation}

\noindent
where ${\Delta}{r}$ is the thickness of the absorber at radius $r$, 
$N_{\rm H}$ is the hydrogen column density,
$v_{\rm wind}$ is the outflow velocity of the X-ray absorber, and $f_{\rm c}$
is the global covering factor of the absorber.

We used a Monte Carlo approach to estimate the errors of  $\dot{M}$, {$\dot{E}$}$_{\rm K}/L_{\rm Bol}$, and  $\dot{p}/(L_{\rm Bol}/c)$.
The values of $v_{\rm wind}$, $N_{\rm Habs}$, $M_{\rm BH}$, and  $L_{\rm Bol}$ were assumed to have normal distributions within their error limits.
The values of $f_{\rm c}$, $r/{\Delta}{r}$, and $r$ were assumed to have uniform distributions within their error limits.
By multiplying these distributions and with the appropriate constants from equations 1, 2, and 3 we obtained the distributions of $\dot{M}$, {$\dot{E}$}$_{\rm K}/L_{\rm Bol}$, and  $\dot{p}/(L_{\rm Bol}/c)$.
We finally determined the mean values of the distributions of 
$\dot{M}$, {$\dot{E}$}$_{\rm K}/L_{\rm Bol}$, and  $\dot{p}/(L_{\rm Bol}/c)$ and estimated the 68\% confidence ranges.

Special relativistic effects in modeling ultrafast outflows were recently presented in \cite{2020A&A...633A..55L}.  As a consequence of these relativistic effects the true hydrogen column densities of the outflowing absorbers are larger than the observed column densities by a velocity depended factor. In Table \ref{tab:outflow} we list the relativistic correction factors as calculated in \cite{2020A&A...633A..55L} for the observed outflow velocities.
The mass-outflow rate, the kinetic power, and the rate of change of momentum of the outflow are all proportional to the column density and therefore these quantities also need to be adjusted by the relativistic correction factor.

The locations of the absorbers is not well constrained with the available CCD resolution spectra. As a conservative approach \citep[see][]{2015MNRAS.451.4169G} we calculate a lower bound of the distance of the absorber from the center of the black hole, $r_{\rm min}$, by equating the observed velocity with the escape velocity at that radius.

\begin{equation}\label{rmin}
r_{min} = R_{s \rm}(c/v_{\rm wind} )^{2}, \label{eq:4}
\end{equation}
where $v_{wind}$ is the observed outflow velocity and $R_{\rm s}$  = $2GM/c^2$, is the Schwarzschild radius. We note that we observe the projected component of the wind velocity. The true 
outflow velocity will be larger than the observed one and the true radius will be smaller than this $r_{\rm min}$ value depending on the angle between our line of sight through the absorber and the velocity of the X-ray absorbing material.  The observed wind velocities are derived from the  best-fit redshift parameters $z_{\rm abs}$  of model 3 in Tables \ref{tab:fit1}, \ref{tab:fit2}, and \ref{tab:fit3}, where the outflowing ionized absorber is modeled with the photoionization software package XSTAR.  

Variability of the properties of the ultrafast outflows has been observed and reported in several quasars of our sample including  APM08279, PG1115, and HS1700 on timescales comparable to the light
crossing time over regions of 10$-$100$r_{\rm g}$ \citep{2003ApJ...595...85C,2009ApJ...706..644C,2011ApJ...737...91S,2012A&A...544A...2L}. The short term variability timescale of the X-ray absorption lines suggest that the distances of the outflowing absorbers from the center of the black holes are consistent with the estimates of $r_{\rm min}$.

In the case that the wind is not continuous but made up of clouds we can define a filling factor $f = V_{\rm gas}/V$. 
Assuming the clouds have a thickness of ${\Delta}r$, the column density is ${ N }_{\rm  H } \sim n(r)(\Delta r)f$
and the distance between the ionizing source and the absorbing cloud is given by

\begin{figure}[ht!]
\plotone{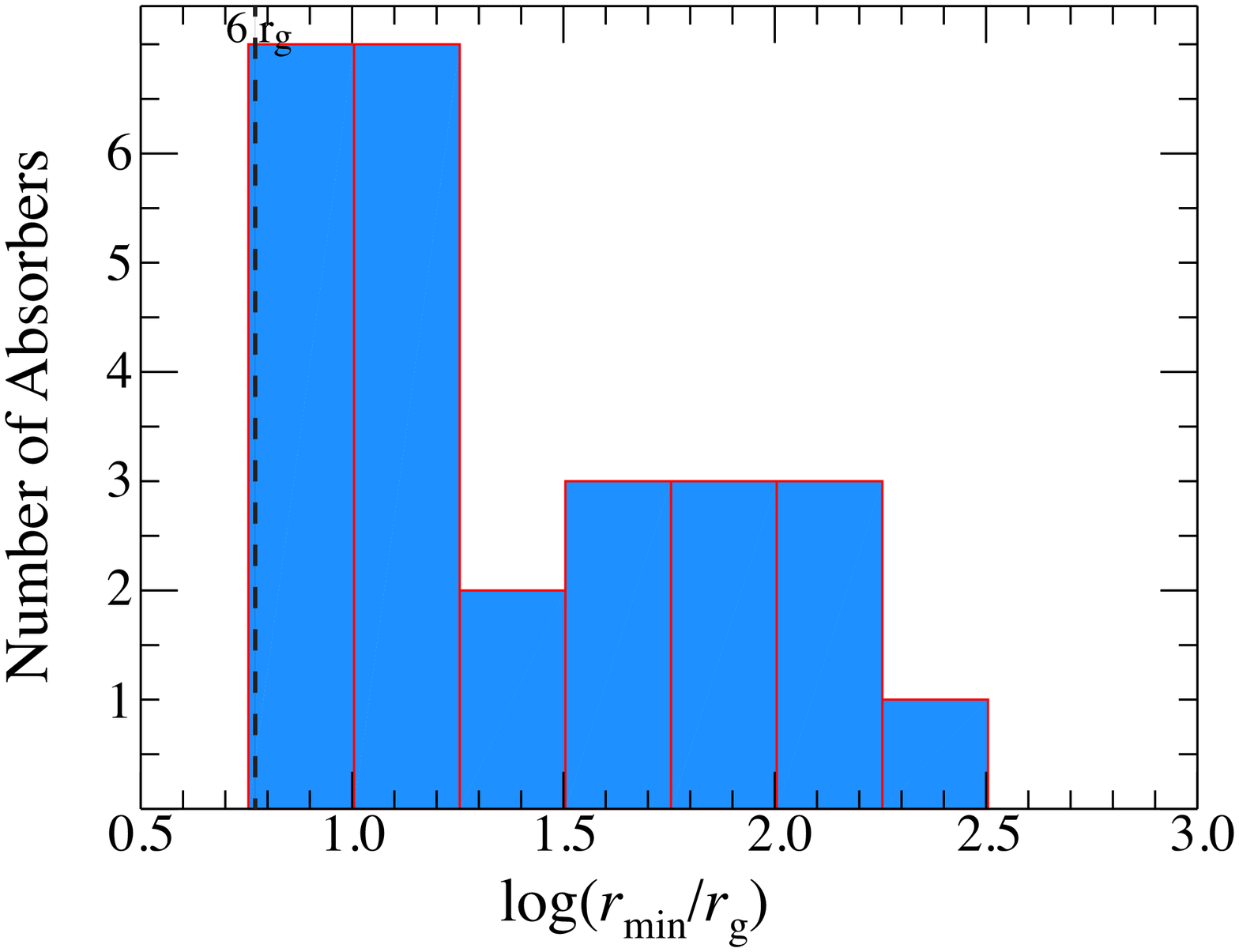}
\caption{Distribution of the distances of the ionized ultrafast absorbers from the central sources of our quasar sample.  \label{fig:radii}}
\end{figure}

\begin{equation}
{ { r }_{\rm  absorber }=\left( \frac { { L }_{ ion }\Delta rf }{ \xi { N }_{ H } }  \right)  }^{ 1/2 }
\end{equation}

In order to obtain an upper limit on the location of the absorber, the following approximations are often used in the literature, ${\Delta}r/r$ = 1 and a filling factor of $f$ = 1 \citep[e.g.,][]{2012MNRAS.422L...1T,2015MNRAS.451.4169G}. These approximations lead to the following upper limit on the location:
 
 \begin{equation}
 { r }_{\rm max }=\frac { { L }_{ ion } }{ \xi { N }_{ H } } \label{eq:rmax}
 \end{equation}

\begin{figure*}[ht!]
\epsscale{1.1}
\plotone{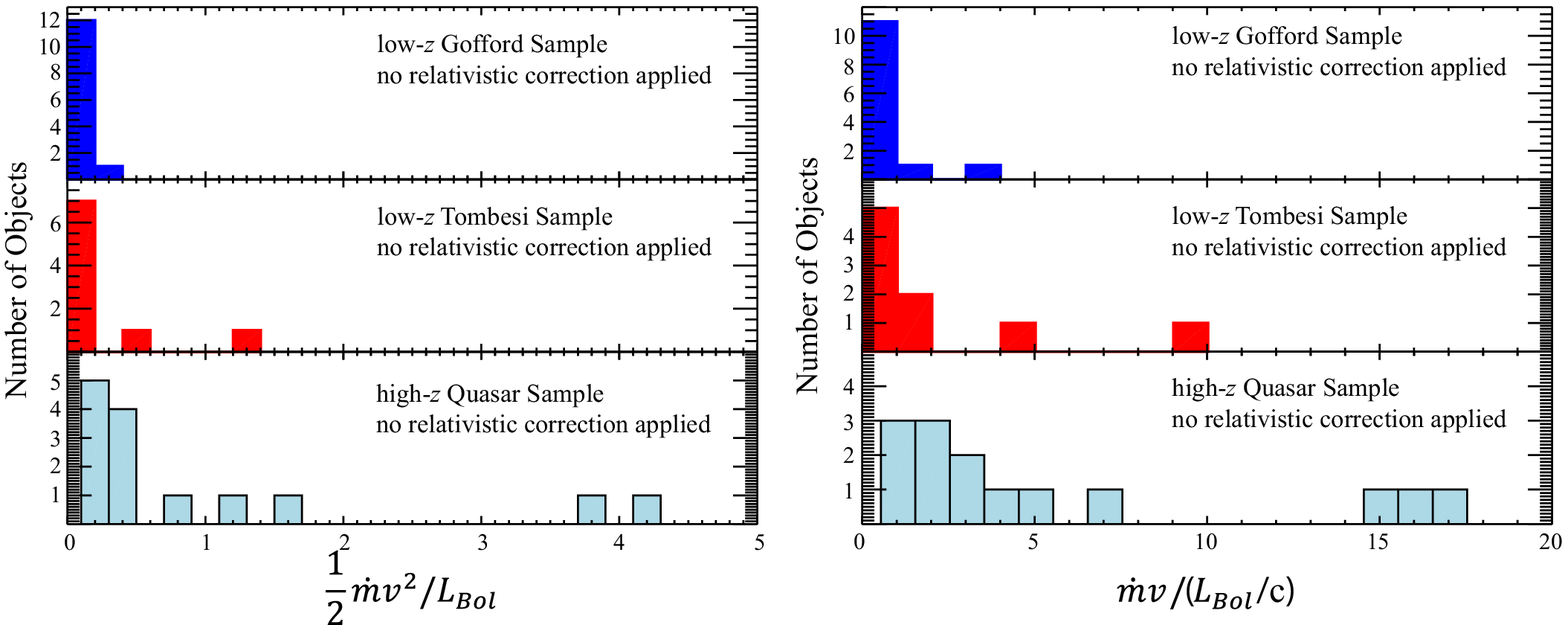}
\caption{Left: Distribution of the fraction of the kinetic luminosity to the bolometric luminosity ({$\dot{E}$}$_{\rm K}/L_{\rm Bol}$) of the ultrafast absorbers of our high-$z$ quasar sample (left-bottom), of the low-$z$ Gofford sample (left-top), and of the low-$z$ Tombesi sample (left-middle). Right: Distribution of the momentum boost ($\dot{p}/(L_{\rm Bol}/c)$) of the ultrafast absorbers of our high-$z$ quasar sample (right-bottom), of the low-$z$ Gofford sample (right-top), and of the low-$z$ Tombesi sample (right-middle). For objects with multiple observations the average values are displayed. No relativistic correction is applied to the outflow efficiency and momentum boost.   \label{fig:effic1}}
\end{figure*}

\begin{figure*}[ht!]
\epsscale{1.1}
\plotone{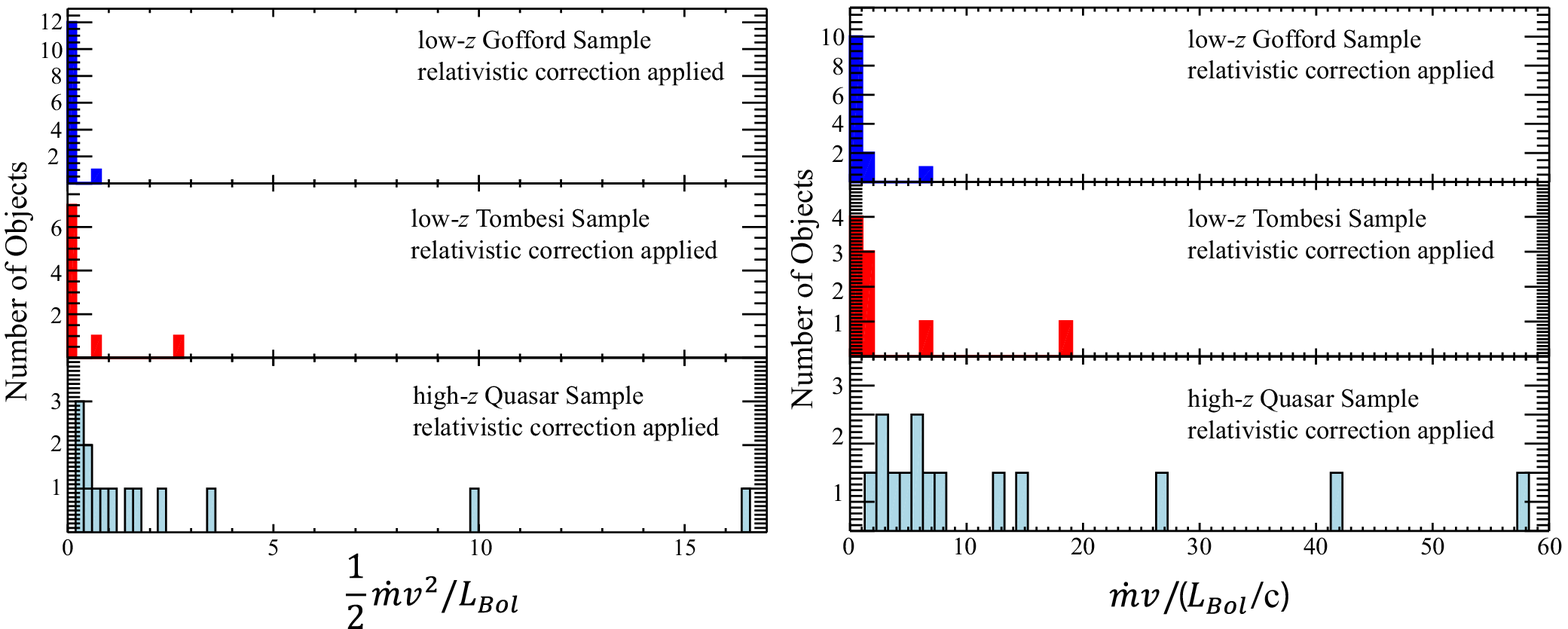}
\caption{Left: Distribution of the fraction of the kinetic luminosity to the bolometric luminosity ({$\dot{E}$}$_{\rm K}/L_{\rm Bol}$) of the ultrafast absorbers of our high-$z$ quasar sample (left-bottom), of the low-$z$ Gofford sample (left-top), and of the low-$z$ Tombesi sample (left-middle). Right: Distribution of the momentum boost ($\dot{p}/(L_{\rm Bol}/c)$) of the ultrafast absorbers of our high-$z$ quasar sample (right-bottom), of the low-$z$ Gofford sample (right-top), and of the low-$z$ Tombesi sample (right-middle). For objects with multiple observations the average values are displayed. A relativistic correction is applied to the outflow efficiency and momentum boost.   \label{fig:effic2}}
\end{figure*}

There are several problems with using ${ r }_{\rm max }$ with ${\Delta}r/r$ = 1 and $f =1$ as a useful upper limit for the location of the absorber. 
Filling factors based on estimates for absorbing clouds can be as small as $\sim$ 1~$\times$~10$^{-6}$ \citep[e.g.,][]{2016MNRAS.457.3896L} . Assuming $f = 1$ and ${\Delta}r = r$ can result in estimated ${ r }_{\rm max }$ values that are several orders of magnitude larger than the true values of  ${ r }_{\rm absorber}$.   Using Equation \ref{eq:rmax} will also result in an overestimate of quantities that are proportional to ${ r }_{\rm  absorber }$ such as the mass-outflow rate, the outflow efficiency and the momentum boost. In Table \ref{tab:outflow} we list the ratio of $r_{\rm max}/r_{\rm min}$.  For our study we are interested in placing conservative constraints on the energetics of ultrafast outflows and determining whether they are powerful enough to produce feedback on their host galaxies based on criteria presented in \cite{2016MNRAS.458..816H}.  We therefore adopt the $r_{\rm min}$ values for estimating the location of the absorbers resulting in lower limits of the energetics of the outflows.

In Table \ref{tab:outflow}, we list the total hydrogen column densities $N_{\rm H}$ of the X-ray absorption lines, the relativistic corrections of the optical depths, the minimum and maximum distances between the ionizing source and the absorbing cloud, the ionization parameters, the outflow velocities of each absorption component, the mass-outflow rates, the efficiency of the outflows and the momentum boosts of the outflows.
In Figure \ref{fig:radii} we present the distribution of the estimated $r_{\rm min}$ values of the quasars of our sample derived from Equation \ref{rmin}. 
Most quasar winds appear to have $r_{\rm min}$ $\simlt$ 100$~r_{\rm g}$, with a significant fraction have $r_{\rm min}$ near $\sim$ 20$~r_{\rm g}$.
This is consistent with detailed general-relativistic radiative magneto-hydrodynamic (GR-rMHD) simulations, showing a continuous production of fast AGN outflows within such a micro-scale region \citep[e.g.,][]{2017MNRAS.468.1398S}.

Insight into the acceleration mechanism of ultrafast outflows is obtained by estimating the fraction of their kinetic luminosity  ({$\dot{E}$}$_{\rm K}$)  to the bolometric luminosity ($L_{\rm Bol}$).  
An efficiency fraction,  {$\dot{E}$}$_{\rm K}/L_{\rm Bol}$  near or greater than the covering fraction would imply that a driving mechanism in addition to radiation pressure must be contributing to the acceleration of ultrafast outflows. Specifically, assuming that the bolometric emission of the high-$z$ sample is approximately isotropic, an outflowing wind with a covering fraction of $f_{\rm c}$ will at most receive a fraction $f_{\rm c}$ of  the total bolometric luminosity. 

The global covering factor of the absorber is often estimated from modeling the P-Cygni profile of the outflowing spectral feature or from the observed fraction of AGN that show ultrafast outflows in their X-ray spectra. Many of the available P-Cygni models used to fit the X-ray spectra of ultrafast quasar winds (e.g., windabs) ignore the presence of the accretion disk and do not include general relativistic effects that are expected to be important for winds launched near the event horizon.  \cite{2009MNRAS.393.1433D} has simulated P-Cygni  profiles produced in the vicinity of quasars, taking into account Doppler and gravitational effects. These simulations indicate that current models that do not include general relativistic effects and an accretion disk are crude approximations and will result in unreliable constraints on the geometry of the wind.  
We therefore assume a global covering factor of $f_c$ = 0.4 with the knowledge that current observations suggest that about 40\% of nearby AGN contain ultrafast outflows  \citep[e.g.,][]{2010A&A...521A..57T,2013MNRAS.430...60G}.
In Figures \ref{fig:effic1} and \ref{fig:effic2} we show the distributions of the efficiency fraction {$\dot{E}$}$_{\rm K}/L_{\rm Bol}$ and momentum boost for our high-$z$ sample without and with relativistic corrections (wrc), respectively.  
About 64\% (86\% wrc) of the high-$z$ sample have {$\dot{E}$}$_{\rm K}/L_{\rm Bol}$~$\simgt$~0.4 and about 86\% (100\% wrc) have  $\dot{p}/(L_{\rm Bol}/c)$~$\simgt$~1.
As a comparison, from the low-$z$ Tombesi sample (see Figures \ref{fig:effic1} and \ref{fig:effic2}) we find that about 20\% (20\% wrc) of local AGN have {$\dot{E}$}$_{\rm K}/L_{\rm Bol}$ $\simgt$ 0.4 and about 40\% (50\% wrc) of them have $\dot{p}/(L_{\rm Bol}/c)$~$\simgt$~1.
For the comparison of the low and high-$z$ AGN samples we have assumed the minimum estimates of {$\dot{E}$}$_{\rm K}$ that assume the distance to the absorber is given by equation  \ref{eq:4}.

Finally, the retrieved high mechanical feedback power ratio (Figures \ref{fig:effic1} and \ref{fig:effic2}, left) is a clear signature that the micro/meso AGN feedback will have a substantial impact on the evolution of the host hot halos estimated to have $T_{\rm x} \sim 1-2$ keV \citep{2019ApJ...884..169G}.

\begin{figure}[h]
\epsscale{1.1}
\plotone{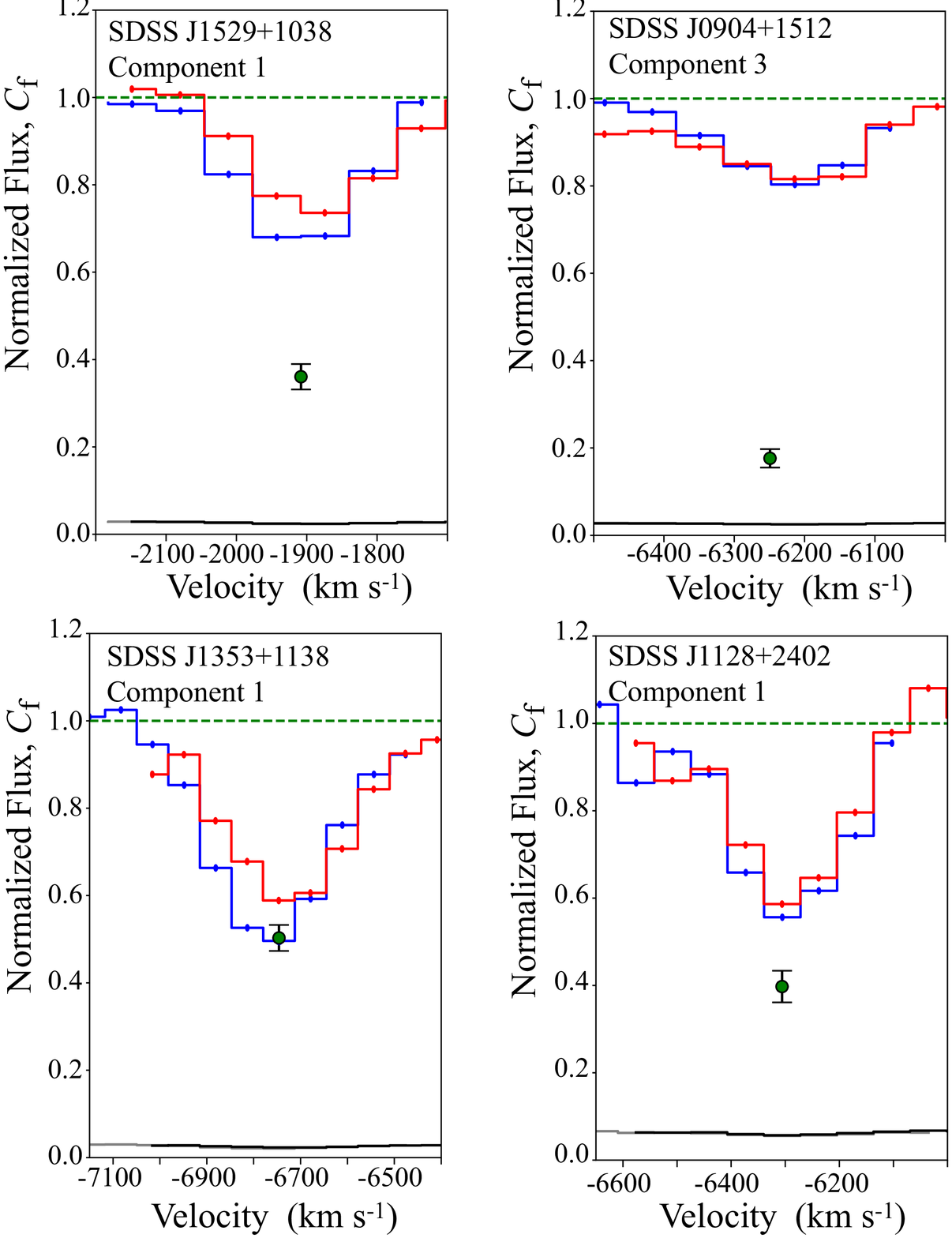}
\caption{Velocity-aligned normalized flux profiles of detected C~IV transitions in NAL systems towards SDSSJ1529, SDSSJ0904, SDSS J1353, and SDSS J1128. 
The blue and red components of the doublet are shown with the blue and red histograms, respectively.
The bottom lines show the normalized 1 $\sigma$ errors.  The coverage fractions evaluated over the central region of the absorption lines are plotted with filled black circles. 
\label{fig:cf}}
\end{figure}

\section{UV Observations and Data Analysis of Subsample} \label{sec:uvobs}

The UV spectra of our subsample of six quasars were taken with the Sloan Digital Sky Survey.  Processed and calibrated spectra were downloaded from the SDSS database\footnote{https://www.sdss.org/dr16/}.  We analyzed each spectrum with a multicomponent fitting code written in Python. Each spectrum was corrected for Galactic extinction using the \cite{2011ApJ...737..103S} map and a Milky Way extinction law of \cite{1999ApJ...525.1011F} with a reddening parameter of $R_{\rm V} = 3.1$. The flux densities were corrected for the gravitational lens magnifications listed in Table \ref{tab:log}. The model fitting function used has the form:
\begin{equation}
{ f }_{ \lambda  }=a+b\lambda +\sum _{ i=1}^{ i=N }{ \frac { 1 }{ \sigma \sqrt { 2\pi  }  }  } { e }^{ -\frac { 1 }{ 2 } \frac { { \left( \lambda-\mu  \right)  }^{ 2 } }{ { \sigma  }^{ 2 } }  }
\end{equation}

The model function was fit to the SDSS spectra using the non-linear least squares Python routine \verb+scipy.optimize.curve_fit+.
The region of interest of the fit covers the C~IV emission line and all significant absorption lines blueward of the emission line.
The C~IV $\lambda\lambda$1548.19,1550.77 doublet ratio is often used to constrain the coverage fraction $C(v)$ and optical depth $\tau(v)$ of the absorbing material in front of the emission source  \citep[e.g.,][]{1997ApJ...478...80H,2004ASPC..311..203H}.
The doublet method of estimating the coverage fraction assumes that the outflowing UV absorbing gas is spatially homogeneous in front of a spatially uniform emission source. The normalized flux densities across the blue and red components of the doublet, assuming a coverage fraction $C(v)$, are given by the following expressions:

\begin{eqnarray}
{ I }_{ B }\left( v \right) =\left( 1-C\left( v \right)  \right) +C\left( v \right) { e }^{ -2\tau (v)  } \\
{ I }_{ R }\left( v \right) =\left( 1-C\left( v \right)  \right) +C\left( v \right) { e }^{ -\tau (v)  } \nonumber
\end{eqnarray}

The solutions to these two equations provide the coverage fraction $C(v)$ and the optical depth $\tau(v)$ as a function of velocity.
A coverage fraction of less than one indicates that the absorber is likely intrinsic and associated with an outflowing wind from the quasar  \citep[e.g.,][]{2007ApJS..171....1M,2020MNRAS.499.3094I}. Intervening absorbers and foreground galaxies have angular sizes considerably larger than the central sources and would produce coverage fractions of $\sim$ 1.

In Table \ref{tab:sdss1} we list the central wavelengths and FWHM of the detected blueshifted absorption lines. We also provide the outflow velocities of the blue-shifted absorption lines. 
We find at least one blueshifted C~IV doublet component in each quasar of the subsample. The FWHM values of all blueshifted C~IV absorption lines in the subsample are $<$ 500 km~s$^{-1}$.  
Based on these velocity widths we classify the blueshifted absorption lines in the subsample as NALs.  In Table \ref{tab:sdss2} we list the equivalent widths of the absorption lines, the coverage fraction $C(v)$, the optical depth $\tau(v)$ averaged over the central region of the absorption lines, and the C~IV ionic column density. 
Several of the absorption line depths are either very shallow in the C~IV doublet or the red component has a slightly larger depth than the blue one within the error bars. In these cases the doublet method yields unphysical values for several velocity bins and we cannot provide estimates for $C_{\rm f}$, $\tau$ and $N_{\rm CIV}$.
We find that 4 of the 6 quasars of the subsample contain an outflowing component with a coverage fraction less than 1 indicating that they are likely intrinsic.
For the two quasars J1442 and J0921, where we cannot constrain the coverage fraction, we find that they contain C~IV doublets with outflow velocities of $\sim$12,700 km~s$^{-1}$ and $\sim$ 5,160~km~s$^{-1}$, respectively. These absorbers are possibly also intrinsic since these outflow speeds are too high for the absorbers to be environmental gas and NAL systems with velocity separations between 3,000 and 12,000~km~s$^{-1}$ are found to be dominated by absorbers intrinsic to and outflowing from the quasar  \citep[e.g.,][]{2019MNRAS.488.5916S}.

We conclude, that based on our estimated coverage fractions and outflow velocities, all the quasars of our subsample contain outflows of UV absorbing gas with velocities ranging between 5,160~km~s$^{-1}$  and 22,740~km~s$^{-1}$.

We have assumed Gaussian optical depth profiles which leads to the following simplified approximation of the 
ionic C~IV column density \citep[e.g.,][]{2017MNRAS.468.4539M}:

\begin{equation}
{ N }_{ CIV }=6.68\times { 10 }^{ 14 }\left( \frac { b{ \tau  }_{ 0,R } }{ { f }_{ R }{ \lambda  }_{ 0,R } }  \right) { cm }^{ -2 }, \label{eq:7}
\end{equation}

\noindent
where $\lambda_{\rm 0,R} $ = 1550.77 \AA\ is the laboratory wavelength of the redder transition of the C~IV doublet,
$f_{\rm R}$ is the line oscillator strength of the $\lambda_{\rm 0,R }$ transition of the doublet, $b$ is the Doppler parameter with units of km~s$^{-1}$,
and $\tau_{\rm 0,R}$ is the line-centre optical depth of the redder transition of the C~IV doublet.

The spectral resolution of the SDSS spectra (1500 at 3800 \AA) is insufficient to resolve the C~IV absorption profiles and many of the absorption lines appear to be saturated. As a result, our estimated values of the ionic column densities of the absorbers listed in Table \ref{tab:sdss2} should be considered as lower limits.    

C~IV $\lambda\lambda$1548.19,1550.77 doublets have intrinsic optical depth ratios of  $\sim$ 2:1.  The observed line depth ratios of $\sim$ 1:1 in most doublets of our subsample indicate strong   saturation, partial covering and large optical depths \citep[e.g.,][]{2019MNRAS.483.1808H}. Saturation may lead to significant underestimates of the column densities of the outflowing UV absorbing gas and estimates of the ionization parameter in these cases are also unreliable. We have thus not attempted to constrain the energetics of the outflowing UV absorbing gas with the available SDSS spectra.

\begin{figure*}[]
\epsscale{1.1}
\plotone{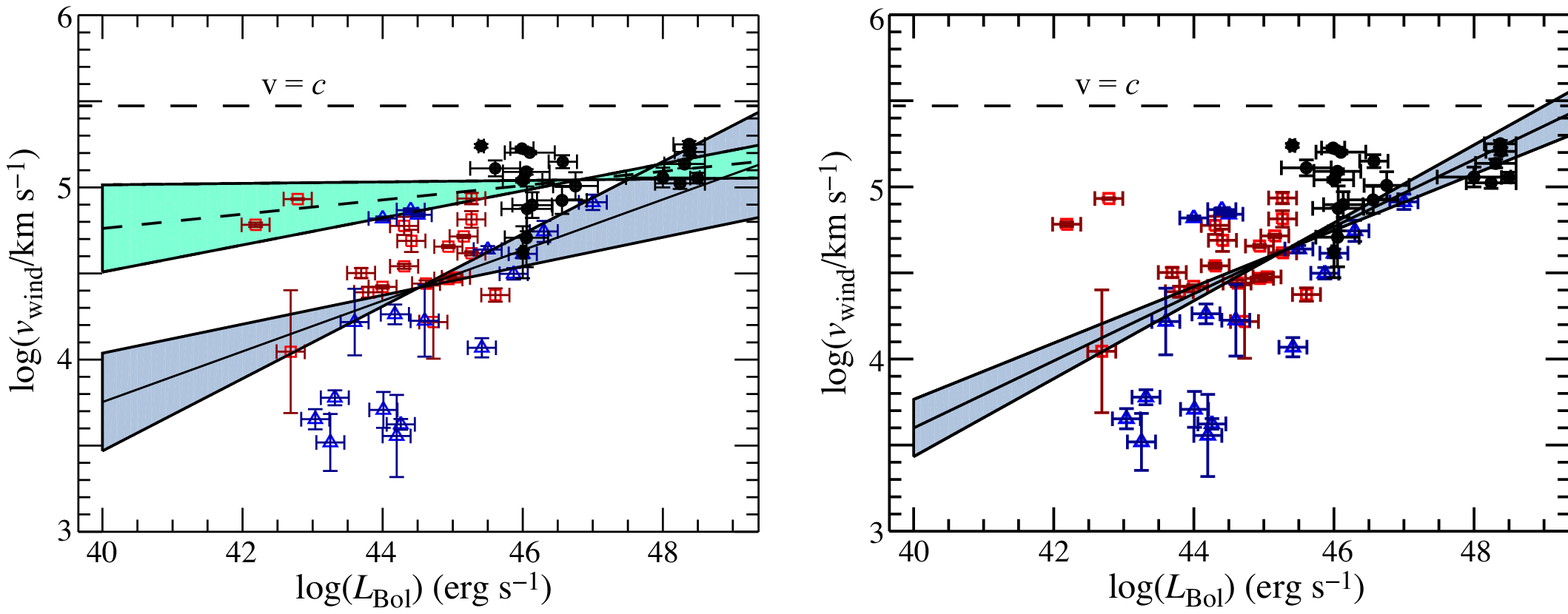}
\caption{The outflow velocity of the ionized absorber versus bolometric luminosity of the low-$z$ Tombesi (red squares),  low-$z$ Gofford (blue triangles) and high-$z$ (black filled circles) AGN samples. (left)  We show the power-law least-squares fits to the combined Tombesi and Gofford samples with the solid line and to the high-$z$ sample with the dashed line. (right)  We show the power-law least-squares fits to the combined Tombesi, Gofford and high-$z$ samples with the solid line. The shaded areas represent the uncertainty of the slopes of our fits to the data. \label{fig:vout}}
\end{figure*}

\section{Discussion}% and Conclusions}
We searched for a possible correlation between the outflow velocity, $v_{\rm wind}$, of X-ray absorbing gas and the bolometric luminosity, $L_{\rm Bol}$, of AGN.
We considered three different AGN samples for our $v_{\rm wind} - L_{\rm Bol}$ correlation analysis. The first sample includes the high-$z$ quasars of our study, with bolometric luminosities and velocities taken from Tables \ref{tab:sample} and \ref{tab:outflow}, respectively. For quasars APM08279, PG1115, HS0810, and SDSSJ1442 that contain two velocity components in the same observation, the largest velocity component is considered in the $v_{\rm wind} - L_{\rm Bol}$ correlation analysis. 
The second sample (referred to as the Tombesi sample) contains the 12 type-1 AGN and 3 type-2 AGN listed in Table 1 of \cite{2012MNRAS.422L...1T}.
The redshifts of the AGN in the Tombesi sample range between 0.00233  and 0.1040. The third comparison sample (referred to as the Gofford sample) includes 15 of the 20 AGN listed in Table 1 of \cite{2015MNRAS.451.4169G}. For the Gofford sample we excluded APM08279 because it is a high-$z$ quasar that is included in our high-$z$ sample, we excluded NGC~3783 and NGC~4395 because their outflow velocities are not constrained, and we excluded MCG-6-30-15 and NGC~3516 because their low outflow velocities  
and relatively low column densities place them close to the warm absorber category of winds.
The Gofford sample is comprised of 3 BLRGs, 9 type-1 AGN and 3 type-2 AGN. The redshifts of the AGN in the Gofford sample range between 0.00234 and 0.18. In cases where multiple observations of an object are available, the outflow velocities and corresponding bolometric luminosities of the individual observations are considered for our 
analysis of the $v_{\rm wind} - L_{\rm Bol}$ data and not the average of these quantities over the observations.

The outflows in the three samples considered for our analysis of the $v_{\rm wind} - L_{\rm Bol}$ data are associated with outflowing absorbers that have ionization parameters larger than $\log (\xi /\rm erg~cm~s^{-1}) = 3$, hydrogen column densities larger than $\log (N_{\rm H} / \rm cm^{-2}) = 22$ and outflow velocities larger than 3,000 km~s$^{-1}$. 
We are not considering outflows commonly associated with warm absorbers that have ionization parameters in the range $\log (\xi /\rm erg~cm~s^{-1}) = -1$ to 3, column densities in the range $\log (N_{\rm H} / \rm cm^{-2})$ = 20--22 and outflow velocities in the range $v$ = 100--2,000 km~s$^{-1}$  \citep[e.g.,][]{2014MNRAS.441.2613L}. We note that the analysis presented in \cite{2010A&A...521A..57T,2012MNRAS.422L...1T}  included outflowing absorbers with velocities above 10,000 km~s$^{-1}$.

\begin{figure}[h]
\epsscale{1.1}
\plotone{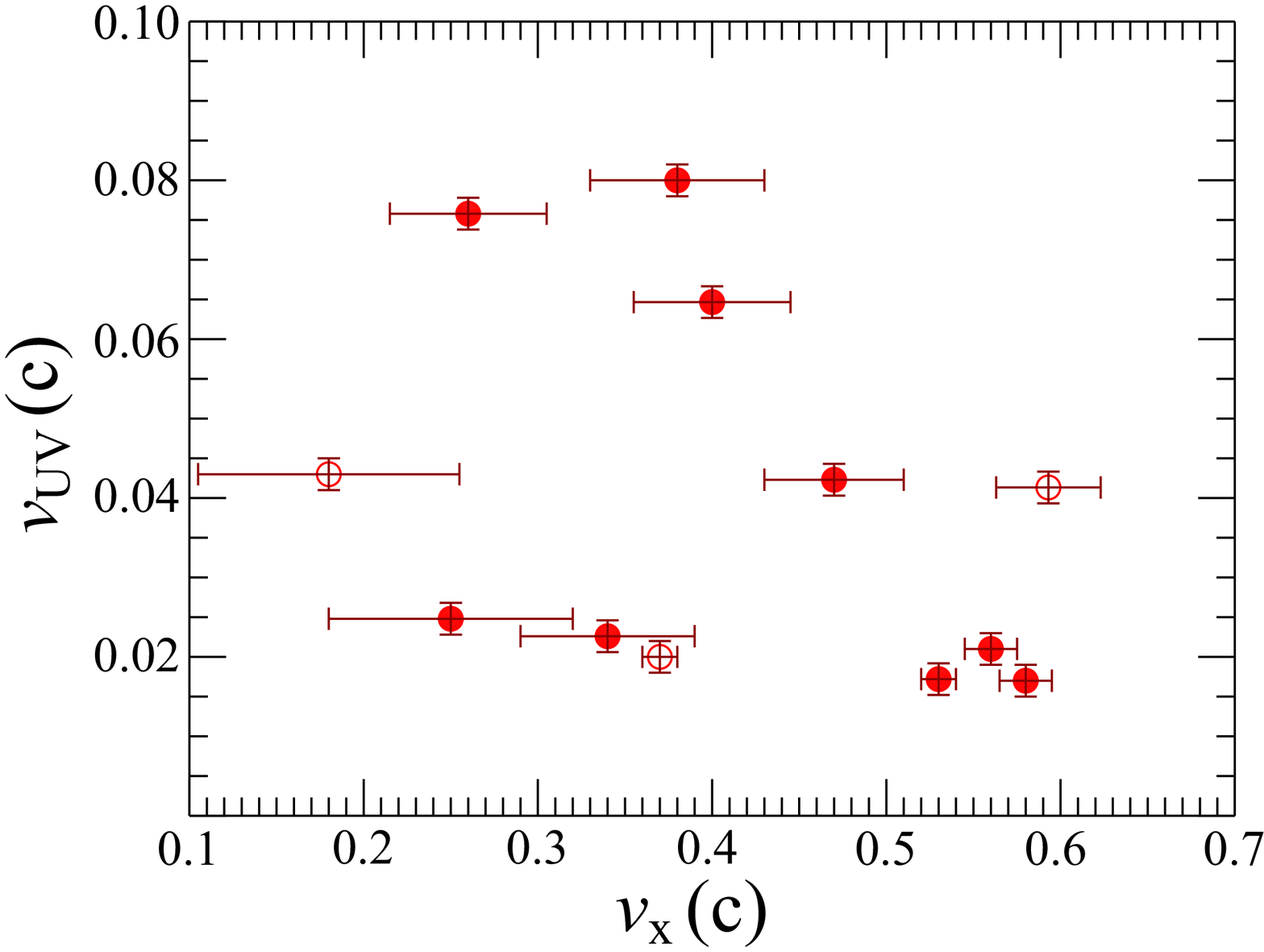}
\caption{The X-ray and UV velocities of the outflowing ionized absorbers of the high-$z$ quasar sample. The NAL quasars are shown with filled red circles and the rest of the sample with open red circles. No significant correlation is found for the entire sample, however, a possible anti-correlation is found for the NAL quasar sample. 
\label{fig:vx_vs_vuv}}
\end{figure}

In Figure \ref{fig:vout} we show the outflow velocities of our high-$z$ AGN sample (in black), the Tombesi low-$z$ sample (in red) and the Gofford low-$z$ sample (in blue)
as a function of bolometric luminosity. For a radiation-driven outflow we expect $v_{wind}$ $\propto$  $L^{1/2}$  \citep[e.g., see equation 1 of][]{2002ApJ...579..169C}.
For the fits to the low-$z$ Tombesi and Gofford samples we find best-fit values of the power-law exponents of $b = 0.02 \pm 0.06$ and $b = 0.29 \pm 0.09$, respectively.
We find a Kendall's rank correlation coefficient of $\tau$~=~0.1 with a null probability  of $P_{\rm null}$~=~0.52 for the low-$z$ Tombesi sample and $\tau$ = 0.45 with a null probability of $P_{\rm null}$~=~8.9~$\times$~10$^{-3}$ 
for the low-$z$ Gofford sample. 
We conclude that the low-$z$ Gofford AGN sample shows a strong and significant correlation between $v_{\rm wind}$ and $L_{\rm Bol}$, in agreement with the result first presented in \cite{2015MNRAS.451.4169G}, whereas no correlation between $v_{\rm wind}$ and $L_{\rm Bol}$ is found for the low-$z$ Tombesi sample. One possible explanation for this difference it that \cite{2012MNRAS.422L...1T} did not include outflows with velocities between 3,000 and 10,000~km~s$^{-1}$.

For the fit to our high-$z$ quasar sample we find a best-fit value of the power-law exponent of $b = 0.03 \pm 0.06$.  We find a Kendall's rank correlation coefficient of $\tau$~=~0.1 with a null probability of $P_{\rm null}$~=~0.55
for the high-$z$ AGN data. For the combined fit to the low-$z$ Tombesi and Gofford samples and our high-$z$ quasar sample we find a best-fit value of the power-law exponent of $b = 0.20  \pm 0.03$, significantly below the value predicted for radiation driving alone.  We find that the fit to the combined samples results in a Kendall's rank correlation coefficient of $\tau$~=~0.51 with a null probability  of $P_{\rm null}$~=~6~$\times$~10$^{-8}$.

In Table \ref{tab:cor} we list the correlation coefficients and best-fit power-law exponents for our analysis of the $v_{\rm wind} - L_{\rm Bol}$ data of the low-$z$ Tombesi and Gofford samples and our high-$z$ quasar sample.  The best-fit value of $b = 0.15 \pm 0.06$ for the combined low-$z$ Tombesi and Gofford samples is also below what would be predicted for radiation driving for the acceleration of accretion disk winds at low-$z$.
The high-$z$ AGN sample alone shows higher outflow velocities and no significant dependence between outflow velocities and bolometric luminosity suggesting that an additional driving mechanism may be contributing to the outflow. Another possibility for the non-dependence between luminosity and outflow velocity for the most luminous objects is that there is a saturation effect in the acceleration process for velocities $\simgt$ 0.1$c$ due to relativistic beaming \citep[e.g.,][]{2007MNRAS.381.1413S,2011ApJ...737...91S,2020A&A...633A..55L}.

\begin{figure}[h]
\epsscale{1.15}
\plotone{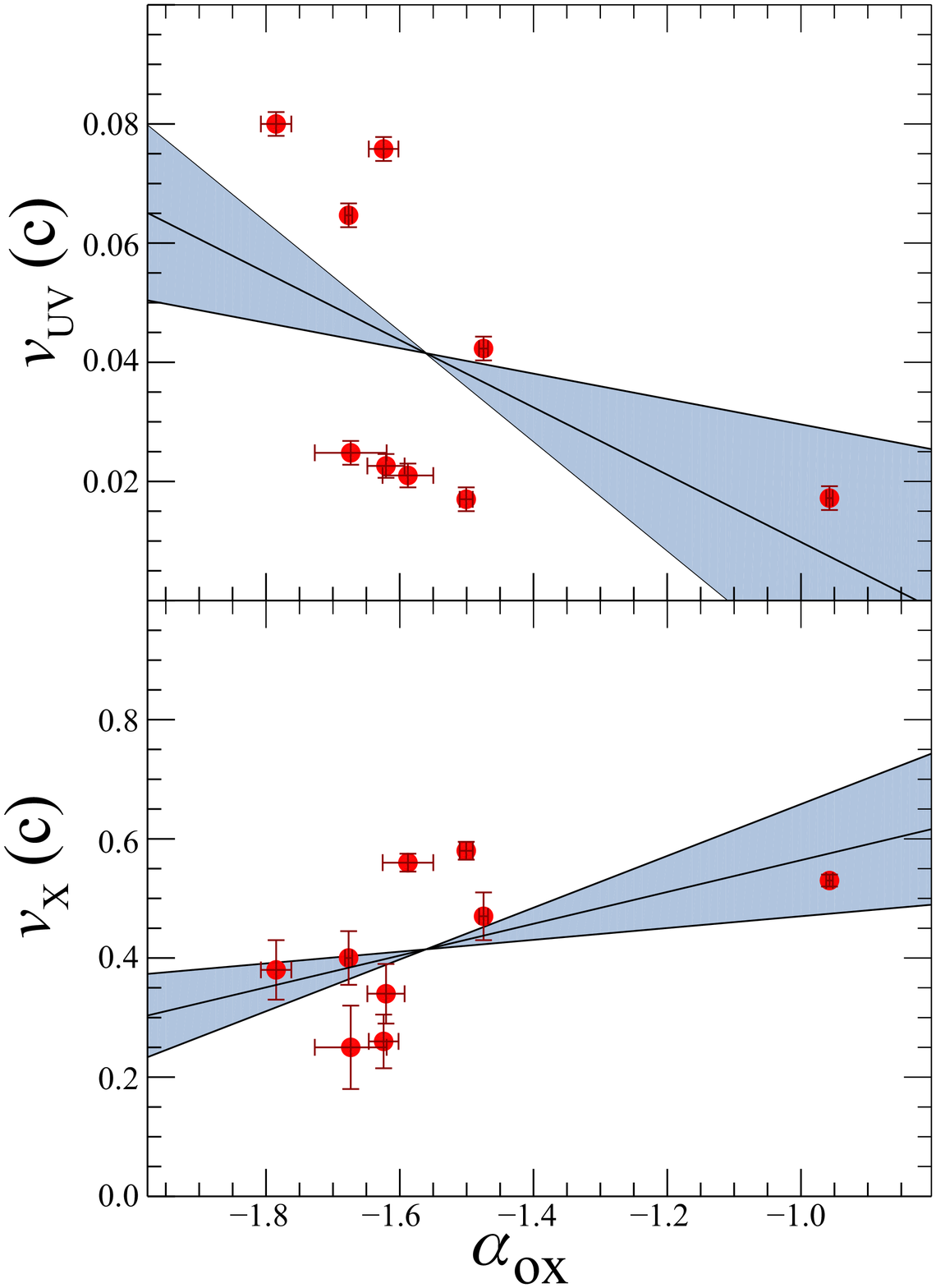}
\caption{The UV (top) and X-ray (bottom) velocities of the outflowing ionized absorbers of the high-$z$ NAL quasars of our sample as a function of the optical-to-X-ray spectral index \aox.  
The solid lines represent the fits to the data and the shaded areas represent the uncertainty of the slopes of our fits to the data.
\label{fig:a_ox}}
\end{figure}

In Figure \ref{fig:vx_vs_vuv} we show the maximum velocities of the UV and X-ray outflowing absorbers of our high-$z$ sample. By including all the quasars in our sample we find no significant correlation between UV and X-ray velocities. If we only include the NAL quasars  (see classification in Table \ref{tab:prop}) we find a possible anti-correlation between the maximum UV and X-ray outflow velocities with Kendall's rank correlation coefficient $\tau$~=~$-$0.5 significant at $>$~94\%. 
To obtain some insight as to the origin of this possible anti-correlation between $v_{\rm UV}$ and $v_{\rm X}$ we calculate the optical-to-Xray spectral slopes
\footnote{\aox\ is defined as the slope of a hypothetical power law extending between 2500\AA\ and 2 keV in the AGN rest frame, i.e. $ \aox ={\rm log} \frac{F_\nu(2keV)}{F_\nu (2500 \Am)}/{\rm log} \frac{\nu(2keV)}{\nu (2500 \Am)}=0.3838~{\rm log} \frac{F_\nu(2keV)}{F_\nu (2500 \Am)} $} 
($\alpha_{\rm ox}$ ) of the NAL quasars of our sample. Several theoretical studies have demonstrated that the spectral energy distribution of the incident flux on the absorbing gas may significantly influence the dynamics of the outflow \citep[e.g.,][]{2011ApJ...737...91S,2013ApJS..206....4K}.  We therefore predict, based on these studies, that the optical-to-X-ray spectral slopes of the NAL quasars of our sample may affect the acceleration of the outflowing absorbers.

In Table \ref{tab:aox} we list the rest-frame 2~keV and 2500~\AA\  flux densities and the calculated $\alpha_{\rm ox}$ values of the NAL quasars.  In Table \ref{tab:aox} we also list the quantity $\Delta\alpha_{\rm ox}$ that represents
the difference between the observed value of $\alpha_{\rm ox}$ and the value expected based on the UV luminosity of the quasar \citep[e.g.,][]{2010A&A...512A..34L}. In all NAL quasars of our sample $\Delta\alpha_{\rm ox}$ is relatively small 
with the exception of SDSSJ0921 with $\Delta\alpha_{\rm ox}$ $\sim$ 0.5. SDSSJ0921 is not significantly absorbed in the X-ray band and the positive value of  $\Delta\alpha_{\rm ox}$  suggests absorption in the UV band.

The X-ray and optical flux densities are corrected for Galactic absorption.
We find that the UV outflow velocities are weakly anti-correlated with $\alpha_{\rm ox}$  (Kendall's rank correlation coefficient of $\tau$~=~$-0.61$ and null probability of $P$~=~0.02) and the X-ray outflow velocities are weakly correlated with $\alpha_{\rm ox}$ (Kendall's rank correlation coefficient of $\tau$~=~$0.44$ and null probability of $P$~=~0.09).  In Figure \ref{fig:a_ox} we show $v_{\rm UV}$ and $v_{\rm X}$ as a function of $\alpha_{\rm ox}$. These trends are suggestive of a possible dependence of the UV and X-ray outflow velocities on the slope of the incident spectral energy distribution and are consistent with an anti-correlation between the maximum UV and X-ray outflow velocities.  The outflows of UV absorbing gas appear to be accelerated to larger velocities in the X-ray-weak NAL quasars of our sample and the outflows of X-ray absorbing gas appear to be accelerated to larger velocities in the X-ray-strong NAL quasars. We caution that the current sample size is relatively small and the correlations between outflow velocities and \aox\ for our sample of NAL quasars are weak. A significant increase of a factor of at least two will be required to confirm these results.

We find no significant correlations between the observed properties of UFO velocity, column density, ionization parameter,  bolometric luminosity, X-ray luminosity, and Eddington ratio in the high-$z$ quasar sample. Conversely,  correlations between $v_{\rm wind}$ versus $L_{\rm X}$ have been reported in individual quasars such as 
 $z=3.91$ APM 08279$+$5255 \citep{2002ApJ...579..169C,2009ApJ...706..644C,2011ApJ...737...91S}, $z=0.184$ PDS~456 \citep{2015Sci...347..860N,2017MNRAS.472L..15M, 2018ApJ...854L...8R}, $z=0.062$ PG~1126$-$041 \citep{2011A&A...536A..49G} and $z=2.7348$ HS~1700+6416 \citep{2012A&A...544A...2L}.
A plausible explanation for this apparent discrepancy is that the outflow velocities of high-$z$ quasars may depend more strongly than low-$z$ AGN on additional driving mechanisms, such as magnetic pressure, and this dependance may vary between quasars. GR-rMHD simulations \citep[e.g.,][]{2017MNRAS.468.1398S} show that both radiative and magnetic driving contribute to accelerating ultrafast outflows.

One possible test of this hypothesis is finding and comparing possible individual correlations between $v_{\rm wind}$ versus $L_{\rm X}$ of quasars in this sample. This test is not yet feasible with the available data.
Differences in the  $v_{\rm wind}$ versus $L_{\rm X}$  correlations of individual objects may lead to a dilution of a single significant correlation in the entire sample.  In support of the presence of a driving mechanism in addition to radiation driving are the observed large fractions of the kinetic luminosity to the bolometric luminosity of the high-$z$ quasars. As reported in Table \ref{tab:outflow}  (column 7), ten out of the fourteen quasars in the sample have fractions of the kinetic luminosity to the bolometric luminosity greater that 0.4, which is not consistent with radiation driving.

Variability of the properties of the ultrafast outflows has been observed and reported in several quasars of our sample including  APM08279, PG1115, HS1700, Q2237,  and HS0810  on timescales comparable to the light crossing time over regions of 10$-$100 $r_{\rm g}$. In APM08729 we found a strong correlation between the maximum outflow velocity and the 2$-$10 keV luminosity 
 (\citep[see Figure 10 of][]{2011ApJ...737...91S}) and between the maximum outflow velocity and the photon index of the X-ray spectrum (steeper spectra resulting in faster outflows). For PG1115 we found \citep[e.g.,][]{2007AJ....133.1849C} that the depths of the X-ray broad absorption features decreased significantly over a period separated by 0.92 yr (proper time)  and detected a marginal decrease over a period separated by 5.9 days (proper time). In \cite{2012A&A...544A...2L} the outflow velocities inferred in HS1700 were found to lie in the range $v = 0.12 - 0.59c$ and vary in energy and width over a timescale of about 5 years.  Long term variability of the relativistic outflow of the $z$=1.51 quasar HS0810 was found between a period of about 10 months \citep[compare figures 6 and 10 of][]{2016ApJ...824...53C}.  
\cite{2020A&A...638A.136B} find clear evidence for intrinsic spectral variability of Q2237 based on  a systematic and comprehensive temporally and spatially resolved X-ray spectral analysis of all the available
{\sl Chandra} and {\sl XMM-Newton} data of this object. They determine the wind duty cycle of Q2237  as $\sim$ 0.31 at 95\% confidence level.

Of the six quasars of our sample selected to contain a UV NAL, all with the exception of SDSS J0921 were observed once and thus do not provide any long-term variability constraints.
SDSS J0921 was observed with the MOS on 2018 May 15 and MOS and pn on 2018 Oct 24. As shown in Figure \ref{fig:spectra} we find a significant  change of the outflow velocity of SDSS J0921 increasing from 0.41$c$ to 0.53$c$. The increase in outflow velocity is associated with a slight increase in the 2$-$10 keV luminosity from $6.1^{+0.2}_{-0.2}$ $\times$ 10$^{45}$ to $7.2^{+0.2}_{-0.2}$ $\times$ 10$^{45}$ erg~s$^{-1}$.  This trend is consistent with that detected in APM 08279, however additional monitoring of SDSS J0921 would be required to determine the strength and significance of any correlations.

Six of the high-$z$ quasars of our sample were selected based on the criteria that they be gravitationally lensed and they contain a NAL in their UV spectra. 
Based on our exploratory XMM-Newton survey of these six NAL quasars, the coexistence of UV NALs and ultrafast outflows is found to be significant in $\simgt$ 83\% of the subsample. Our current subsample of NAL quasars is small and at least a doubling of its size would be required for our conclusion to be considered significant in the general population of quasars. For comparison, the fraction of AGN that contain ultrafast outflows in samples that are unbiased with respect to the presence of UV winds is about 40\% \citep{2010A&A...521A..57T,2013MNRAS.430...60G}.
Theoretical simulations of MHD and radiative accretion disk winds find stratified outflows with higher velocity components of the wind launched from smaller distances from the black hole 
\citep[e.g.,][]{2015ApJ...805...17F,2004ApJ...616..688P}. These simulations predict that certain lines of sight will produce detectable outflows of both UV and X-ray outflowing absorbing material. One possible explanation for the large detection rate of UFOs in our NAL quasar sample is that UV and X-ray observations of intrinsic NAL quasars sample different parts of the outflow along the same line of sight, with the X-ray absorbers located closer to the continuum source, thus having a higher ionization level than the UV absorbers, which are likely located further out.

\cite{2017ApJ...837..149G} show that for UFOs to be effective in providing feedback to their host galaxy halos they must be able to produce entrainment and local condensation at macro-scales ($\simgt$kpc), 
leading to multiphase outflows.  \cite{2017ApJ...837..149G} also find typical velocities of several  $\sim$10$^{3}$~km~s$^{-1}$ for the outflows of UV and optical absorbing ionized gas, similar in value to the outflow velocities of the UV absorbing gas of our NAL quasar sample.

\begin{figure}[h]
\epsscale{1.15}
\plotone{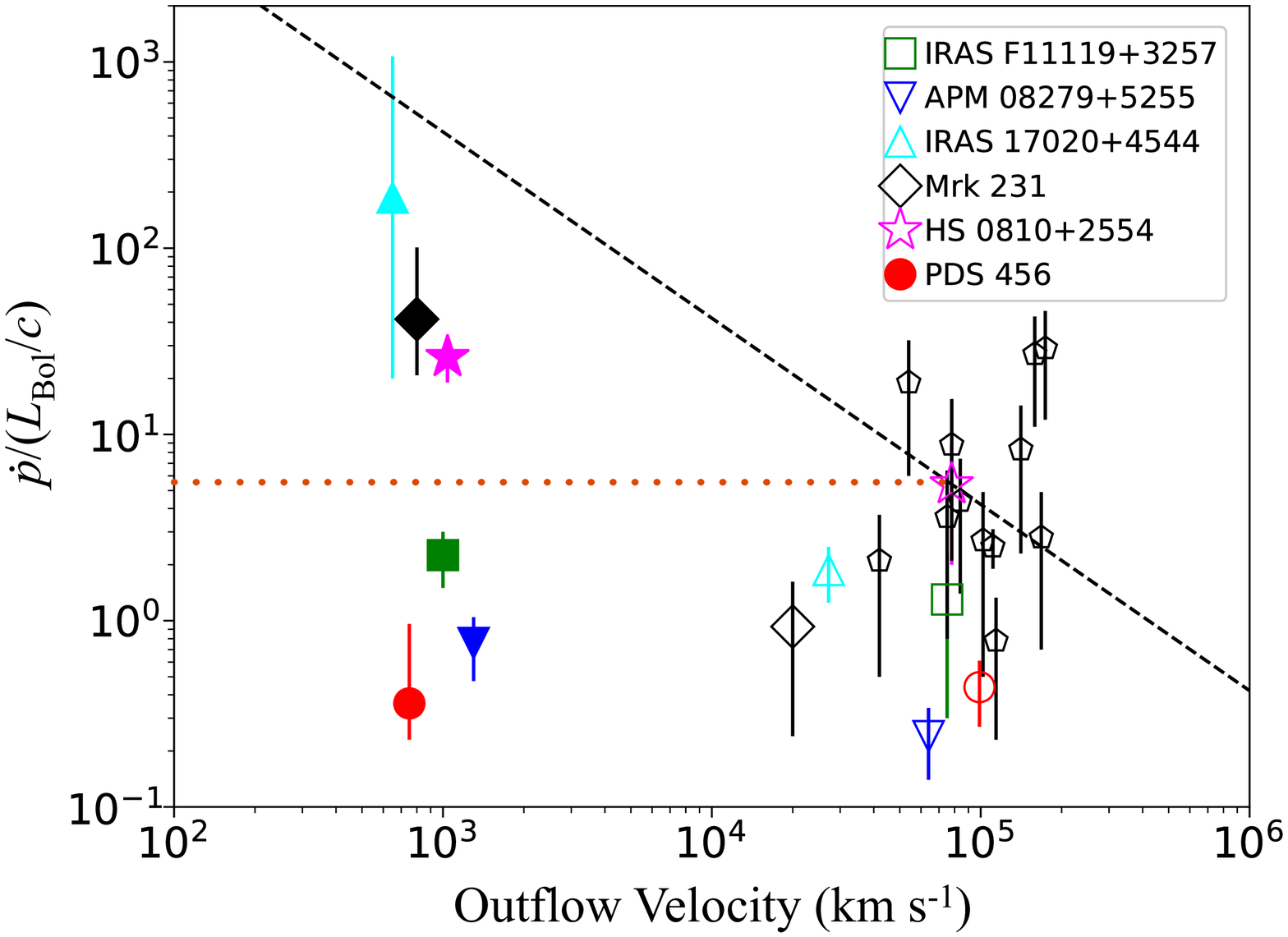}
\caption{The momentum boost $\dot { { p } }/(L_{Bol}/c)$ as a function of outflow velocity. 
The filled and unfilled symbols correspond to the molecular and ultrafast outflows, respectively. The dashed and dotted lines represent the dependence of the momentum boost with outflow velocity for energy-conserving and momentum-conserving outflows, respectively for \hs. The data from this work are the open pentagon symbols.
The data for AGN other than this work have been obtained from the literature 
\citep{2019A&A...628A.118B,2015Natur.519..436T,2017ApJ...843...18V,2017A&A...608A..30F,2015A&A...583A..99F,2020MNRAS.496..598C}.
\label{fig:momentum_boost}}
\end{figure}

An indicator of the impact of an ultrafast outflow on the interstellar medium is the momentum boost, $\dot { { p } }_{\rm mo}/(L_{Bol}/c)$.
In Figure \ref{fig:momentum_boost} we show the momentum boost plotted against the outflow velocity for the ultrafast outflows of the high-z quasars presented in this work.  For comparison, we also show the momentum boosts and velocities of the ultrafast and molecular outflows of several other AGNs based on published results \citep[see][and references therein]{2020MNRAS.496..598C}.
Half of our sample of high-$z$ quasars have momentum boosts that are considerably larger than those presently known to have both ultrafast and molecular outflows. 
If the small-scale relativistic outflows drive large-scale molecular outflows we predict that a large fraction of the quasars in our sample will also show molecular outflows with considerably larger momentum boosts than the comparison AGN sample shown in Figure \ref{fig:momentum_boost}. We note, however, that this prediction assumes that the energetics of the ultrafast winds have not varied significantly over the past $10^{6 }- 10^{7}$ years, the time for the impact of these winds to travel to the observed distances ($1-10$ kpc) of molecular outflows.

\section{Conclusions}
We presented results from a comprehensive study of UFOs detected in a heterogeneous sample of fourteen quasars in a redshift range of 1.41 -- 3.91. 
A unique and unifying characteristic of our quasar sample is that all the X-ray observations display high quality spectra, to our knowledge, the best for quasars with $z \simgt 1$.
Seven of these 14 quasars have reported ultrafast outflows in the literature, and one, SDSS~J1029$+$2623, was previously studied but not searched for UFOs.
Six of the 14 quasars of our sample, referred to as our subsample, were selected to contain a UV NAL without prior knowledge of the existence of a UFO. This subsample is therefore unbiased towards UFO detection and is used to infer the fraction of $z > 1$ NAL quasars that contain UFOs.

The main conclusions of our spectral analyses of a sample of 14 high-$z$ quasars are the following:

\begin{enumerate}

\item{Ultrafast outflows of X-ray absorbing material are a common property of our sample of 14 high-$z$ quasars (1.44 $< z < $3.91).
%UFOs are detected at $>$ 99\% confidence in 13 of the14 quasars and possibly detected at  $>$ 90\% confidence in the remaining one.
Interestingly, all objects but one show UFO signatures. We have detected a UFO in the lensed quasar SDSS J1029+2623, that was previously studied but not searched for UFOs.  Our current study has almost doubled the number of detected UFOs in quasars at $z > 1.4$. The presence of ultrafast outflows is also supported by the detection of P-Cygni profiles in a large fraction of the sample. Specifically, we find that the X-ray spectra of 5(3) of the 14(6) quasars in our sample(subsample) contain P-Cygni profiles.}

\item{{The estimated momentum boosts of 12 of the 14 quasars in our sample are $\dot{p}/(L_{\rm Bol}/c)$~$>$~1.}
Numerical calculations  \citep[e.g.,][]{2016MNRAS.458..816H,2020NatAs...4...10G} have demonstrated that high-velocity winds with momentum fluxes of  $\sim L/c$ suppress the star formation rate and black hole accretion rate in the galactic nucleus. The estimated momentum boosts for our sample confirm that relativistic winds of high-$z$ quasars have a dramatic effect on the circum-BH ISM. }

\item{Relativistic outflows are detected (at $>$ 99\% confidence) in 5 of the 6 lensed quasars of our subsample of quasars that were selected to contain a UV NAL. 
Based on our estimated coverage fractions and UV outflow velocities, all the quasars of our subsample contain outflows of UV absorbing gas with velocities ranging between 5,160~km~s$^{-1}$ and 22,740~km~s$^{-1}$.}
The coexistence of UV NALs and ultrafast outflows is found to be significant in $>$ 83\% of the six quasars selected to contain a UV NAL.
For comparison, the fraction of AGN that contain ultrafast outflows in samples that are unbiased with respect to the presence of UV winds is about 40\% \citep{2010A&A...521A..57T,2013MNRAS.430...60G}.  Despite the variable AGN feedback duty cycle involved and the small subsample size, our findings suggest a key link between multiphase AGN feedback properties of the micro-scale UFOs and meso-scale UV outflows, as predicted by theoretical models \citep[e.g.,][]{2020NatAs...4...10G}.

\item{We find a possible anti-correlation between the maximum velocities of the outflowing UV and X-ray absorbers of the NAL quasars in our sample.  
This anti-correlation also supports a possible link between the micro-scale UFOs and meso-scale UV outflows.
An increase of the number of high-$z$ NAL quasars by at least a factor of two will be required to confirm this anti-correlation.}

\item{The large kinematic luminosities of the ultrafast outflows compared to their bolometric luminosities (see Figures \ref{fig:effic1} and \ref{fig:effic2}) implies that radiation driving alone cannot explain the acceleration of these winds. We propose that magnetic driving may be a significant contributor to their acceleration as predicted by numerical simulations \citep[][]{2017MNRAS.468.1398S}.}

\item{Our high-$z$ quasar sample has outflow velocities ranging between $\sim$ 0.1 and $\sim$ 0.6$c$, significantly higher than those found in low-redshift AGN (see Figure \ref{fig:vout}).  We find no significant dependence between outflow velocities and bolometric luminosity for the high-$z$ quasar sample suggesting that an additional driving mechanism may be contributing to the outflow and/or that relativistic beaming is producing a saturation effect in the acceleration process for velocities $\simgt$ 0.1$c$. For the combined fit to the Tombesi and Gofford samples of low-$z$ AGN and our high-$z$ quasar sample we find a best-fit value of the power-law exponent of $b = 0.20  \pm 0.03$.  
This power-law exponent is significantly below the predicted value of $\sim$ 0.5 for radiation driving suggesting that another driving mechanism, such as magnetic driving, may be contributing to the acceleration of the outflow. This possibility is also suggested by the extreme kinetic luminosities and power boosts of a large fraction of the quasars in our sample.}

\end{enumerate}

\facility{CXO, XMM-Newton, SDSS}

\software{glafic \citep[v1.1.6;][]{2010ascl.soft10012O}, SAS  \citep[v18;][]{2004ASPC..314..759G}, CIAO  \citep[v4.12;][]{2006SPIE.6270E..1VF}, XSPEC  \citep[v12;][]{1996ASPC..101...17A}, XSTAR photoionization model warmabs \citep{2001ApJS..133..221K,1996ApJ...465..994K}, Astropy \citep{2013A&A...558A..33A,2018AJ....156..123A}, SciPy \citep{2020NatMe..17..261V}, NumPy \citep{2020Natur.585..357H}}

\acknowledgments
We acknowledge financial support from NASA via the grants 80NSSC19K095 and NNX16AH33G. 
M.Ga. acknowledges partial support by NASA Chandra GO8-19104X/GO9-20114X and HST GO-15890.020-A.
M.Gi. is supported by the ``Programa de Atracci\'on de Talento'' of the Comunidad de Madrid, grant number 2018-T1/TIC-11733.
We thank Daniel Proga, Bing Zhang, Cristian Saez, Monika Moscibrodzka, and Tim Waters for useful comments and suggestions.
We greatly appreciate the useful comments made by the referee. Scientific results reported in this article are based partly on observations made by the {\sl Chandra X-ray Observatory} (CXO) and {\sl XMM-Newton}, an ESA science mission with instruments and contributions directly funded by ESA Member States and NASA.
This publication makes use of data from the Sloan Digital Sky Survey (SDSS).

\clearpage
\begin{deluxetable*}{lcllcl}
\tablecaption{Properties of Quasar Sample \label{tab:prop}}
\tablecolumns{6}
\tablenum{1}
\tablewidth{0pt}
\tablehead{
\colhead{Object}    & \colhead{Classification}    & \colhead{$z_{\rm s}$}     & \colhead{$z_{\rm l}$}     & \colhead{$N^{Gal}_{\rm H}$~\tablenotemark{a}}  & \colhead{$\log(M_{\rm BH})$ } \\
\colhead{}              & \colhead{}                         & \colhead{}                        & \colhead{}                       & \colhead{($10^{20}~{\rm cm}^{-2}$)}                    & \colhead{(M$_\odot$)}  
}
\startdata
& & & & & \\ 
APM~08279+5255  & BALQSO          & 3.91         & 1.01                            & 3.84  & 10.0$^{+0.1}_{-0.1}${}$^{c}$  \\ 
HS~1700+6416       & NALQSO         & 2.735        & --\tablenotemark{b}    &2.66  & 10.2$^{+0.2}_{-0.2}${}$^{d}$\\ 
MG~J0414+0534    & QSO                 & 2.64         &  0.9584                        &  11.4  & 9.0$^{+0.2}_{-0.2}${}$^{e}$    \\              
SDSS~J1442+4055  & NALQSO  &  2.593      & $\sim$0.4            & 1.30         &  $9.7^{+0.2}_{-0.2}${}$^{d}$  \\
SDSS~J1029+2623 & NALQSO         &  2.197     & 0.58              &1.78   & 8.8$^{+0.2}_{-0.2}${}$^{d}$\\ 
SDSS~J1529+1038 & NALQSO  & 1.984    &   $\sim$0.4                     & 2.72      &  8.9$^{+0.2}_{-0.2}${}$^{d}$        \\
SDSS~J0904+1512  & NALQSO  & 1.826       & $\sim$0.3  & 3.69   & 9.3$^{+0.2}_{-0.2}${}$^{d}$ \\  
PG~1115+080         & mini-BALQSO  & 1.72         & 0.31           & 3.53  & 8.8$^{+0.2}_{-0.2}${}$^{f}$\\
Q~2237+0305         & QSO                & 1.695        & 0.0386        &5.43  & 9.1$^{+0.4}_{-0.4}${}$^{f}$\\ 
SDSS~J1353+1138 & NALQSO  & 1.627        & $\sim$0.25 &1.86 & 9.4$^{+0.2}_{-0.2}${}$^{d}$ \\ 
SDSS~J1128+2402  & NALQSO  &   1.608         &   --{}$^{h}$                &  1.15        & 8.7$^{+0.2}_{-0.2}${}$^{d}$\\
PID352                    & QSO                & $\sim$1.6 & --\tablenotemark{b}               &  0.70    & 8.7$^{+0.4}_{-0.4}${}$^{g}$ \\ 
HS~0810+2554       & NALQSO         & 1.51         & 0.08           & 3.94   & 8.6$^{+0.2}_{-0.2}${}$^{f}$\\ 
SDSS~J0921+2854  & NALQSO  &  1.41          &  0.445           &    2.30      & 8.9$^{+0.2}_{-0.2}${}$^{d}$  \\
\enddata
\tablenotetext{a}{Galactic absorption due to neutral gas.}
\tablenotetext{b}{This quasar is not known to be gravitationally lensed.}
\tablenotetext{c}{Based  on reverberation mapping of the Si IV and C IV emission lines published in Saturni et al. (2016). }
\tablenotetext{d}{Based  on our C IV virial black hole mass estimates using SDSS spectra  corrected using the prescription published in \cite{2017MNRAS.465.2120C}.}
\tablenotetext{e}{Based on $H_{\beta}$ virial black hole mass estimates published in \cite{2007ApJ...661...19P}.}
\tablenotetext{f}{Based on $H_{\beta}$ virial black hole mass estimates published in \cite{2011ApJ...742...93A}.}
\tablenotetext{g}{Based on SED fitting published in Vignali et al. (2015).}
\tablenotetext{h}{Lens redhshift is currently unknown.}
\end{deluxetable*}

%\clearpage
\begin{deluxetable*}{llcccc}
\tablecaption{Log of Observations of {\it Chandra} and \xmm\  Quasar Sample \label{tab:log}}
\tablecolumns{6}
\tablenum{2}
\tablewidth{0pt}
\tablehead{
\colhead{}              & \colhead{}                       & \colhead{}                     & \colhead{}                           & \colhead{Effective}                                        & \colhead{} \\
\colhead{Object}   &\colhead{Observation}     & \colhead{Observatory}  & \colhead{Observational}    &  \colhead{Exposure Time\tablenotemark{a}}  & \colhead{$N_{sc}$\tablenotemark{b}}   \\
\colhead{}              & \colhead{Date}               & \colhead{}                        & \colhead{ID}                     & \colhead{(ks)}                                                  & \colhead{(counts)} 
}
\startdata
APM~08279+5255                   & 2002 Feb 24   &   {\it Chandra}          & 2979              &  88.8                                &   5,636$\pm$75\\
APM~08279+5255                   & 2002 April 28          &   {\it XMM-Newton}  & 0092800201   &  83.46/72.59/73.32    &  13,053$\pm$140/3,307$\pm$59/3,283$\pm$59  \\
APM~08279+5255                   & 2007 Oct  6     &   {\it XMM-Newton}   & 0502220201  &  56.38/59.85/43.74     & 11,400$\pm$114/3,547$\pm$62/2,418$\pm$51 \\
APM~08279+5255                   & 2007 Oct  22   &    {\it XMM-Newton}  & 0502220301  &  60.37/72.70/72.14    & 16,698$\pm$133/5,847$\pm$78/5,589$\pm$77 \\
APM~08279+5255                   & 2008 Jan 14   &   {\it Chandra}            & 7684              &  88.1                                &   6,938$\pm$83\\
HS~1700+6416                        & 2000 Oct 31    &    {\it Chandra}          & 547                &  49.5                                & 380$\pm$20  \\
%HS~1700+6416                        & 2002 May 31      &   {\it XMM-Newton}  & 0107860301   &  10.8/12.9/13.2           & 312$\pm$18/145$\pm$12/155$\pm$13 \\
MG~J0414+0534                      & 2017 Mar 11  &   {\it XMM-Newton}  & 0781210301  &  59.4/71.52/71.55     &    5,319$\pm$77/1,814$\pm$44/1,991$\pm$46 \\
SDSS~J1442+4055                 & 2019 Jan 18  &  {\it XMM-Newton}    & 0822530101 &  51.2/ 86.2/85.4 &  2,190$\pm$48 / 720$\pm$29 / 920 $\pm$ 32 \\
SDSS~J1029+2623                 & 2010 Mar 11     &  {\it Chandra}          & 11755             & 111.5                              & 2,280$\pm$48 \\
SDSS~J1529+1038                 & 2015 July 19     &{\it XMM-Newton}   & 0762520201 & 80.3/91.1/91.3                                 &  1,826$\pm$45/460$\pm$22/499$\pm$23 \\
SDSS J0904+1512                  & 2015 Nov 8 &{\it XMM-Newton}       &  0762520101   & 49.8/72.9/74.7                            & 1,493$\pm$41/677$\pm$27/795$\pm$30 \\
PG~1115+080                          & 2001 Nov 25         &    {\it XMM-Newton}    & 0082340101 &  52.37/60.40/60.04           &10,783$\pm$109/3,369$\pm$60/3,299$\pm$60 \\
PG~1115+080                          & 2004 Jun 10         &    {\it XMM-Newton}   & 0203560201 &  51.47/66.07/66.13           & 10,814$\pm$107/3,855$\pm$64/3,544$\pm$61 \\
PG~1115+080                          & 2004 Jun 26         &    {\it XMM-Newton}   & 0203560401 &  54.92/81.88/81.88           & 9,867$\pm$105/4,247$\mp$65/3,943$\pm$65\\
Q~2237+0305                           & 2017 Jan 4     &    {\it Chandra}          & 19639            & 32.92                              & 577$\pm$24 \\
SDSS~J1353+1138                  & 2016  Jan 13   &   {\it XMM-Newton}  & 0762520101  & 31.80/51.08/50.29     & 1,788$\pm$44/748$\pm$29/845$\pm$31  \\
SDSS~J1128+2402                 & 2018 May 28  &  {\it XMM-Newton}    &0822530201 &  29.27/35.12/35.14 &    1,994$\pm$47/409$\pm$22/547$\pm$ 25 \\
PID352                                     & ${}^c$                 &     {\it XMM-Newton}   & ${}^c$            & 3450                                       &   2,560/690/1,250                                   \\
%SDSS J1529+1038                   & 2015  July   19       &   {\it XMM-Newton}   & 0762520201  & (80.67)(91.12)(91.30)    &  1824$\pm$45/613$\pm$30/594$\pm$30  \\ 
%APM~08279+5255                   & 2001 October 30    &    {\it XMM-Newton}  & 0092800101  &      & \\
HS~0810+2554                        & 2013 Dec 16 &   {\it Chandra}           &16110           &  97.73                               & 5,776$\pm$76 \\
HS~0810+2554                        & 2016 Dec 23 &   {\it Chandra}           & 18204          &   96.76                              & 8,525$\pm$92  \\
HS~0810+2554                        &2014 Oct 04   &   {\it XMM-Newton}     &  0728990101  &  46.4/54.97/54.99     & 11,811$\pm$114/3,131$\pm$59/3,009$\pm$56 \\
SDSS~J0921+2854                 & 2018 May 15  &  {\it XMM-Newton}    & 0822530301 & \tablenotemark{d}/32.75/32.76&  \tablenotemark{d}/2,425$\pm$50/2,625$\pm$52 \\
SDSS~J0921+2854                 & 2018 Oct 24  &  {\it XMM-Newton}    & 0822530501 & 25.34/30.69/30.73       &9,991$\pm$105/2,885$\pm$55/2,978$\pm$55 \\
\enddata
\tablenotetext{a}{Exposure time is the effective exposure time remaining after the application of good time-interval (GTI) tables and the removal of portions of the
observation that were severely contaminated by background flaring. When multiple exposure times are provided they correspond to the exposure time of the pn, MOS1, and MOS2  instruments and are listed in this order.} 
\tablenotetext{b}{Background-subtracted source counts including events with energies within the 0.2$-$10 keV band, except for PID352 where we list the counts in the 0.3$-$7~keV band. When multiple source counts are provided they correspond to the counts of the pn, MOS1, and MOS2  instruments and are listed in this order.}
\tablenotetext{c}{PID352 lies within the {\it Chandra} Deep Field South that was observed with {\sl XMM-Newton} with 33 exposures over the years 2001--2002 and 2008--2010, for a total exposure time of $\sim$3.45 Ms.}
\tablenotetext{d}{Due to an operational problem, EPIC-pn was not functional during the observation of SDSS J0921+2854 in revolution 3376.}
\end{deluxetable*}

%\clearpage
\begin{deluxetable*}{llcccclcc}
\tablecaption{Luminosities of Quasar Sample \label{tab:sample}}
\tablecolumns{9}
\tablenum{3}
\tablewidth{0pt}
\tablehead{
\colhead{Object} & \colhead{Observation} &\colhead{$f_{\rm 2-10~keV}$\tablenotemark{a}} & \colhead{$L_{\rm 2-10~keV}$\tablenotemark{b}} & \colhead{$L_{\rm ion}$\tablenotemark{c}} & \colhead{$\log({\lambda}L_{\lambda}/\rm erg~s^{-1})${}$^{d}$} & \colhead{$\log(L_{\rm Bol}/\rm erg~s^{-1})$\tablenotemark{e}} & \colhead{$L_{\rm Bol}/L_{\rm Edd}$\tablenotemark{f}} & \colhead{$\mu$ }  \\
\colhead{} & \colhead{Date} & \colhead{(erg~s$^{-1}$) } &  \colhead{(erg~s$^{-1}$) } & \colhead{(erg~s$^{-1}$) } & \colhead{$1450~{\rm \AA}$} & \colhead{2$-$10~keV/$1450~{\rm \AA}$ } & \colhead{} & \colhead{}
}
\startdata
APM~08279+5255       &2002 Feb 24  &4.3$^{+0.2}_{-0.1}$  & $43^{+2}_{-1}$       & $12^{+0.6}_{-0.3}$        & --               & 47.9/47.83$_{}^{}${}$^{g}$      & 0.57$_{-0.13}^{+0.16}$  & 4\\
APM~08279+5255       &2002 April 28 & 4.1$^{+0.2}_{-0.2}$ & $53^{+3}_{-3}$       & $18^{+1.0}_{-1.0}$        & --               & 48.0/47.83$_{}^{}${}$^{g}$      & 0.57$_{-0.16}^{+0.19}$ & 4\\
APM~08279+5255       &2007 Oct  6   & 3.9$^{+0.2}_{-0.2}$ & $67^{+3}_{-3}$       & $32^{+1.4}_{-1.4}$         & --               & 48.2/47.83$_{}^{}${}$^{g}$      &0.57$_{-0.26}^{+0.28}$ & 4\\
APM~08279+5255       &2007 Oct  22 & 5.0$^{+0.2}_{-0.2}$ & $95^{+3}_{-3}$       & $56^{+1.8}_{-1.8}$         &--                & 48.4/47.83$_{}^{}${}$^{g}$      &0.57$_{-0.35}^{+0.38}$ & 4\\
APM~08279+5255       &2008 Jan  14 & 4.5$^{+0.2}_{-0.2}$ & $65^{+3}_{-3}$       & $25^{+1.2}_{-1.2}$         & --               & 48.2/47.83$_{}^{}${}$^{g}$      &0.57$_{-0.26}^{+0.28}$ & 4\\
HS~1700+6416            &2000 Oct 31  & 0.55$^{+0.1}_{-0.05}$ &2.3$^{+0.5}_{-0.2}$ & $0.62^{+0.1}_{-0.1}$     & 47.67       &   47.0/48.15$_{}^{}$                  &0.8$_{-0.5}^{+1.3}$ & 1\\
MG~J0414+0534         & 2017 Mar 11  & 3.3$^{+0.02}_{-0.02}$ & $15.0^{+0.1}_{-0.2}$  &  $3.9^{+0.4}_{-0.4}$      &  --            & 45.9/46.56$^{h}$               &0.26$_{-0.15}^{+0.35}$     & 47 \\
SDSS~J1442+4055     & 2019 Jan 18   & 2.5$^{+0.1}_{-0.1}$ & $3.7^{+0.2}_{-0.2}$  & $1.2^{+0.2}_{-0.2}$ &  46.10             &    46.2/46.6                    &0.06$_{-0.03}^{+0.06}$    & 6.3\\
SDSS~J1029+2623     &2010 Mar 11   & 1.7$^{+0.1}_{-0.1}$ & $3.4^{+0.2}_{-0.2}$ &  $0.68^{+0.04}_{-0.04}$ & 44.82       & 45.3/45.40                 &0.03$_{-0.01}^{+0.02}$ & 30 \\
SDSS~J1529+1038      &2015 July 19 & 0.31$^{+0.10}_{-0.10}$ & $0.8^{+0.3}_{-0.3}$  & $0.21^{+0.08}_{-0.08}$ &45.54        & 45.33/46.06&  0.12$_{-0.07}^{+0.14}$&  7\\
SDSS~J0904+1512      &2015 Nov 8  & 0.75$^{+0.1}_{-0.1}$ & $1.6^{+0.3}_{-0.3}$ & $0.66^{+0.09}_{-0.09}$& 45.62       &  45.6/46.13                 &0.06$_{-0.03}^{+0.06}$ & 9\\
PG~1115+080              &2001 Nov 25    & 4.0$^{+0.1}_{-0.1}$ & $6.6^{+0.2}_{-0.2}$  & $1.94^{+0.06}_{-0.06}$ &45.47        & 45.4/46.0              &0.16$_{-0.09}^{+0.18}$ & 46\\
PG~1115+080              &2004 Jun 26    & 4.0$^{+0.1}_{-0.1}$ & $7.6^{+0.2}_{-0.2}$ & $2.76^{+0.07}_{-0.07}$ &45.47        & 45.5/46.0                &0.16$_{-0.09}^{+0.18}$ &46 \\
PG~1115+080              &2004 Jun 10    & 3.5$^{+0.2}_{-0.2}$ & $6.3^{+0.3}_{-0.3}$ & $2.14^{+0.09}_{-0.09}$ &45.47       &  45.4/46.0                 &0.16$_{-0.09}^{+0.18}$ & 46\\
Q~2237+0305              &2017 Jan 4     & 1.9$^{+0.1}_{-0.1}$ & $2.2^{+0.1}_{-0.1}$  & $0.43^{+0.02}_{-0.02}$ & 45.53      & 45.4/46.05                &0.08$_{-0.04}^{+0.10}$ & 16\\
SDSS~J1353+1138      &2016  Jan 13  & 1.2$^{+0.2}_{-0.2}$ & $2.1^{+0.3}_{-0.3}$  & $0.8^{+0.1}_{-0.1}$        &  46.32     & 46.1/46.75                &0.20$_{-0.11}^{+0.22}$  & 4 \\
SDSS~J1128+2402     & 2018 May 28  & 1.1$^{+0.3}_{-0.2}$ & $1.3^{+0.4}_{-0.3}$ &  $0.23^{+0.03}_{-0.03}$   &  45.46          &  45.7/46.0                & 0.15$_{-0.07}^{+0.13}$  & 6\\
PID352                         &                      & 0.28$^{+0.03}_{-0.03}$ & 2.8$^{+0.3}_{-0.3}$                  & --         &    --    &   45.9/$46.0$   &      0.13$_{-0.07}^{+0.17}$     & 1\\
HS~0810+2554            &2013 Dec 16   & 3.3$^{+0.1}_{-0.2}$ & $4.8^{+0.4}_{-0.4}$ & $1.9^{+0.2}_{-0.2}$ & 45.04             & 44.9/45.6                   &0.08$_{-0.04}^{+0.10}$ & 103 \\
HS~0810+2554            &2016 Dec 23   & 5.2$^{+0.5}_{-0.5}$ & $7.8^{+0.8}_{-0.8}$ & $3.0^{+0.3}_{-0.3}$  & 45.04             & 45.1/45.6                  &0.08$_{-0.04}^{+0.10}$ & 103\\
HS~0810+2554            & 2014 Oct 04   & 3.1$^{+0.1}_{-0.1}$ & $4.6^{+0.4}_{-0.4}$  & $3.6^{+0.3}_{-0.3}$  & 45.04             & 44.9/45.6                 &0.08$_{-0.04}^{+0.10}$ &103 \\
SDSS~J0921+2854     & 2018 May 15  & 6.8$^{+0.2}_{-0.2}$ & $6.2^{+0.2}_{-0.2}$  & $1.6^{+0.2}_{-0.2}$ &  44.80             &   46.7/45.4                     &0.03$_{-0.02}^{+0.04}$    & 4\\
SDSS~J0921+2854     & 2018 Oct 24   & 6.9$^{+0.2}_{-0.2}$ & $7.3^{+0.2}_{-0.2}$  & $1.3^{+0.2}_{-0.2}$ &  44.80             &   46.8/45.4                     &0.03$_{-0.02}^{+0.04}$    & 4\\
\enddata
\tablenotetext{a}{2--10~keV unabsorbed fluxes (units of 10$^{-13}$) not corrected for lensing magnification.}
\tablenotetext{b}{2--10~keV unabsorbed luminosities (units of 10$^{45}$) not corrected for lensing magnification.}
\tablenotetext{c}{13.6~eV--13.6~keV unabsorbed luminosities (units of 10$^{46}$) not corrected for lensing magnification.}
\tablenotetext{d}{The luminosity densities at 1450{\AA} are corrected for lensing magnification.}
\tablenotetext{e}{The bolometric luminosities are corrected for lensing magnification. The bolometric luminosities were calculated using two different methods, first from the 2$-$10~keV luminosities using X-ray bolometric correction factors and second from the luminosity density at 1450{\AA}. See text for details.}
\tablenotetext{f}{The ratios of bolometric to Eddington luminosities use the  bolometric luminosities derived from the observed optical/UV flux densities.}
\tablenotetext{g}{This bolometric luminosity is derived from the observed flux density at 3000 {\AA} (see Saturni et al. 2016) assuming a bolometric correction factor of 5 
(Richards et al. 2006). }
\tablenotetext{h}{This bolometric luminosity is derived from the observed flux density at 5100 {\AA} (see \cite{2007ApJ...661...19P}) assuming a bolometric correction factor of 9. }
\end{deluxetable*}

%% This command is needed to show the entire author+affilation list when
%% the collaboration and author truncation commands are used.  It has to
%% go at the end of the manuscript.
%\allauthors
%% Include this line if you are using the \added, \replaced, \deleted
%% commands to see a summary list of all changes at the end of the article.
%\listofchanges

%\clearpage
\begin{deluxetable*}{ccccc}
\tablecaption{Results from Fits  to the Spectra of  SDSS~J1442+4055, SDSS~J1029+2623,  and SDSS~J1529+1038. \label{tab:fit1}}
\tablecolumns{6}
\tablenum{4}
\tablewidth{0pt}
\tablehead{
\colhead{Model$^{a}$} & \colhead{Parameter$^{b}$} & \colhead{Fitted Values${}^{c}$} & \colhead{Fitted Values${}^{c}$} & \colhead{Fitted Values${}^{c}$}    \\
\colhead{}                     & \colhead{}                             & \colhead{SDSS~J1442}          &  \colhead{SDSS~J1029}             & \colhead{SDSS~J1529}             
}
\startdata
${1}$ &$\Gamma$          & 1.80$_{-0.12}^{+0.12}$    &  1.54$_{-0.06}^{+0.06}$                             &    2.0$_{-0.1}^{+0.1}$                                                  \\
&   $N_{\rm H}$               & ---                                     &  $<$ 0.4 $\times$10$^{22}$~cm$^{-2}$     &   ---     \\
&$\chi^2/{\nu}$                & 55.07/58                          & 60.1/82                                                       &  33.3/39                          \\
&$P(\chi^2/{\nu})$$^{d}$  & 5.8~$\times$~10$^{-1}$  & 9.7~$\times$~10$^{-1}$                             & 7.3~$\times$~10$^{-1}$                  \\  
 &  &                     &    &         \\
${2}$ &  $\Gamma$            & 1.80$_{-0.14}^{+0.08}$ & 1.49$_{-0.06}^{+0.06}$                              &    1.85$_{-0.15}^{+0.19}$      \\
  &   $N_{\rm H}$                 & --- & $<$ 0.4 $\times$10$^{22}$~cm$^{-2}$    &  ---   \\
  &  E$_{\rm 1}$                   & 9.14$_{-0.2}^{+0.2}$~keV & 13.3$_{-0.9}^{+0.7}$~keV                        &    9.0$_{-1.5}^{+1.5}$~keV              \\
  &  $\sigma_{\rm 1}$           & $ < $ 0.1~keV                   & 0.7$_{-0.7}^{+1.2}$~keV                          &    $ < $ 2.8~keV                                   \\
  &  EW$_{\rm 1}$                & $-$194$_{-160}^{+150}$~eV & $-$610$_{-245}^{+240}$~eV                  &  $-$1.8$_{-348}^{+404}$~keV           \\
  &  E$_{\rm 2}$                   & 11.3$_{-0.6}^{+0.6}$~keV  &  ---                                                            &      ---                  \\
  &  $\sigma_{\rm 2}$           & $ < $ 0.2~keV                     & ---                                                             &      ---             \\
  &  EW$_{\rm 2}$                & $-$546$_{-200}^{+220}$~eV & ---                                                              &   ---       \\
  &  E$_{\rm 3}$                   & 16.44$_{-0.2}^{+0.2}$~keV &  ---                                                              &    ---           \\
  &  $\sigma_{\rm 3}$           & $ < $ 0.1~keV                       & ---                                                               &   ---         \\
  &  EW$_{\rm 3}$                & $+$1.3$_{-0.5}^{+0.6}$~keV &  ---                                                              &    ---         \\
  & $\chi^2/{\nu}$                 & 45.00/54                                 &   53.8/80                                                      & 26.8/36                  \\
  &$P(\chi^2/{\nu})$$^{d}$   & 8.03~$\times$~10$^{-1}$         & 9.9~$\times$~10$^{-1}$                              &  8.7~$\times$~10$^{-1}$            \\
 %    & $F{}^{e}$                          &      3.6                                  &    5.08                                                           & 5.13  \\
 %   & $P_{F}{}^{f}$                        &       1$\times$10$^{-2}$     &     6.7$\times$10$^{-3}$                              &   1$\times$10$^{-2}$   \\
     & $F_{1}/F_{2}/F_{3}{}^{e}$                                         &  7.1/7.4/13.8                                                                                          &    5.08/--/--                                                          & 5.13/--/--  \\
    & $P_{\rm MC,1}/P_{\rm MC,2}/P_{\rm MC,3}{}^{f}$    &   1.2$\times$10$^{-2}$/8.0$\times$10$^{-3}$/1.0$\times$10$^{-3}$    &     6.7$\times$10$^{-3}$/--/--                              &   1$\times$10$^{-2}$/--/--   \\
  &  &                      &             & \\
  ${3}$  &$\Gamma$           & 1.79$_{-0.14}^{+0.14}$              & 1.47$_{-0.06}^{+0.07}$                                 &  1.90$_{-0.07}^{+0.07}$            \\
&   $N_{\rm H}$                  & ---                                                & $<$ 0.4$\times$10$^{22}$~cm$^{-2}$                       &  ---        \\
  &  E$_{\rm 1}$                  & 16.35$_{-0.2}^{+0.2}$~keV          &  ---                                                                               & ---                \\
  &  $\sigma_{\rm 1}$          & $ < $ 0.1~keV  & ---                                                                               &    ---              \\
  &  EW$_{\rm 1}$               & $+$1.5$_{-0.1}^{+0.1}$~keV  & ---                                                                               &    ---         \\
  &   $N_{\rm Habs1}$        & 3.5$_{-1.7}^{+1.5}$ $\times$ 10$^{23}$~cm$^{-2}$(68\%) & 7.4$_{-2.4}^{+2.6}$ $\times$ 10$^{23}$~cm$^{-2}$(68\%) &   3.2$_{-2}^{+5}$ $\times$ 10$^{23}$~cm$^{-2}$(68\%)        \\
  &    $\log\xi_{\rm abs1}$   & 3.2$_{-0.1}^{+0.1}$~erg~cm~s$^{-1}$(68\%)           & 2.8$_{-0.1}^{+0.1}$~erg~cm~s$^{-1}$(68\%)            &  3.6$_{-0.4}^{+0.4}$~erg~cm~s$^{-1}$(68\%)                                   \\
  & $v_{\rm turb,abs1}$       & 15,000 km~s$^{-1}$                                                 &   19,000 km~s$^{-1}$          &             36,000 km~s$^{-1}$                    \\
   &   $z_{\rm abs1}$           & 1.16$_{-0.10}^{+0.12}$                                             &  0.66$_{-0.04}^{+0.05}$                                             &  1.31$_{-0.2}^{+0.2}$              \\
&   $\chi^2/{\nu}$                & 36.79/56                                                                     & 55.13/79                                                              &  26.7/37                           \\
&  $P(\chi^2/{\nu})$$^{d}$  & 9.7~$\times$~10$^{-1}$                                              & 9.8~$\times$~10$^{-1}$                                             &   9.0~$\times$~10$^{-1}$              \\ 
 &  &                         &      &  \\
\enddata
\tablenotetext{a}{Model 1 consists of a power law and neutral absorption at the source. Model 2  consists of a power law, neutral absorption at the source, and Gaussian absorption and/or emission lines at the source.
Model 3 consists of a power law, neutral absorption at the source and one or two outflowing ionized absorbers at the source, and a number of emission lines.
All model fits include the Galactic absorption toward the source  (HI4PI Collaboration et al. 2016).}
\tablenotetext{b}{All absorption-line parameters are calculated for the rest frame.}
\tablenotetext{c}{Spectral fits were performed using the $\chi^{2}$ statistic and all errors are for 90\% confidence unless mentioned otherwise.}
\tablenotetext{d}{$P(\chi^2/{\nu})$ is the probability of exceeding $\chi^{2}$ for ${\nu}$ degrees of freedom if the model is correct.}
\tablenotetext{e}{$F_{\rm i}$-statistic between the null and alternative model for line component $i$. }
\tablenotetext{f}{$P_{\rm MC,i}$ is the probability of exceeding this $F_{\rm i}$-value for line component $i$ as determined from the Monte-Carlo simulations.}
\end{deluxetable*}

%\clearpage
\begin{deluxetable*}{ccccc}
\tablecaption{Results from Fits  to the Spectra of  SDSS~J1353+1138,  SDSS~J0904+1512, and Q~2237+0305. \label{tab:fit2}}
\tablecolumns{6}
\tablenum{5}
\tablewidth{0pt}
\tablehead{
\colhead{Model$^{a}$} & \colhead{Parameter$^{b}$} & \colhead{Fitted Values${}^{c}$} & \colhead{Fitted Values${}^{c}$} & \colhead{Fitted Values${}^{c}$}    \\
\colhead{}                     & \colhead{}                             & \colhead{SDSS~J1353}          &  \colhead{SDSS~J0904}            & \colhead{Q~2237}              
}
\startdata
${1}$ &$\Gamma$          &     2.16$_{-0.12}^{+0.15}$                                                        &      2.02$_{-0.14}^{+0.15}$                                                &  1.76 $_{-0.24}^{+0.26}$  \\
&   $N_{\rm H}$               &  0.61 $_{-0.26}^{+0.30}$$\times$10$^{22}$~cm$^{-2}$     &    0.25 $_{-0.25}^{+0.36}$$\times$10$^{22}$~cm$^{-2}$  & 1.42$_{-1.34}^{+1.55}$$\times$10$^{22}$~cm$^{-2}$   \\
&$\chi^2/{\nu}$                &  61.14/76                                                          &     49.6/75                                                                           &     42.16/33                                                                    \\
&$P(\chi^2/{\nu})$$^{d}$  &    8.9~$\times$~10$^{-1}$                                                       &    9.9~$\times$~10$^{-1}$                                                  &      1.3~$\times$~10$^{-1}$                                       \\  
 &  &                     &         &     \\
${2}$ &  $\Gamma$            &  2.12$_{-0.19}^{+0.16}$         &     1.96$_{-0.13}^{+0.16}$                                                  &    1.64 $_{-0.24}^{+0.26}$                                            \\
  &   $N_{\rm H}$                 &  0.56 $_{-0.27}^{+0.30}$$\times$10$^{22}$~cm$^{-2}$    &   0.16 $_{-0.16}^{+0.36}$$\times$10$^{22}$~cm$^{-2}$  &   1.51$_{-1.21}^{+1.50}$$\times$10$^{22}$~cm$^{-2}$\\
  &  E$_{\rm 1}$                   &   8.6$_{-0.3}^{+0.2}$~keV     &      9.0$_{-1.6}^{+0.7}$~keV                                              &    6.58$_{-0.09}^{+0.14}$~keV                                              \\
  &  $\sigma_{\rm 1}$           &  0.33$_{-0.33}^{+0.45}$~keV     &      $ < $ 1.5~keV                                                               &     $ < $ 0.2~keV                                                 \\
  &  EW$_{\rm 1}$                &  $+$1.6$_{-1.1}^{+1.1}$~keV     &     $-$650$_{-230}^{+290}$~eV                                          &    {+805}$_{-230}^{+260}$~eV                                                          \\
  &  E$_{\rm 2}$                   &   9.3$_{-0.6}^{+0.6}$~keV       &    ---                                                                                    &      7.3$_{-1.2}^{+1.2}$~keV                                                 \\
  &  $\sigma_{\rm 2}$           &  0.78 $_{-0.78}^{+1.4}$~keV           &  ---                                                                                    &    1.9 $_{-0.9}^{+1.7}$~keV                                                  \\
  &  EW$_{\rm 2}$                &   $-$947$_{-375}^{+462}$~eV                                                                 &     ---                                                                                  &   $-$1620$_{-460}^{+810}$~eV                                                  \\
  &  E$_{\rm 3}$                   &    ---                                                                  &      ---                                                                                  &  ---  \\
  &  $\sigma_{\rm 3}$           &    ---                                                                   &    ---                                                                                   &  --- \\
  &  EW$_{\rm 3}$                &   ---                                                                    &    ---                                                                                  &  ---\\
  & $\chi^2/{\nu}$                 &   52.02/71                                                                 &    43.8/72                                                                            &      25.61 /27                                                                                   \\
  &$P(\chi^2/{\nu})$$^{d}$   &   9.56~$\times$~10$^{-1}$                                                                   &   9.96~$\times$~10$^{-1}$                                               &       5.4~$\times$~10$^{-1}$                                                       \\
%    & $F{}^{e}$                      & 6.81                                                                    &     4.80 & 5.82 \\
%    & $P_{F}{}^{f}$              &  3.33$\times$10$^{-3}$      &     3.5$\times$10$^{-2}$  & 4$\times$10$^{-3}$ \\
    &  $F_{1}/F_{2}/F_{3}{}^{e}$                                         & 13.6/13.8/--                                                                    &     4.80/--/-- & 12.5 /12.5 /-- \\
    & $P_{\rm MC,1}/P_{\rm MC,2}/P_{\rm MC,3}{}^{f}$    &  3.0$\times$10$^{-3}$/1.0$\times$10$^{-2}$/--              &     3.5$\times$10$^{-2}$/--/--  & 4$\times$10$^{-3}$/2$\times$10$^{-3}$/-- \\ 
  &  &                                 && \\
  ${3}$  &$\Gamma$           & 2.10$_{-0.12}^{+0.11}$              &      1.97$_{-0.13}^{+0.14}$                                             &     1.59$_{-0.22}^{+0.25}$                          \\
&   $N_{\rm H}$                  & 0.56 $_{-0.25}^{+0.30}$$\times$10$^{22}$~cm$^{-2}$    &   0.15$_{-0.15}^{+0.15}$$\times$10$^{22}$~cm$^{-2}$    &      0.94 $_{-0.94}^{+0.14}$$\times$10$^{22}$~cm$^{-2}$      \\
  &  E$_{\rm 1}$                  &   8.5$_{-0.6}^{+0.5}$~keV   &  ---                                                                                 &    6.59$_{-0.09}^{+0.14}$~keV                                                      \\
  &  $\sigma_{\rm 1}$          &    $ < $ 0.1~keV                                                                &   ---                                                                              &    $ < $ 0.2~keV                               \\
  &  EW$_{\rm 1}$               &    $+$460$_{-380}^{+430}$~eV                                                                   &     ---                                                                              &     {+480}$_{-180}^{+190}$~eV                                          \\
  &   $N_{\rm Habs1}$        &     3.9$_{-2.3}^{+2.4}$ $\times$ 10$^{23}$~cm$^{-2}$(68\%)  &    3.3$_{-2.0}^{+3.0}$ $\times$ 10$^{23}$~cm$^{-2}$(68\%) &    1.0$_{-0.4}^{+0.4}$  $\times$ 10$^{24}$~cm$^{-2}$         \\
  &    $\log\xi_{\rm abs1}$   &   3.6$_{-0.1}^{+0.2}$~erg~cm~s$^{-1}$(68\%)  &   3.6$_{-0.1}^{+0.4}$~erg~cm~s$^{-1}$(68\%)              &    3.10$_{-0.3}^{+0.2}$~erg~cm~s$^{-1}$(68\%)                        \\
  & $v_{\rm turb,abs1}$       &  10,000 km~s$^{-1}$                                                                    &     11,000 km~s$^{-1}$                                                     &   30,000 km~s$^{-1}$                                                            \\
   &   $z_{\rm abs1}$           &   0.85$_{-0.05}^{+0.20}$                                                                   &    1.15$_{-0.12}^{+0.10}$                                                &   1.27$_{-0.12}^{+0.18}$                                                       \\
&   $\chi^2/{\nu}$                & 56.5/73                                              &   45.4/71                                                                          &    28.6/29                                                                            \\
&  $P(\chi^2/{\nu})$$^{d}$  &  9.24~$\times$~10$^{-1}$            &     9.92~$\times$~10$^{-1}$                                            &    4.9~$\times$~10$^{-1}$                                                    \\ 
 &  &                                && \\
\enddata
\tablenotetext{a}{Model 1 consists of a power law and neutral absorption at the source. Model 2  consists of a power law, neutral absorption at the source, and Gaussian absorption and/or emission lines at the source.
Model 3 consists of a power law, neutral absorption at the source and one or two outflowing ionized absorbers at the source, and a number of emission lines.
All model fits include the Galactic absorption toward the source (HI4PI Collaboration et al. 2016).}
\tablenotetext{b}{All absorption-line parameters are calculated for the rest frame.}
\tablenotetext{c}{Spectral fits were performed using the $\chi^{2}$ statistic and all errors are for 90\% confidence unless mentioned otherwise.}
\tablenotetext{d}{$P(\chi^2/{\nu})$ is the probability of exceeding $\chi^{2}$ for ${\nu}$ degrees of freedom if the model is correct.}
\tablenotetext{e}{$F_{\rm i}$-statistic between the null and alternative model for line component $i$. }
\tablenotetext{f}{$P_{\rm MC,i}$ is the probability of exceeding this $F_{\rm i}$-value for line component $i$ as determined from the Monte-Carlo simulations.}
\end{deluxetable*}

%\clearpage
\begin{deluxetable*}{ccccc}
\tablecaption{Results from Fits  to the X-ray Spectra of SDSS~J1128+2402 and SDSS~J0921+2854 \label{tab:fit3}}
\tablecolumns{5}
\tablenum{6}
\tablewidth{0pt}
\tablehead{
\colhead{Model$^{a}$} & \colhead{Parameter$^{b}$} & \colhead{Fitted Values${}^{c}$} & \colhead{Fitted Values${}^{c}$}  & \colhead{Fitted Values${}^{c}$}   \\
\colhead{}                     & \colhead{}                            &  \colhead{SDSS~J1128}             & \colhead{SDSS~J0921 (OBS1)}     & \colhead{SDSS~J0921 (OBS2)}  
}
\startdata
${1}$ &$\Gamma$            &   1.92$_{-0.08}^{+0.09}$               &  1.69$_{-0.07}^{+0.08}$                    &  1.72$_{-0.13}^{+0.13}$ \\
&   $N_{\rm H}$                 &     ---                                              & ---                                                      & --- \\
&$\chi^2/{\nu}$                  &        68.53/85                                 &    81.9/79 	                                  & 68.21/88 \\
&$P(\chi^2/{\nu})$$^{d}$   &      9.03~$\times$~10$^{-1}$         &  4.10~$\times$~10$^{-1}$                & 9.42~$\times$~10$^{-1}$ \\ 
 &  &                        &           \\
${2}$ &  $\Gamma$             &         1.88$_{-0.09}^{+0.1}$         &  1.63$_{-0.08}^{+0.08}$                    &   1.68$_{-0.13}^{+0.13}$\\
  &   $N_{\rm H}$                  &    ---                                                  & ---                                                      &   --- \\
  &  E$_{\rm 1}$                    &     11.7$_{-0.19}^{+0.15}$~keV        &   10.3$_{-0.05}^{+0.08}$~keV          &   10.45$_{-0.21}^{+0.31}$~keV\\
  &  $\sigma_{\rm 1}$            &   $ < $ 1~keV                                  &   $ < $ 0.1~keV                                &  $ < $ 0.1~keV \\
  &  EW$_{\rm 1}$                 &       $+$2.3$_{-0.5}^{+0.3}$~keV    &  $+$0.75$_{-0.31}^{+0.39}$~keV    &  $+$143$_{-50}^{+60}$~eV \\
  &  E$_{\rm 2}$                    &           12.4$_{-1.1}^{+1.2}$~keV    & 10.45$_{-0.42}^{+0.29}$~keV        &   12.56$_{-0.16}^{+0.13}$~keV  \\
  &  $\sigma_{\rm 2}$            &       $+$1.1$_{-0.5}^{+1.2}$~keV    &  $+$0.52$_{-0.47}^{+0.52}$~keV  &  $ < $ 0.1~keV  \\
  &  EW$_{\rm 2}$                 &      $-$880$_{-1280}^{+80}$~eV       &   $-$0.45$_{-0.19}^{+0.12}$~keV  &     $-$217$_{-120}^{+10}$~eV  \\
  &  E$_{\rm 3}$                    &            ---                                            &   ---   & 6.39$_{-0.99}^{+0.86}$~keV \\
  &  $\sigma_{\rm 3}$            &            ---                                            &   ---  &  $ < $ 0.1~keV \\
  &  EW$_{\rm 3}$                 &            ---                                            &   ---  &     $+$100$_{-60}^{+55}$~eV \\
  & $\chi^2/{\nu}$                  &    61.52/84                                          &   64.11/73 &  54.54/82 \\
  &$P(\chi^2/{\nu})$$^{d}$    &   9.69~$\times$~10$^{-1}$                  &  7.62~$\times$~10$^{-1}$    & 9.92~$\times$~10$^{-1}$ \\
%      & $F{}^{e}$                          &      6.52                                           &  4.54 & 4.02\\
 %   & $P_{F}{}^{f}$              &       6.67$\times$10$^{-3}$                      &  6.00$\times$10$^{-3}$  &  8.00$\times$10$^{-3}$ \\
          & $F_{1}/F_{2}/F_{3}{}^{e}$                                   &      10.4/11.3/--                                                                &  11.7/11.8/--                                                              & 10.8/8.4/8.2\\
    & $P_{\rm MC,1}/P_{\rm MC,2}/P_{\rm MC,3}{}^{f}$   &       5.3$\times$10$^{-3}$/8.0$\times$10$^{-3}$/--        &  8.0$\times$10$^{-3}$/9.0$\times$10$^{-3}$/--  &  1.5$\times$10$^{-3}$/ 6.0$\times$10$^{-3}$/9.3$\times$10$^{-3}$ \\
  &  &                                & \\
  ${3}$  &$\Gamma$            &      1.88$_{-0.04}^{+0.04}$                    &  1.58$_{-0.14}^{+0.14}$  &  1.63$_{-0.08}^{+0.09}$  \\
&   $N_{\rm H}$                   &      ---                                                        &   ---  & --- \\
  &  E$_{\rm 1}$                   &      11.4$_{-0.2}^{+0.1}$~keV                  &  10.3$_{-0.11}^{+0.09}$~keV  &  10.4$_{-0.18}^{+0.21}$~keV \\
  &  $\sigma_{\rm 1}$           &             $ < $ 0.1~keV                             &   $ < $ 0.1~keV &   $ < $ 0.1~keV \\
  &  EW$_{\rm 1}$                &      $+$1.2$_{-0.8}^{+0.9}$~keV             &  $+$315$_{-265}^{+280}$~eV &     $+$124$_{-120}^{+100}$~eV  \\
  &   $N_{\rm Habs1}$         &     2.3$_{-1.7}^{+1.3}$ $\times$ 10$^{23}$~cm$^{-2}$(68\%)        &   3.7$_{-1.5}^{+2.4}$ $\times$ 10$^{23}$~cm$^{-2}$(68\%)                    &  6$_{-3}^{+3}$ $\times$ 10$^{23}$~cm$^{-2}$(68\%)    \\
  &    $\log\xi_{\rm abs1}$    &      3.15$_{-0.15}^{+0.15}$~erg~cm~s$^{-1}$(68\%)                      &   3.1$_{-0.3}^{+0.2}$~erg~cm~s$^{-1}$(68\%) &   3.14$_{-0.04}^{+0.1}$~erg~cm~s$^{-1}$(68\%)  \\
  & $v_{\rm turb,abs1}$        &        9,000 km~s$^{-1}$                                                                   &  3,000 km~s$^{-1}$  &  6,000 km~s$^{-1}$ \\
   &   $z_{\rm abs1}$            &    0.39$_{-0.03}^{+0.03}$                                                                 &   0.56$_{-0.02}^{+0.03}$ &   0.33$_{-0.02}^{+0.02}$ \\
&   $\chi^2/{\nu}$                 &                  30.42/46                                                                           &   65.15/72 &   86.63/120 \\
&  $P(\chi^2/{\nu})$$^{d}$   &    9.63~$\times$~10$^{-1}$                                                                  &  7.03~$\times$~10$^{-1}$ &  9.90~$\times$~10$^{-1}$  \\ 
 &  &                                & \\
\enddata
\tablenotetext{a}{Model 1 consists of a power law and neutral absorption at the source. Model 2  consists of a power law, neutral absorption at the source, and Gaussian absorption and/or emission lines at the source. Model 3 consists of a power law, neutral absorption at the source and one or two outflowing ionized absorbers at the source, and a number of emission lines. All model fits include the Galactic absorption toward the source (HI4PI Collaboration et al. 2016).}
\tablenotetext{b}{All absorption-line parameters are calculated for the rest frame.}
\tablenotetext{c}{Spectral fits were performed using the $\chi^{2}$ statistic and all errors are for 90\% confidence unless mentioned otherwise.}
\tablenotetext{d}{$P(\chi^2/{\nu})$ is the probability of exceeding $\chi^{2}$ for ${\nu}$ degrees of freedom if the model is correct.}
\tablenotetext{e}{$F_{\rm i}$-statistic between the null and alternative model for line component $i$. }
\tablenotetext{f}{$P_{\rm MC,i}$ is the probability of exceeding this $F_{\rm i}$-value for line component $i$ as determined from the Monte-Carlo simulations.}
\end{deluxetable*}

\begin{deluxetable*}{llcrcccr}
\tablecaption{Results from Fits to the SDSS Spectra of the Subsample \label{tab:sdss1}}
\tablecolumns{8}
\tablenum{7}
\tablewidth{0pt}
\tablehead{
\colhead{Object} & \colhead{Plate-MJD-Fiber$^{a}$} & \colhead{Comp$^{b}$} & \colhead{$\lambda_{\rm CIV,B}$$^{c}$} & \colhead{FWHM$_{\rm CIV,B}$} & \colhead{$\lambda_{\rm CIV,R}$$^{c}$} & \colhead{FWHM$_{\rm CIV,R}$} & \colhead{$v_{\rm UV}$}  \\
\colhead{} & \colhead{} &  \colhead{} & \colhead{(\AA)} & \colhead{(km~s$^{-1}$)} & \colhead{(\AA)} & \colhead{(km~s$^{-1}$)} & \colhead{$(c)$} 
}
\startdata
SDSS~J1442+4055 &6061-56076-0132& 1&1482.68$\pm$0.09 & 227$\pm$45 & 1484.89$\pm$0.26 & 322$\pm$159& 0.0423$\pm$0.0001 \\
%SDSS~J1529+1038 &5493-56009-0900&1 &1511.13$\pm$0.18 & 620$\pm$97  & 1513.57$\pm$0.09 & 122$\pm$59 & 0.0239$\pm$0.0001\\
SDSS~J1529+1038 &5493-56009-0900&1 &1509.79$\pm$0.17 & 188$\pm$62  & 1512.33$\pm$0.17 & 188$\pm$62 & 0.0248$\pm$0.0001\\
SDSS~J1529+1038 &5493-56009-0900&2 &1511.06$\pm$0.17 & 203$\pm$63  & 1513.78$\pm$0.17 & 203$\pm$63 & 0.0239$\pm$0.0001\\
SDSS~J1529+1038 &5493-56009-0900&3 &1538.32$\pm$0.03 & 192$\pm$15  & 1540.95$\pm$0.04 & 201$\pm$22 & 0.0063$\pm$0.0001\\
SDSS~J0904+1512 &5295-55978-0976&1 &1430.86$\pm$0.07 & 244$\pm$37 & 1433.17$\pm$0.12 &  289$\pm$61 & 0.0758$\pm$0.0001\\
SDSS~J0904+1512 &5295-55978-0976&2 &1452.87$\pm$0.06 & 244$\pm$29 & 1455.04$\pm$0.06 &  306$\pm$29 & 0.0616$\pm$0.0001\\
SDSS~J0904+1512 &5295-55978-0976&3 &1516.10$\pm$0.10 & 269$\pm$51 & 1518.47$\pm$0.10 &  280$\pm$51 & 0.0207$\pm$0.0001\\
SDSS~J1353+1138 &1702-53144-0397& 1 &1513.26$\pm$0.04 & 284$\pm$18 & 1515.92$\pm$0.06 &  310$\pm$25 & 0.0226$\pm$0.0001\\
SDSS~J1128+2402 &2497-54154-0113& 1 &1515.71$\pm$0.15 & 230$\pm$73 & 1518.14$\pm$0.11 &  203$\pm$50 &0.0210$\pm$0.0001 \\
SDSS~J0921+2854 &2353-53794-0139& 1&1521.55$\pm$0.20 & 139$\pm$87 & 1523.69$\pm$0.18 & 125$\pm$77 & 0.0172$\pm$0.0001\\
\enddata
\tablenotetext{a}{Plate is the SDSS plug plate used to collect the spectrum, MJD is the Modified Julian Date of the night when the observation was carried out and Fiber is the fiber number used for the object.}
\tablenotetext{b}{Comp is the assigned number to an C~IV doublet component.}
\tablenotetext{c}{$\lambda_{CIV,B}$ and $\lambda_{CIV,R}$ are the fitted rest-frame wavelengths of the blue and red components of the C~IV doublet, respectively.}
\end{deluxetable*}
%\tablenotetext{a}{Plate is the SDSS plug plate used to collect the spectrum, MJD is the Modified Julian Date of the night when the observation was carried out
%and Fiber is the fiber number used for the object.}
%Plate is the SDSS plug plate used to collect the spectrum, MJD is the Modified Julian Date of the night when the observation was carried out
%and Fiber is the fiber number used for the object.
%$\lambda_{CIV,B}$ and $\lambda_{CIV,R}$ are the blue and red lines of the CIV doublet

%\clearpage
\begin{deluxetable*}{llcrcccc}
\tablecaption{Properties of Outflowing C~IV Absorbers of the Subsample \label{tab:sdss2}}
\tablecolumns{8}
\tablenum{8}
\tablewidth{0pt}
\tablehead{
\colhead{Object} & \colhead{Plate-MJD-Fiber} & \colhead{Comp} & \colhead{$W_{\rm CIV,B}$$^{a}$} & \colhead{$W_{\rm CIV,R}$$^{a}$} & \colhead{$C_{\rm f}$$^{b}$} & \colhead{$\tau$$^{c}$} & \colhead{$N_{\rm CIV}$$^{d}$}  \\
\colhead{} & \colhead{} &  \colhead{} & \colhead{(\AA)} & \colhead{(\AA)} & \colhead{} & \colhead{} & \colhead{($\times~10^{14}$~cm$^{-2})$} 
}
\startdata
SDSS~J1442+4055 &6061-56076-0132& 1&0.46$\pm$0.10  & 0.23$\pm$0.10 & --  & -- & -- \\
%SDSS~J1442+4055 &6061-56076-0132& 1&0.46$\pm$0.10  & 0.23$\pm$0.10 & 1  & 0.07$\pm$0.02& 0.3$\pm$0.1 \\
%SDSS~J1529+1038 &5493-56009-0900&1 &1.7$\pm$0.2 & 0.30$\pm$0.10  & 1  & 0.10$\pm$0.20 & --\\
SDSS~J1529+1038 &5493-56009-0900&1 &0.4$\pm$0.1 & 0.4$\pm$0.1  & --  & -- & --\\
SDSS~J1529+1038 &5493-56009-0900&2 &0.5$\pm$0.1 & 0.5$\pm$0.1  & 0.21$\pm$0.06  & 4$\pm$3 & 3$\pm$2\\
SDSS~J1529+1038 &5493-56009-0900&3 &1.07$\pm$0.08 & 0.83$\pm$0.14  & 0.36$\pm$0.03 & 1.20$\pm$0.40 & 1.3$\pm$0.3\\
SDSS~J0904+1512 &5295-55978-0976&1 &1.5$\pm$0.2 & 1.2$\pm$0.3 & 0.85$\pm$0.12 &  0.36$\pm$0.05 & 1.3$\pm$0.1\\
SDSS~J0904+1512 &5295-55978-0976&2 &1.0$\pm$0.2 & 1.2$\pm$0.4 & -- &  -- & --\\
SDSS~J0904+1512 &5295-55978-0976&3 &0.7$\pm$0.2 & 0.7$\pm$0.2 & 0.18$\pm$0.02 &  3.0$\pm$0.3 & 2.3$\pm$0.4\\
SDSS~J1353+1138 &1702-53144-0397& 1 &1.80$\pm$0.23 & 1.70$\pm$0.29 & 0.51$\pm$0.03 & 1.4$\pm$0.4 & 3.2$\pm$0.1\\
SDSS~J1128+2402 &2497-54154-0113& 1 &1.27$\pm$0.30 & 1.14$\pm$0.30 & 0.40$\pm$0.04 &  2.2$\pm$0.5 &2.9$\pm$0.2 \\
SDSS~J0921+2854 &2353-53794-0139& 1&0.93$\pm$0.04 & 0.96$\pm$0.03 & -- & -- & --\\
\enddata
\tablenotetext{a}{$W_{\rm CIV,B}$ and $W_{\rm CIV,R}$ are the observed-frame equivalent widths of the blue and red components of the C~IV doublet, respectively. }
\tablenotetext{b}{$C_{\rm f}$ is the coverage factor averaged over the central region of the C~IV absorption lines. 
In several cases the absorption depths are either very shallow in the C~IV doublet or the red component has a slightly larger depth than the blue one within the error bars. In these cases
the doublet method yields unphysical values for several velocity bins and we cannot provide estimates for $C_{\rm f}$, $\tau$ and $N_{\rm CIV}$.}
\tablenotetext{c}{$\tau$ is the optical depth averaged over the central region of the C~IV absorption lines.}
\tablenotetext{d}{$N_{\rm CIV}$ is the C~IV ionic column density calculated using equation \ref{eq:7}.}
\end{deluxetable*}

\begin{deluxetable*}{llrrrrrcrrr}
\tablecaption{Quasar Outflow Properties \label{tab:outflow}}
\tablecolumns{11}
\tablenum{9}
\tablewidth{0pt}
\tablehead{
\colhead{Object} & \colhead{Observation} & \colhead{$N_{\rm H}$$^{a}$} & \colhead{ $C_{\rm SR}$ }  & \colhead{$r_{\rm min}$$^{a}$} & \colhead{$r_{\rm max}/r_{\rm min}$ } &  \colhead{$\log\xi$$^{a}$} & \colhead{$v_{\rm wind}$} & \colhead{$\dot{M}$} & \colhead{ {$\dot{E}$}$_{\rm K}/L_{\rm Bol}$} & \colhead{$\dot{p}/(L_{\rm Bol}/c)$}  \\
\colhead{} & \colhead{Date} & \colhead{} &  &  &  &  \colhead{} & \colhead{($c$)} & \colhead{(M$_{\odot}~{\rm yr}^{-1}$)} & \colhead{} & \colhead{} 
}
\startdata
APM~08279+5255       & 2002 Feb 24 &  10$^{+6.0}_{-6.0}$   &1.44  &121$^{+43}_{-38}$& 24 &   $4.0^{+0.8}_{-0.1}$     &   $0.16^{+0.02}_{-0.02}$      & $101^{+59}_{-53}$          &  $0.11^{+0.07}_{-0.06}$     & $1.35^{+0.84}_{-0.72}$ \\
APM~08279+5255       & 2002 Feb 24 & 1.0$^{+0.8}_{-0.5}$ &2.21   &25$^{+7}_{-7}$ & 7,756 &   $3.2^{+0.2}_{-0.3}$   &     $0.35^{+0.03}_{-0.03}$        &  $5.0^{+3.2}_{-2.6}$         &  $0.026^{+0.018}_{-0.014}$     & $0.15^{+0.10}_{-0.08}$ \\
APM~08279+5255       & 2002 April 28& 1.7$^{+0.2}_{-0.2}$ &1.59   & 76$^{+24}_{-22}$&  2,926  &  $3.1^{+0.1}_{-0.7}$   &      $0.20^{+0.02}_{-0.02}$   &  $12.7^{+3.9}_{-3.6}$         &   $0.022^{+0.009}_{-0.007}$    & $0.22^{+0.08}_{-0.07}$ \\
APM~08279+5255       & 2002 April 28& 1.4$^{+0.4}_{-0.8}$   &2.77  & 14$^{+4}_{-4}$ &  7,208  &  $3.5^{+0.8}_{-0.4}$    &     $0.46^{+0.03}_{-0.03}$    & $4.6^{+2.3}_{-2.1}$          &  $0.04^{+0.02}_{-0.02}$     & $0.19^{+0.10}_{-0.09}$ \\
APM~08279+5255       & 2007 Oct  6& 3.9$^{+1.1}_{-0.9}$  &1.51  & 94$^{+32}_{-28}$ & 557 &   $3.6^{+0.9}_{-0.3}$   &     $0.18^{+0.02}_{-0.02}$    &  $33^{+13}_{-11}$         &  $0.043^{+0.017}_{-0.016}$     &  $0.61^{+0.42}_{-0.30}$\\
APM~08279+5255       & 2007 Oct  6& 8.3$^{+1.0}_{-1.0}$  & 3.39 & 10$^{+3}_{-3}$ & 1,190&    $3.9^{+0.2}_{-0.3}$   &     $0.54^{+0.03}_{-0.03}$    &  $23^{+7}_{-6}$         &  $0.28^{+0.08}_{-0.08}$     & $1.3^{+0.9}_{-0.6}$ \\
APM~08279+5255       & 2007 Oct  22& 1.2$^{+0.2}_{-0.3}$  &1.33  & 178$^{+51}_{-47}$ & 5,477 &   $3.1^{+0.1}_{-0.2}$   &     $0.13^{+0.02}_{-0.02}$  &  $14^{+4}_{-4}$       &  $0.013^{+0.011}_{-0.007}$     & $0.20^{+0.17}_{-0.11}$ \\
APM~08279+5255       & 2007 Oct  22& 1.0$^{+0.2}_{-0.2}$  & 2.34 &  21$^{+6}_{-6}$ & 53,891 &   $3.1^{+0.1}_{-0.1}$   &      $0.38^{+0.03}_{-0.03}$   &  $4^{+1}_{-1}$         &  $0.031^{+0.029}_{-0.017}$     & $0.16^{+0.15}_{-0.09}$ \\
APM~08279+5255       &2008 Jan  14 & 0.9$^{+0.4}_{-0.4}$  & 1.67 & 63$^{+2}_{-2}$ & 5,815 &   $3.3^{+0.7}_{-0.6}$   &      $0.22^{+0.02}_{-0.02}$   &  $6^{+3}_{-3}$         &  $0.016^{+0.012}_{-0.009}$     & $0.14^{+0.11}_{-0.08}$ \\
APM~08279+5255       &2008 Jan  14 & 4.0$^{+4.0}_{-3.2}$  &3.95   & 8.5$^{+2.1}_{-2.1}$ & 2,317 &   $3.9^{+0.5}_{-0.5}$   &      $0.59^{+0.03}_{-0.03}$   &  $13^{+8}_{-7}$       &  $0.23^{+0.20}_{-0.14}$     &  $0.77^{+0.62}_{-0.46}$\\
HS~1700+6416            & 2000 Oct 31&   5.6$^{+4.4}_{-0}$   & 2.34 & 3.1$^{+1.7}_{-1.5}$ & 55 &    $3.9^{+1.0}_{-0.4}$   &     $0.38^{+0.05}_{-0.05}$  &  $33^{+21}_{-18}$          &  $0.10^{+0.09}_{-0.07}$      & $0.54^{+0.41}_{-0.36}$\\
MG~J0414+0534        & 2017 Mar 11 &    9.8$^{+5.7}_{-3.6}$  & 1.91   &4.7$^{+2.9}_{-2.4}$   &206 &   $3.0^{+0.1}_{-0.1}$    & $0.28^{+0.05}_{-0.05}$      &  $6.2^{+4.2}_{-3.6}$         &  $0.49^{+0.40}_{-0.32}$   & $3.5^{+2.3}_{-2.3}$ \\
SDSS~J1442+4055     & 2019 Jan 18&   3.5$^{+1.5}_{-1.7}$ & 2.83 &6.6$^{+3.1}_{-2.9}$  &1,155 & $6.6^{+3.1}_{-2.9}$&   $0.47^{+0.03}_{-0.05}$               & $5.5^{+3.7}_{-3.2}$        &    $1.2^{+0.9}_{-0.8}$       & $5.1^{+4.0}_{-3.2}$\\
SDSS~J1029+2623    & 2010 Mar 11 &     7.4$^{+2.6}_{-2.4}$  &3.82  & 0.6$^{+0.2}_{-0.2}$ & 880 &  $2.8^{+0.1}_{-0.1}$           &$0.58^{+0.01}_{-0.02}$      & $0.9^{+0.6}_{-0.5}$         &   $4.3^{+2.9}_{-2.5}$      & $15^{+10}_{-8}$\\
SDSS~J1529+1038      & 2015 July 19&      3.2$^{+4.9}_{-1.9}$    & 1.79 &5.6$^{+4.5}_{-3.4}$  &62  &$3.6^{+0.4}_{-0.4}$          & $0.25^{+0.06}_{-0.08}$     & $2.0^{+1.6}_{-1.2}$        & $0.37^{+0.26}_{-0.26}$  & $3.0^{+2.3}_{-2.0}$\\
SDSS~J0904+1512      & 2015 Nov 8&   4.0$^{+2.3}_{-2.4}$   &1.83  & 9.1$^{+5.5}_{-4.5}$ &59 & $3.84^{+0.16}_{-0.19}$     &$0.26^{+0.05}_{-0.04}$       & $4.8^{+2.6}_{-2.9}$         &   $0.9^{+0.7}_{-0.6}$      &$6.9^{+5.8}_{-4.7}$ \\
PG~1115+080              & 2001 Nov 25&    1.9$^{+0.5}_{-0.5}$  &1.17  & 23$^{+12}_{-11}$ &63 & $3.23^{+0.07}_{-0.03}$       &$0.08^{+0.02}_{-0.02}$       & $1.7^{+1.0}_{-0.9}$        &   $0.045^{+0.039}_{-0.028}$      & $1.1^{+0.8}_{-0.6}$\\
PG~1115+080              & 2001 Nov 25&    1.6$^{+0.4}_{-0.3}$  &2.30 & 1.4$^{+0.5}_{-0.5}$ &1913 & $3.14^{+1.2}_{-0.03}$       & $0.37^{+0.01}_{-0.01}$      & $0.31^{+0.16}_{-0.14}$       &    $0.17^{+0.12}_{-0.11}$      &$0.9^{+0.2}_{-0.1}$ \\
Q~2237+0305               & 2017 Jan 4&   10$^{+4}_{-4}$         &1.51  & 30$^{+25}_{-24}$ &18 & $3.1^{+0.2}_{-0.3}$                &$0.18^{+0.06}_{-0.09}$       & $16^{+13}_{-10}$       &    $1.58^{+1.24}_{-1.10}$      &$17.5^{+10}_{-11}$ \\
SDSS~J1353+1138       & 2016  Jan 1&   3.9$^{+2.4}_{-2.3}$  &2.16 & 7.3$^{+4.1}_{-3.4}$ &213 & $3.6^{+0.2}_{-0.1}$       &$0.34^{+0.02}_{-0.09}$       & $11.2^{+3.7}_{-3.0}$       &    $0.18^{+0.24}_{-0.23}$      &$2.0^{+1.8}_{-1.4}$ \\
SDSS~J1128+2402      &  2018 May 28&    2.3$^{+1.7}_{-1.3}$ &3.60 & 0.5$^{+0.2}_{-0.2}$ & 1,949 & $3.15^{+0.15}_{-0.15}$       &$0.56^{+0.01}_{-0.02}$  & $0.35^{+0.25}_{-0.21}$       &    $0.41^{+0.34}_{-0.27}$       &$1.5^{+1.3}_{-1.0}$ \\
PID352                           &                      &   2.2$^{+3.1}_{-1.6}$  &1.36 & 12$^{+9}_{-7}$ && $2.9^{+1.47}_{-0.38}$&          $0.14^{+0.02}_{-0.06}$       &$1.9^{+1.7}_{-1.2}$       &    $0.14^{+0.11}_{-0.09}$      &1.95$^{+1.65}_{-1.23}$ \\
HS~0810+2554             &  2014 Oct 04      &   2.1$^{+1.0}_{-1.1}$  &1.26 & 11$^{+7}_{-5}$ & 1.5 & $5.0^{+1.0}_{-0.4}$&       $0.11^{+0.05}_{-0.03}$       & $1.2^{+0.9}_{-0.7}$      &     $0.15^{+0.14}_{-0.10}$      & $2.6^{+2.2}_{-1.9}$\\
HS~0810+2554             & 2014 Oct 04     &   1.4$^{+0.3}_{-0.5}$  &2.59 & 0.7$^{+0.3}_{-0.3}$ & 317 & $4.1^{+0.4}_{-0.2}$&        $0.43^{+0.04}_{-0.05}$      & $0.21^{+0.12}_{-0.11}$       &   $0.35^{+0.23}_{-0.22}$       & $1.7^{+1.1}_{-1.1}$\\
SDSS~J0921+2854      & 2018 May 15    &   3.7$^{+2.4}_{-1.5}$  & 2.49 & 1.3$^{+0.6}_{-0.6}$ &6,381  & $3.1^{+0.2}_{-0.3}$&      $0.41^{+0.02}_{-0.01}$        &  $1.1^{+0.8}_{-0.6}$      &    $2.6^{+2.4}_{-1.9}$       &13$^{+12}_{-9}$ \\
SDSS~J0921+2854      & 2018 Oct 24     &   6.0$^{+2.7}_{-2.4}$  & 3.30& 0.8$^{+0.3}_{-0.3}$ & 62 & $3.14^{+0.1}_{-0.04}$&   $0.53^{+0.01}_{-0.01}$        & $1.3^{+0.9}_{-0.7}$       &    $5.2^{+3.9}_{-3.7}$       & $20^{+13}_{-14}$\\
\enddata
%\tablenotetext{a}{2--10~keV unabsorbed luminosities not corrected for lensing magnification.}
\tablenotetext{a}{The hydrogen column density $N_{\rm H}$ has units of $\times$10$^{23}{\rm cm}^{-2}$, the minimum radius of the absorber has units of $\times$10$^{13}{\rm m}$, and the ionization parameter $\xi$ has units of erg~cm~s$^{-1}$.}
\tablecomments{The properties of the outflowing X-ray absorbing gas are obtained from the photoionization model XSTAR.  $\dot{M}$, {$\dot{E}$}$_{\rm K}$, and $\dot{p}$ are derived from equations 1, 2, and 3, respectively. 
To correct for relativistic effects proposed in Luminary et al. 2020, the values of $N_{\rm H}$,  $\dot{M}$, {$\dot{E}$}$_{\rm K}$, and $\dot{p}$ need to be multiplied by the factor $C_{\rm SR}$.
$N_{\rm H}$,  $\dot{M}$, {$\dot{E}$}$_{\rm K}$, and $\dot{p}$ have been calculated assuming that the distance between the outflowing absorber and the center of the black hole is $r_{\rm min}$ given by Equation \ref{rmin}.
For APM~08279+5255, PG~1115+080, and HS~0810+2554 we list the properties of the two detected outflowing absorbers per observation. For SDSS~J0921+2854 we list the outflow properties of one detected component in two observations.}
\end{deluxetable*}

%\clearpage
\begin{deluxetable*}{lcrr}
\tablecaption{Correlation Table of $v_{\rm wind}$ versus $L_{\rm Bol}$ \label{tab:cor}}
\tablecolumns{4}
\tablenum{10}
\tablewidth{0pt}
\tablehead{
\colhead{Samples\tablenotemark{a}}    & \colhead{$\tau$\tablenotemark{b}}    & \colhead{$P$\tablenotemark{c}}     & \colhead{$b$\tablenotemark{d}}    \\
%\colhead{}              & \colhead{}                         & \colhead{}                        & \colhead{}                                      
}
\startdata
%& & & \\ 
Tombesi                                    & 0.11         &  0.52                                     & 0.02 $\pm$ 0.06       \\ 
Gofford                                      & 0.45        &  8.9~$\times$~10$^{-3}$       & 0.29  $\pm$ 0.09     \\ 
high-$z$                                     & 0.10       &  0.55                                       & 0.04  $\pm$ 0.04    \\              
Tombesi + high-$z$                    & 0.50       &  8.9~$\times$~10$^{-6}$         & 0.13  $\pm$ 0.03    \\
Gofford + high-$z$                      &  0.55      &  2.0~$\times$~10$^{-6}$        &  0.26  $\pm$ 0.04   \\ 
Tombesi + Gofford                       &  0.25     &    2.9~$\times$~10$^{-2}$         &  0.15  $\pm$ 0.06\\ 
Tombesi + Gofford + high-$z$    &  0.51       &  6.0~$\times$~10$^{-8}$   & 0.20  $\pm$ 0.03   \\
\enddata
\tablenotetext{a}{high-$z$ represents the high-$z$ quasar sample of our study. Tombesi and Gofford are the low-$z$ AGN samples described in the text. A $+$ sign between samples indicates that the samples are combined.}
\tablenotetext{b}{$\tau$ is the Kendall rank correlation coefficient of the $v_{\rm wind}$ versus $L_{\rm Bol}$ data.}
\tablenotetext{c}{The null probability of the $v_{\rm wind}$ versus $L_{\rm Bol}$ correlation.}
\tablenotetext{d}{The best-fit  values of the power-law exponent $b$, where the $v_{\rm wind}$ vs. $L_{\rm Bol}$ samples are fit with the function $v_{wind}$ =  $aL_{\rm Bol}^{b}$. }
\end{deluxetable*}

%\clearpage
\begin{deluxetable*}{llrrrr}
\tablecaption{Flux Densities and $\alpha_{\rm ox}$ of the NAL Quasars of our Sample \label{tab:aox}}
\tablecolumns{6}
\tablenum{11}
\tablewidth{0pt}
\tablehead{
\colhead{Object}    & \colhead{Date\tablenotemark{a}}   & \colhead{$f_{\rm 2~keV}$\tablenotemark{b}}     & \colhead{$f_{\rm 2500~\AA}$\tablenotemark{c}}   & \colhead{$\alpha_{\rm ox}$ \tablenotemark{d}}  & \colhead{$\Delta\alpha_{\rm ox}$ \tablenotemark{e}}    \\                                    
%\colhead{}              & \colhead{}                         & \colhead{($\rm \times~10^{-31}~erg~s^{-1}~cm^{-2}~Hz^{-1}$)}                 & \colhead{($\rm \times~10^{-27}~erg~s^{-1}~cm^{-2}~Hz^{-1}$)}     & \colhead{}
}
\startdata
HS~1700+6416              &2000 Oct 31         &  4.34     & 19.44   & $-$1.79 $\pm$ 0.02  & 0.049  \\ 
SDSS~J1442+4055        &2019 Jan 18        &  6.70     &  4.66  &   $-$1.48 $\pm$ 0.01 & 0.014 \\ 
SDSS~J1029+2623        &2010 Mar 11        &  2.32     & 1.88   &  $-$1.50 $\pm$  0.01 & $-$0.076 \\              
SDSS~J1529+1038        &2015 July 19       &  1.42      & 3.25   &  $-$1.67 $\pm$  0.05 &  $-$0.127\\
SDSS~J0904+1512        &2015 Nov 8         &  2.48      &  4.23 &  $-$1.62 $\pm$ 0.02   & $-$0.087 \\ 
SDSS~J1353+1138        &2016 Jan 1         &   4.55     &   7.60 &  $-$1.62 $\pm$ 0.03  & 0.004 \\ 
SDSS~J1128+2402        &2018 May 28      &  1.84      & 2.53   &    $-$1.59 $\pm$ 0.04 & $-$0.074\\
HS~0810+2554              &2014 Oct 04       & 10.88      &  25.41  &  $-$1.68 $\pm$ 0.01 & $-$0.206\\
SDSS~J0921+2854        &2018 Oct 24      &  22.29     &  0.70   & $-$0.96 $\pm$ 0.01 & 0.481\\
\enddata
\tablenotetext{a}{Date of X-ray observation of NAL quasar}
\tablenotetext{b}{The 2~keV flux density in the rest frame corrected for Galactic absorption with units of $\rm \times~10^{-31}~erg~s^{-1}~cm^{-2}~Hz^{-1}$. For the lensed quasars, the 2~keV flux density has not been corrected for magnification. }
\tablenotetext{c}{The 2500~${\rm \AA}$ flux density in the rest frame corrected for Galactic absorption with units of $\rm \times~10^{-27}~erg~s^{-1}~cm^{-2}~Hz^{-1}$. For the lensed quasars, the 2500~${\rm \AA}$ flux density has not been corrected for magnification. }
\tablenotetext{d}{The optical-to-X-ray spectral index. For the lensed quasars, we have assumed that the lensing magnifications at  2~keV and 2500~${\rm \AA}$ are equal.}
\tablenotetext{e}{The difference between the observed value of $\alpha_{\rm ox}$  and the value expected based on the UV luminosity of the quasar (e.g., Lusso et al. 2010)}
\end{deluxetable*}


\newpage
\begin{thebibliography}{}
\bibitem[Abramowicz et al.(1991)]{1991ApJ...369..175A} Abramowicz, M.~A., Novikov, I.~D., \& Paczynski, B.\ 1991, \apj, 369, 175. doi:10.1086/169748
\bibitem[Arnaud(1996)]{1996ASPC..101...17A} Arnaud, K.~A.\ 1996, Astronomical Data Analysis Software and Systems V, 101, 17 
\bibitem[Assef et al.(2011)]{2011ApJ...742...93A} Assef, R.~J., Denney, K.~D., Kochanek, C.~S., et al.\ 2011, \apj, 742, 93. doi:10.1088/0004-637X/742/2/93
\bibitem[Astropy Collaboration et al.(2013)]{2013A&A...558A..33A} Astropy Collaboration, Robitaille, T.~P., Tollerud, E.~J., et al.\ 2013, \aap, 558, A33. doi:10.1051/0004-6361/201322068
\bibitem[Astropy Collaboration et al.(2018)]{2018AJ....156..123A} Astropy Collaboration, Price-Whelan, A.~M., Sip{\H{o}}cz, B.~M., et al.\ 2018, \aj, 156, 123. doi:10.3847/1538-3881/aabc4f
\bibitem[Bertola et al.(2020)]{2020A&A...638A.136B} Bertola, E., Dadina, M., Cappi, M., et al.\ 2020, \aap, 638, A136
\bibitem[Bischetti et al.(2019)]{2019A&A...628A.118B} Bischetti, M., Piconcelli, E., Feruglio, C., et al.\ 2019, \aap, 628, A118. doi:10.1051/0004-6361/201935524
\bibitem[Carniani et al.(2016)]{2016A&A...591A..28C} Carniani, S., Marconi, A., Maiolino, R., et al.\ 2016, \aap, 591, A28 
\bibitem[Bischetti et al.(2019)]{2019A&A...630A..59B} Bischetti, M., Maiolino, R., Carniani, S., et al.\ 2019, \aap, 630, A59. doi:10.1051/0004-6361/201833557
\bibitem[Brusa et al.(2015)]{2015A&A...578A..11B} Brusa, M., Feruglio, C., Cresci, G., et al.\ 2015, \aap, 578, A11. doi:10.1051/0004-6361/201425491
\bibitem[Cappi(2006)]{2006AN....327.1012C} Cappi, M.\ 2006, Astronomische Nachrichten, 327, 1012. doi:10.1002/asna.200610639
\bibitem[Chartas et al.(2002)]{2002ApJ...579..169C} Chartas, G., Brandt, W.~N., Gallagher, S.~C., \& Garmire, G.~P.\ 2002, \apj, 579, 169 
\bibitem[Chartas(2000)]{2000ApJ...531...81C} Chartas, G.\ 2000, \apj, 531, 81. doi:10.1086/308441
\bibitem[Chartas et al.(2003)]{2003ApJ...595...85C} Chartas, G., Brandt, W.~N., \& Gallagher, S.~C.\ 2003, \apj, 595, 85
\bibitem[Chartas et al.(2007)]{2007AJ....133.1849C} Chartas, G., Brandt, W.~N., Gallagher, S.~C., et al.\ 2007, \aj, 133, 1849
\bibitem[Chartas et al.(2009)]{2009NewAR..53..128C} Chartas, G., Charlton, J., Eracleous, M., et al.\ 2009, \nar, 53, 128. doi:10.1016/j.newar.2009.07.010
\bibitem[Chartas et al.(2009)]{2009ApJ...706..644C} Chartas, G., Saez, C., Brandt, W.~N., et al.\ 2009, \apj, 706, 644
\bibitem[Chartas et al.(2014)]{2014ApJ...783...57C} Chartas, G., Hamann, F., Eracleous, M., et al.\ 2014, \apj, 783, 57
\bibitem[Chartas et al.(2016)]{2016ApJ...824...53C} Chartas, G., Cappi, M., Hamann, F., et al.\ 2016, \apj, 824, 53
\bibitem[Chartas et al.(2017)]{2017ApJ...837...26C} Chartas, G., Krawczynski, H., Zalesky, L., et al.\ 2017, \apj, 837, 26. doi:10.3847/1538-4357/aa5d50
\bibitem[Chartas et al.(2020)]{2020MNRAS.496..598C} Chartas, G., Davidson, E., Brusa, M., et al.\ 2020, \mnras, 496, 598
\bibitem[Chartas \& Canas(2018)]{2018ApJ...867..103C} Chartas, G. \& Canas, M.~H.\ 2018, \apj, 867, 103. doi:10.3847/1538-4357/aae438
\bibitem[Coatman et al.(2017)]{2017MNRAS.465.2120C} Coatman, L., Hewett, P.~C., Banerji, M., et al.\ 2017, \mnras, 465, 2120. doi:10.1093/mnras/stw2797
\bibitem[Cresci et al.(2015)]{2015ApJ...799...82C} Cresci, G., Mainieri, V., Brusa, M., et al.\ 2015, \apj, 799, 82
\bibitem[Dadina et al.(2018)]{2018A&A...610L..13D} Dadina, M., Vignali, C., Cappi, M., et al.\ 2018, \aap, 610, L13
\bibitem[de La Calle P{\'e}rez et al.(2010)]{2010A&A...524A..50D} de La Calle P{\'e}rez, I., Longinotti, A.~L., Guainazzi, M., et al.\ 2010, \aap, 524, A50. doi:10.1051/0004-6361/200913798
\bibitem[Dickey \& Lockman(1990)]{1990ARA&A..28..215D} Dickey, J.~M. \& Lockman, F.~J.\ 1990, \araa, 28, 215. doi:10.1146/annurev.aa.28.090190.001243
\bibitem[Di Matteo et al.(2005)]{2005Natur.433..604D} Di Matteo, T., Springel, V., \& Hernquist, L.\ 2005, \nat, 433, 604 
\bibitem[Duras et al.(2020)]{2020A&A...636A..73D} Duras, F., Bongiorno, A., Ricci, F., et al.\ 2020, \aap, 636, A73. doi:10.1051/0004-6361/201936817
\bibitem[Dorodnitsyn(2009)]{2009MNRAS.393.1433D} Dorodnitsyn, A.~V.\ 2009, \mnras, 393, 1433. doi:10.1111/j.1365-2966.2008.14171.x
\bibitem[Ebrero et al. (2020)]{2020XMM-Newton..Users..Handbook} Ebrero, J.\ 2020, XMM-Newton Users Handbook, (ESA: XMM-Newton SOC)", Issue 2.18 
\bibitem[Everett(2007)]{2007Ap&SS.311..269E} Everett, J.~E.\ 2007, \apss, 311, 269. doi:10.1007/s10509-007-9536-2
\bibitem[Faucher-Gigu{\`e}re \& Quataert(2012)]{2012MNRAS.425..605F} Faucher-Gigu{\`e}re, C.-A., \& Quataert, E.\ 2012, \mnras, 425, 605 
\bibitem[Feruglio et al.(2015)]{2015A&A...583A..99F} Feruglio, C., Fiore, F., Carniani, S., et al.\ 2015, \aap, 583, A99 
\bibitem[Feruglio et al.(2017)]{2017A&A...608A..30F} Feruglio, C., Ferrara, A., Bischetti, M., et al.\ 2017, \aap, 608, A30 
\bibitem[Fitzpatrick \& Massa(1999)]{1999ApJ...525.1011F} Fitzpatrick, E.~L. \& Massa, D.\ 1999, \apj, 525, 1011. doi:10.1086/307944
\bibitem[Fruscione et al.(2006)]{2006SPIE.6270E..1VF} Fruscione, A., McDowell, J.~C., Allen, G.~E., et al.\ 2006, \procspie, 6270, 62701V. doi:10.1117/12.671760
\bibitem[Fukumura et al.(2010)]{2010ApJ...715..636F} Fukumura, K., Kazanas, D., Contopoulos, I., et al.\ 2010, \apj, 715, 636. doi:10.1088/0004-637X/715/1/636
\bibitem[Fukumura et al.(2014)]{2014ApJ...780..120F} Fukumura, K., Tombesi, F., Kazanas, D., et al.\ 2014, \apj, 780, 120. doi:10.1088/0004-637X/780/2/120
\bibitem[Fukumura et al.(2015)]{2015ApJ...805...17F} Fukumura, K., Tombesi, F., Kazanas, D., et al.\ 2015, \apj, 805, 17. doi:10.1088/0004-637X/805/1/17
 \bibitem[Fukumura et al.(2018)]{2018ApJ...864L..27F} Fukumura, K., Kazanas, D., Shrader, C., et al.\ 2018, \apjl, 864, L27. doi:10.3847/2041-8213/aadd10
 \bibitem[Fukumura \& Tombesi(2019)]{2019ApJ...885L..38F} Fukumura, K. \& Tombesi, F.\ 2019, \apjl, 885, L38. doi:10.3847/2041-8213/ab5193
\bibitem[Gabriel et al.(2004)]{2004ASPC..314..759G} Gabriel, C., Denby, M., Fyfe, D.~J., et al.\ 2004, Astronomical Data Analysis Software and Systems (ADASS) XIII, 314, 759
\bibitem[Gibson et al.(2009)]{2009ApJ...696..924G} Gibson, R.~R., Brandt, W.~N., Gallagher, S.~C., et al.\ 2009, \apj, 696, 924. doi:10.1088/0004-637X/696/1/924
\bibitem[Giustini et al.(2011)]{2011A&A...536A..49G} Giustini, M., Cappi, M., Chartas, G., et al.\ 2011, \aap, 536, A49
\bibitem[Gaspari et al.(2020)]{2020NatAs...4...10G} Gaspari, M., Tombesi, F., \& Cappi, M.\ 2020, Nature Astronomy, 4, 10. doi:10.1038/s41550-019-0970-1
\bibitem[Gaspari et al.(2019)]{2019ApJ...884..169G} Gaspari, M., Eckert, D., Ettori, S., et al.\ 2019, \apj, 884, 169. doi:10.3847/1538-4357/ab3c5d
%\bibitem[Gaspari \& S{\k{a}}dowski(2017)]{2017ApJ...837..149G} Gaspari, M. \& S{\k{a}}dowski, A.\ 2017, \apj, 837, 149. doi:10.3847/1538-4357/aa61a3
\bibitem[Gaspari \& S{{a}}dowski(2017)]{2017ApJ...837..149G} Gaspari, M. \& S{{a}}dowski, A.\ 2017, \apj, 837, 149. doi:10.3847/1538-4357/aa61a3
\bibitem[Gofford et al.(2013)]{2013MNRAS.430...60G} Gofford, J., Reeves, J.~N., Tombesi, F., et al.\ 2013, \mnras, 430, 60 
\bibitem[Gofford et al.(2015)]{2015MNRAS.451.4169G} Gofford, J., Reeves, J.~N., McLaughlin, D.~E., et al.\ 2015, \mnras, 451, 4169. doi:10.1093/mnras/stv1207
\bibitem[Goodrich(1997)]{1997ApJ...474..606G} Goodrich, R.~W.\ 1997, \apj, 474, 606. doi:10.1086/303481
\bibitem[Hamann et al.(1997)]{1997ApJ...478...80H} Hamann, F., Barlow, T.~A., Junkkarinen, V., et al.\ 1997, \apj, 478, 80. doi:10.1086/303781
\bibitem[Hamann \& Sabra(2004)]{2004ASPC..311..203H} Hamann, F. \& Sabra, B.\ 2004, AGN Physics with the Sloan Digital Sky Survey, 311, 203
\bibitem[Hamann et al.(2011)]{2011MNRAS.410.1957H} Hamann, F., Kanekar, N., Prochaska, J.~X., et al.\ 2011, \mnras, 410, 1957. doi:10.1111/j.1365-2966.2010.17575.x
\bibitem[Hamann et al.(2013)]{2013MNRAS.435..133H} Hamann, F., Chartas, G., McGraw, S., et al.\ 2013, \mnras, 435, 133 
\bibitem[Hamann et al.(2019)]{2019MNRAS.483.1808H} Hamann, F., Herbst, H., Paris, I., et al.\ 2019, \mnras, 483, 1808. doi:10.1093/mnras/sty2900
\bibitem[Harris et al.(2020)]{2020Natur.585..357H} Harris, C.~R., Millman, K.~J., van der Walt, S.~J., et al.\ 2020, \nat, 585, 357. doi:10.1038/s41586-020-2649-2
\bibitem[HI4PI Collaboration et al.(2016)]{2016A&A...594A.116H} HI4PI Collaboration, Ben Bekhti, N., Floer, L., et al.\ 2016, \aap, 594, A116. doi:10.1051/0004-6361/201629178
\bibitem[Hopkins \& Elvis(2010)]{2010MNRAS.401....7H} Hopkins, P.~F., \& Elvis, M.\ 2010, \mnras, 401, 7 
\bibitem[Hopkins et al.(2016)]{2016MNRAS.458..816H} Hopkins, P.~F., Torrey, P., Faucher-Gigu{\`e}re, C.-A., et al.\ 2016, \mnras, 458, 816. doi:10.1093/mnras/stw289
\bibitem[Inada et al.(2012)]{2012AJ....143..119I} Inada, N., Oguri, M., Shin, M.-S., et al.\ 2012, \aj, 143, 119. doi:10.1088/0004-6256/143/5/119
\bibitem[Inada et al.(2014)]{2014AJ....147..153I} Inada, N., Oguri, M., Rusu, C.~E., et al.\ 2014, \aj, 147, 153. doi:10.1088/0004-6256/147/6/153
\bibitem[Inoue et al.(2007)]{2007ApJ...662..860I} Inoue, H., Terashima, Y., \& Ho, L.~C.\ 2007, \apj, 662, 860. doi:10.1086/517995
\bibitem[Itoh et al.(2020)]{2020MNRAS.499.3094I} Itoh, D., Misawa, T., Horiuchi, T., et al.\ 2020, \mnras, 499, 3094. doi:10.1093/mnras/staa2793
\bibitem[Kakkad et al.(2017)]{2017MNRAS.468.4205K} Kakkad, D., Mainieri, V., Brusa, M., et al.\ 2017, \mnras, 468, 4205. doi:10.1093/mnras/stx726
\bibitem[Kallman et al.(1996)]{1996ApJ...465..994K} Kallman, T.~R., Liedahl, D., Osterheld, A., et al.\ 1996, \apj, 465, 994. doi:10.1086/177485
\bibitem[Kallman \& Bautista(2001)]{2001ApJS..133..221K} Kallman, T., \& Bautista, M.\ 2001, \apjs, 133, 221
\bibitem[King(2010)]{2010MNRAS.402.1516K} King, A.~R.\ 2010, \mnras, 402, 1516. doi:10.1111/j.1365-2966.2009.16013.x 
\bibitem[King \& Pounds(2015)]{2015ARA&A..53..115K} King, A., \& Pounds, K.\ 2015, \araa, 53, 115 
\bibitem[Konigl \& Kartje(1994)]{1994ApJ...434..446K} Konigl, A. \& Kartje, J.~F.\ 1994, \apj, 434, 446. doi:10.1086/174746
\bibitem[Krawczyk et al.(2013)]{2013ApJS..206....4K} Krawczyk, C.~M., Richards, G.~T., Mehta, S.~S., et al.\ 2013, \apjs, 206, 4. doi:10.1088/0067-0049/206/1/4
\bibitem[Laha et al.(2014)]{2014MNRAS.441.2613L} Laha, S., Guainazzi, M., Dewangan, G.~C., et al.\ 2014, \mnras, 441, 2613. doi:10.1093/mnras/stu669
\bibitem[Laha et al.(2016)]{2016MNRAS.457.3896L} Laha, S., Guainazzi, M., Chakravorty, S., et al.\ 2016, \mnras, 457, 3896. doi:10.1093/mnras/stw211
\bibitem[Lamers \& Cassinelli(1999)]{1999isw..book.....L} Lamers, H.~J.~G.~L.~M. \& Cassinelli, J.~P.\ 1999, Introduction to Stellar Winds, by Henny J. G. L. M. Lamers and Joseph P. Cassinelli, pp. 452. ISBN 0521593980. Cambridge, UK: Cambridge University Press, June 1999., 452
\bibitem[Lanzuisi et al.(2012)]{2012A&A...544A...2L} Lanzuisi, G., Giustini, M., Cappi, M., et al.\ 2012, \aap, 544, A2 
\bibitem[Luminari et al.(2020)]{2020A&A...633A..55L} Luminari, A., Tombesi, F., Piconcelli, E., et al.\ 2020, \aap, 633, A55. doi:10.1051/0004-6361/201936797
\bibitem[Lusso et al.(2010)]{2010A&A...512A..34L} Lusso, E., Comastri, A., Vignali, C., et al.\ 2010, \aap, 512, A34. doi:10.1051/0004-6361/200913298
\bibitem[Lusso et al.(2012)]{2012MNRAS.425..623L} Lusso, E., Comastri, A., Simmons, B.~D., et al.\ 2012, \mnras, 425, 623. doi:10.1111/j.1365-2966.2012.21513.x
\bibitem[Madau \& Dickinson(2014)]{2014ARA&A..52..415M} Madau, P. \& Dickinson, M.\ 2014, \araa, 52, 415. doi:10.1146/annurev-astro-081811-125615
\bibitem[Matzeu et al.(2017)]{2017MNRAS.472L..15M} Matzeu, G.~A., Reeves, J.~N., Braito, V., et al.\ 2017, \mnras, 472, L15 
\bibitem[Misawa et al.(2007)]{2007ApJS..171....1M} Misawa, T., Charlton, J.~C., Eracleous, M., et al.\ 2007, \apjs, 171, 1. doi:10.1086/513713
\bibitem[Moravec et al.(2017)]{2017MNRAS.468.4539M} Moravec, E.~A., Hamann, F., Capellupo, D.~M., et al.\ 2017, \mnras, 468, 4539. doi:10.1093/mnras/stx775
\bibitem[More et al.(2016)]{2016MNRAS.456.1595M} More, A., Oguri, M., Kayo, I., et al.\ 2016, \mnras, 456, 1595. doi:10.1093/mnras/stv2813
\bibitem[Murray et al.(1995)]{1995ApJ...451..498M} Murray, N., Chiang, J., Grossman, S.~A., et al.\ 1995, \apj, 451, 498. doi:10.1086/176238
\bibitem[Nardini et al.(2015)]{2015Sci...347..860N} Nardini, E., Reeves, J.~N., Gofford, J., et al.\ 2015, Science, 347, 860 
\bibitem[O'Dowd et al.(2011)]{2011MNRAS.415.1985O} O'Dowd, M., Bate, N.~F., Webster, R.~L., et al.\ 2011, \mnras, 415, 1985. doi:10.1111/j.1365-2966.2010.18119.x
\bibitem[Oguri(2010)]{2010ascl.soft10012O} Oguri, M.\ 2010, glafic: Software Package for Analyzing Gravitational Lensing, ascl:1010.012
\bibitem[Ota et al.(2012)]{2012ApJ...758...26O} Ota, N., Oguri, M., Dai, X., et al.\ 2012, \apj, 758, 26
\bibitem[Planck Collaboration et al.(2016)]{2016A&A...594A..13P} Planck Collaboration, Ade, P.~A.~R., Aghanim, N., et al.\ 2016, \aap, 594, A13 
\bibitem[Pooley et al.(2007)]{2007ApJ...661...19P} Pooley, D., Blackburne, J.~A., Rappaport, S., et al.\ 2007, \apj, 661, 19. doi:10.1086/512115
\bibitem[Pounds et al.(2003)]{2003MNRAS.345..705P} Pounds, K.~A., Reeves, J.~N., King, A.~R., et al.\ 2003, \mnras, 345, 705 
\bibitem[Proga et al.(2000)]{2000ApJ...543..686P} Proga, D., Stone, J.~M., \& Kallman, T.~R.\ 2000, \apj, 543, 686. doi:10.1086/317154
\bibitem[Proga \& Kallman(2004)]{2004ApJ...616..688P} Proga, D. \& Kallman, T.~R.\ 2004, \apj, 616, 688. doi:10.1086/425117
\bibitem[Protassov et al.(2002)]{2002ApJ...571..545P} Protassov, R., van Dyk, D.~A., Connors, A., Kashyap, V.~L., \& Siemiginowska, A.\ 2002, \apj, 571, 545 
\bibitem[Reeves et al.(2003)]{2003ApJ...593L..65R} Reeves, J.~N., O'Brien, P.~T., \& Ward, M.~J.\ 2003, \apjl, 593, L65 
\bibitem[Reeves et al.(2018)]{2018ApJ...854L...8R} Reeves, J.~N., Braito, V., Nardini, E., et al.\ 2018, \apjl, 854, L8 
\bibitem[Runnoe et al.(2012)]{2012MNRAS.422..478R} Runnoe, J.~C., Brotherton, M.~S., \& Shang, Z.\ 2012, \mnras, 422, 478
\bibitem[Saez \& Chartas(2011)]{2011ApJ...737...91S} Saez, C., \& Chartas, G.\ 2011, \apj, 737, 91 
\bibitem[S{{a}}dowski \& Gaspari(2017)]{2017MNRAS.468.1398S} S{{a}}dowski, A. \& Gaspari, M.\ 2017, \mnras, 468, 1398. doi:10.1093/mnras/stx543
\bibitem[Saturni et al.(2016)]{2016A&A...587A..43S} Saturni, F.~G., Trevese, D., Vagnetti, F., et al.\ 2016, \aap, 587, A43. doi:10.1051/0004-6361/201527152
\bibitem[Savage \& Sembach(1991)]{1991ApJ...379..245S} Savage, B.~D. \& Sembach, K.~R.\ 1991, \apj, 379, 245. doi:10.1086/170498
\bibitem[Schlafly \& Finkbeiner(2011)]{2011ApJ...737..103S} Schlafly, E.~F. \& Finkbeiner, D.~P.\ 2011, \apj, 737, 103. doi:10.1088/0004-637X/737/2/103 
\bibitem[Schurch \& Done(2007)]{2007MNRAS.381.1413S} Schurch, N.~J. \& Done, C.\ 2007, \mnras, 381, 1413. doi:10.1111/j.1365-2966.2007.12336.x
\bibitem[Silk \& Rees(1998)]{1998A&A...331L...1S} Silk, J. \& Rees, M.~J.\ 1998, \aap, 331, L1
\bibitem[Sim et al.(2010)]{2010MNRAS.404.1369S} Sim, S.~A., Miller, L., Long, K.~S., et al.\ 2010, \mnras, 404, 1369. doi:10.1111/j.1365-2966.2010.16396.x
\bibitem[Sim et al.(2012)]{2012MNRAS.426.2859S} Sim, S.~A., Proga, D., Kurosawa, R., et al.\ 2012, \mnras, 426, 2859. doi:10.1111/j.1365-2966.2012.21816.x
\bibitem[Stone \& Richards(2019)]{2019MNRAS.488.5916S} Stone, R.~B. \& Richards, G.~T.\ 2019, \mnras, 488, 5916. doi:10.1093/mnras/stz2111
\bibitem[Str{\"u}der et al.(2001)]{2001A&A...365L..18S} Str{\"u}der, L., Briel, U., Dennerl, K., et al.\ 2001, \aap, 365, L18 
\bibitem[Tarter \& Salpeter(1969)]{1969ApJ...156..953T} Tarter, C.~B. \& Salpeter, E.~E.\ 1969, \apj, 156, 953. doi:10.1086/150027
\bibitem[Tombesi et al.(2010)]{2010A&A...521A..57T} Tombesi, F., Cappi, M., Reeves, J.~N., et al.\ 2010, \aap, 521, A57
\bibitem[Tombesi et al.(2012)]{2012MNRAS.422L...1T} Tombesi, F., Cappi, M., Reeves, J.~N., et al.\ 2012, \mnras, 422, L1. doi:10.1111/j.1745-3933.2012.01221.x
\bibitem[Tombesi et al.(2015)]{2015Natur.519..436T} Tombesi, F., Mel{\'e}ndez, M., Veilleux, S., et al.\ 2015, \nat, 519, 436
\bibitem[Turner et al.(2001)]{2001A&A...365L..27T} Turner, M.~J.~L., Abbey, A., Arnaud, M., et al.\ 2001, \aap, 365, L27. doi:10.1051/0004-6361:20000087
\bibitem[Wagner et al.(2013)]{2013ApJ...763L..18W} Wagner, A.~Y., Umemura, M., \& Bicknell, G.~V.\ 2013, \apjl, 763, L18. doi:10.1088/2041-8205/763/1/L18
\bibitem[Woo et al.(2017)]{2017ApJ...839..120W} Woo, J.-H., Son, D., \& Bae, H.-J.\ 2017, \apj, 839, 120
\bibitem[Zubovas \& King(2012)]{2012ApJ...745L..34Z} Zubovas, K., \& King, A.\ 2012, \apjl, 745, L34 
\bibitem[Veilleux et al.(2017)]{2017ApJ...843...18V} Veilleux, S., Bolatto, A., Tombesi, F., et al.\ 2017, \apj, 843, 18 
\bibitem[Vietri et al.(2018)]{2018A&A...617A..81V} Vietri, G., Piconcelli, E., Bischetti, M., et al.\ 2018, \aap, 617, A81. doi:10.1051/0004-6361/201732335
\bibitem[Vignali et al.(2015)]{2015A&A...583A.141V} Vignali, C., Iwasawa, K., Comastri, A., et al.\ 2015, \aap, 583, A141
\bibitem[Virtanen et al.(2020)]{2020NatMe..17..261V} Virtanen, P., Gommers, R., Oliphant, T.~E., et al.\ 2020, Nature Methods, 17, 261. doi:10.1038/s41592-019-0686-2
\end{thebibliography}
\end{document}